\documentclass{aa}  

\usepackage{multicol}
\usepackage{graphicx}
\usepackage{txfonts}
\usepackage{xcolor}
\usepackage{caption}
\usepackage{comment}
\usepackage{multicol}  

\newcommand\ppre{\mathrm{pre}}
\newcommand\post{\mathrm{post}}
\newcommand\Pm{\mathcal{P}_m} 
\newcommand{\change}[1]{#1} 
\begin{document}

   \title{Probing the nature of dissipation in compressible MHD turbulence}


   \author{Thibaud RICHARD
          \inst{1}
          \and
          Pierre LESAFFRE\inst{1}
          \and
          Edith FALGARONE\inst{1}
          \and
          Andrew LEHMANN\inst{1}
          }

   \institute{Laboratoire de Physique de l’Ecole Normale Sup\'{e}rieure, ENS, Universit\'{e} PSL, CNRS, Sorbonne Universit\'{e}, Universit\'{e} de Paris, Paris, France\\
              \email{thibaud.richard@phys.ens.fr}
             }


 
  \abstract 
  {An essential facet of turbulence is the space-time intermittency of the cascade of energy that leads to coherent structures of high dissipation. 
} 
  {In this work, we attempt to investigate systematically the physical nature of the intense dissipation regions in decaying isothermal magnetohydrodynamical (MHD) turbulence.}
  {We probe the turbulent dissipation with grid based simulations of compressible isothermal decaying MHD turbulence. We take unprecedented care at resolving and controlling dissipation: we design methods to locally recover the dissipation due to the numerical scheme. We locally investigate the geometry of the gradients of the fluid state variables. We develop a method to assess the physical nature of the largest gradients in simulations and to estimate  their travelling velocity. Finally we investigate their statistics.}
  {We find that intense dissipation regions mainly correspond to sheets: locally, density, velocity and magnetic fields vary primarily across one direction. 
We identify these highly dissipative regions as fast/slow shocks or Alfvén discontinuities (Parker sheets or rotational discontinuities). On these structures, we find the main deviation from 1D planar steady-state is mass loss in the plane of the structure. We investigate the effect of initial conditions which yield different imprints at early time on the relative distributions between these four categories. However, these differences fade out after about one turnover time, when they become dominated by weakly compressible Alfvén discontinuities. We show that the magnetic Prandtl number has little influence on the statistics of these discontinuities, but it controls the Ohmic vs viscous heating rates within them.  Finally, we find the entrance characteristics of the structures (such as entrance velocity and magnetic pressure) are strongly correlated.} 
{These new methods allow to consider developed compressible turbulence as a statistical collection of intense dissipation structures. This can be used to post-process 3D turbulence with detailed 1D models apt for comparison with observations. It could also reveal useful as a framework to formulate new dynamical properties of turbulence.}


   \maketitle
%

\section{Introduction}

Gravity drives the evolution of the universe, but the gas dissipative dynamics is a central, yet unsolved, issue in the theories of galaxy and star formation \citep[e.g.][]{WhiteRees1978}. An emergent scenario is that a large fraction of the gas internal energy is stored and eventually dissipated in turbulent motions of the coldest phases instead of being radiated away, and therefore lost, by the warmest phases \citep[e.g.][]{Guillard2012,Appleton2013,Falgarone2017}. Turbulence however adds a colossal level of complexity to the gas dynamics because cosmic turbulence is supersonic, involves magnetic fields, exhibits plasma facets, and pervades all the thermal phases. Moreover its dissipation is known to occur in bursts localized in time and space, i.e. the space-time intermittency of turbulence \citep{LandauLifschitz1959,Kolmogorov1962,MeneveauSreenivasan1991}.

A valuable and unexpected guidance in the investigation of the intermittent dissipation of interstellar turbulence is provided by a number of molecular observations, including the existence in the cold neutral medium (CNM) of specific molecules that require large inputs of supra-thermal energy to form \citep{Nehme2008,Godard2012} and of molecules more excited than what an equilibrium at the ambient temperature would predict \citep{Falgarone2005,Gry2002,Ingalls2011}. 
The mere existence of large amounts of CO molecules surviving in irradiated diffuse media requires a formation route which is not controlled only by photons and cosmic rays \citep{Levrier2012}. This is in line with the large observed abundances of HCO$^+$ in diffuse gas \citep{LucasLiszt1996,LisztLucas1998} that is now recognized observationally as a signature of supra-thermal chemistry \citep{Gerin2021}.

Supra-thermal chemistry can be driven by several processes that do not appeal to turbulent dissipation bursts, such as 
the ion-neutral drift in Alfv\'en waves \citep{Federman1996}, conduction at interfaces between the warm (WNM) and cold neutral medium \citep{Lesaffre2007}, transport between the WNM and CNM \citep{Valdivia2017}. These latter processes tap the reservoir of thermal energy of the WNM and are able to drive a warm chemistry in the CNM but they fall short of reproducing the observed abundances of molecules with highly endothermic formation.

The channels linked to dissipation bursts, such as the ion-neutral drift in C-type shocks \citep{Flower1985,Flower1998,DraineKatz1986,Lesaffre2013} and in magnetised vortices \citep{Godard2009,Godard2014}, dissipative heating in shear layers  \citep{Falgarone1995,Joulain1998}, shock heating and compression \citep{CHEMSES} tap the mechanical energy reservoir of the CNM, which is roughly of the same magnitude as the thermal energy reservoir of the WNM. However, they
are naturally more successful because they can be much more concentrated in space, thus leading to potentially very strong effective temperature bursts. Out-of-equilibrium chemical and excitation signatures have been modelled for 
all these channels, related to specific localised structures where turbulent dissipation is enhanced. 
This detailed modelling is hard to reconcile with a coherent description of the energy cascade from the large scales of turbulence down to the dissipation scales, including
intermittency. It has been attempted for the first time by chemical post-processing of state-of-the-art numerical simulations of MHD turbulence, including ion-neutral drift  \citep{Myers2015,Moseley2021}.  The smallest scales reached in these simulations are however far above the dissipation scales but the results are promising. 
 It is the subject of the present paper to explore the nature, topology and statistics of the dissipation structures that form in magnetised turbulence.

 Turbulent dissipation has been extensively studied in incompressible media. In hydrodynamical (HD) turbulence, \cite{Moisy2004} have examined the geometrical properties of sites of extreme vorticity and shear. 
 \cite{Uritsky2010} examined the statistical properties of sites of strong dissipation in incompressible magnetohydrodynamical (MHD) turbulence, and \cite{Momferratos2014} extended their work to include ambipolar diffusion (i.e. ion-neutral drifts). \cite{Zhdankin2013,Zhdankin2014,Zhdankin2015,Zhdankin2016} studied extensively the statistics and dynamics of current sheets in reduced MHD. For example \cite{Zhdankin2013}  confirm the Sweet-Parker view of reconnection, although they note that not all current sheets are involved in reconnection.

All the above studies are performed in an incompressible framework while the interstellar medium is known to be extremely compressible.  We want here to examine dissipation in the extreme case of isothermal turbulence, where thermal effects cannot help pressure to resist against compression. In the incompressible framework \citep[see][for example]{Momferratos2014}, the physical nature of a dissipation structure (current sheet or shearing sheet) is directly linked to the nature of the dissipation within this structure (it is either purely Ohmic for current sheets or purely viscous for shearing sheets). The situation however is much more complicated in compressible HD turbulence, where shocks and shear can both lead to viscous dissipation, and even worse in compressible MHD, where dissipation structures can lead to viscous and resistive dissipation at the same place (as in a fast shock, see \citet{Lehmann2016} or our appendix \ref{sec:numerical-dissipation}).

Previous studies have attempted to characterise various individual types of structures.  \cite{Smith2000b} and \cite{Smith2000a} investigate velocity jumps in the three main directions as a proxy to shocks. \cite{Yang_2015} were able to  single out and study the formation of one rotational discontinuity in a simulation of MHD turbulence. \cite{Lehmann2016} introduced the SHOCKFIND algorithm which investigates an MHD snapshot to systematically extract every fast and slow shocks. In the present study, we attempt to characterise the physical nature of all intense dissipation structures: we present a new improved method able to characterise fast and slow shocks as well as Alfvén discontinuities.  

We want to examine the statistics of the various physical structures and their parameters and possibly assess how much dissipation is due to each category of dissipation structure. To this effect, we examine grid based simulations of decaying isothermal MHD turbulence which we present in sections 2.1 and 2.2. Because grid-based simulations are known to be more dissipative than pseudo-spectral simulations (which are however ill-suited to compressible fluids due to the Gibbs phenomenon), we devise and test a new method to retrieve the local dissipation intrinsic to the scheme (see appendix \ref{sec:numerical-dissipation}).
\cite{Stone1998} investigate dissipation in driven and decaying MHD turbulence and conclude about half is due to shocks. More precisely, they measure that 50\% of the total dissipation is due to their artificial viscosity term. However, they do not account for implicit numerical dissipation, and they did not check whether their artificial viscosity was indeed located in shocks. Similar studies by \cite{Smith2000b}  and \cite{Smith2000a}  (see their table 1) also use artificial viscosity  and suffer from the same uncertainties. \cite{Porter2015} and \cite{Park2019} did a much better job at detecting shocks and assigning dissipation to them but their method still suffers from uncertainty when the shocks are not aligned with the grid, and it is restricted to shocks (it would not work for Alfvén discontinuities because they focus on density jumps). In the present work, thanks to our method to recover everywhere the local dissipation (including the losses implicitly incurred by the numerical scheme), and because we carefully analyse the nature of intense dissipation structures, we hope to make more robust claims. For example,  \cite{CHEMSES}  performed a 2D HD simulation to such small scales that they were able to fully resolve the dissipation length scale, and to characterise almost all dissipation structures.

High dissipation is necessarily associated with strong variations of some of the variables controlling the physical state of the gas. We design a technique to assess locally the main direction of the gradients of the physical state of the gas (section 2.3). We observe that the regions of highest dissipation have their gradients locally along one direction primarily (in other words, intense dissipation structures are sheet-like). We show how to decompose the gradients along this direction using a basis of MHD waves (section 2.4). In section 3 we examine the connected sets of pixels above a large threshold of dissipative heating and we locally assess the nature of the physical profiles obtained scanning along the main direction of the gradient. We test whether the physical nature of these profiles agrees with the celebrated Rankine-Hugoniot (RH) relations \citep{Rankine1870} and perform various consistency checks which confirm the physical nature of these scans. In section 4 we examine the statistical properties of the scans we find. We discuss our results in section 5 and conclude in section 6.

\section{Numerical method}

   \subsection{Simulation}\label{sec:simu}
    In the present study, we run a set of simulations of decaying magnetohydrodynamics (MHD) turbulence. 
    \subsubsection{Numerical method}
    We solve the evolution equations of resistive and viscous isothermal MHD which we write here in conservative form:
    
       \begin{align}
        0=\,&\partial_{t}\rho+\boldsymbol{\nabla}\cdot(\rho \boldsymbol{u})\label{eq:continuity}\\
        0=\,&\partial_{t}\rho \boldsymbol{u}+\boldsymbol{\nabla}\cdot(\rho \boldsymbol{uu}-\nu\rho \boldsymbol{S}[\boldsymbol{u}])+\boldsymbol{\nabla}p-\boldsymbol{J\times B}\\
        0=\,&\partial_{t}\boldsymbol{B}-\boldsymbol{\nabla}\times\left(\boldsymbol{u\times B}-\eta\boldsymbol{\nabla\times B}\right)
    \end{align}

    where $\rho$ is the mass density, $\boldsymbol{u}$ is the fluid velocity vector, $p=\rho c^2$ is the thermal pressure with $c$ the isothermal sound speed, $\boldsymbol{B}$ is the magnetic field and $\boldsymbol{J}=\frac{1}{4\pi}\boldsymbol{\nabla \times B}$ is the  current vector. $\nu$ and $\eta$ are respectively the viscous and resistive coefficients. The components of the viscous stress tensor $\boldsymbol{S}$ are expressed as :
\begin{equation}
S_{ij}[\boldsymbol{u}]=\partial_{i}u_{j}+\partial_{j}u_{i}-\frac{2}{3}\partial_{k}u_{k}\delta_{ij}
\end{equation}
where $\partial_i$ denotes the derivative with respect to the space coordinate $i$.

  To integrate these equations, we use the code CHEMSES \citep{CHEMSES}, which originates from DUMSES \citep{DUMSES}, a version of RAMSES \citep{RAMSES} without adaptive mesh refinement. The ideal part of the evolution step is evolved thanks to a Godunov scheme with a Lax-Friedrichs Riemann solver and a minmod slope limiter function \citep[see][for details]{Toro1999}. The magnetic field is evolved with constrained transport to preserve its zero divergence  \citep{DUMSES}. This ideal MHD step is sandwiched between two half dissipation steps to preserve the second order accuracy of the time integration  \citep[see][for more details]{CHEMSES}. CHEMSES inherits the centering of the RAMSES code, with densities and velocity components at the center of cells and magnetic fields components at the center of their respective cell interface \citep{DUMSES}. The resistive and viscous stresses are centered accordingly, and a diffusion estimate (for both viscous and resistive dissipation) replaces the reference Courant time step whenever it is shorter. For example, the viscous diffusion time step constraint is $\Delta\tau=(\Delta x)^2/(6\nu)$ where $\Delta x$ is the pixel size. We set the Courant number\footnote{By Courant number we mean here the ratio between the used time step and the shortest numerically unstable time step.} at the value of 0.7 throughout all the simulations of the present paper. \change{For an isothermal gas, the viscous coefficient $\nu$ should be such that $\mu=\rho\nu$ is a constant : indeed, $\nu$ scales as the sound speed times the mean free path, which itself scales as $1/\rho$. However, we still use  a constant kinematic viscous coefficient $\nu$ as in \cite{Federrath2016} rather than a constant dynamical viscosity $\mu=\rho \nu$, as this allows easier numerical convergence for shocks (see appendix \ref{sec:numerical-dissipation}).}

\subsubsection{Initial conditions}
\label{sec:initial-conditions}
\begin{table*}
    \centering
   \begin{tabular}{|c|c|c|c|c|c|c|c|}
      \hline
      Init. cond. &N&$\nu$&$\eta$& $\mathcal{R}{\rm e}$&$\mathcal{P}_m$&$\mathcal{H}_c$&$\mathcal{H}_m$\\
      \hline
       ABC &512 &$7\times10^{-4}$& $7\times10^{-4}$ &$9\times10^{3}$& 1 & \change{$-2.5\times10^{-2}$ } & 0.2 \\
       ABC &1024&$7\times10^{-4}$& $7\times10^{-4}$ &$9\times10^{3}$& 1 & \change{$-2.5\times10^{-2}$ }& 0.2 \\
       ABC &1024&$2.8\times10^{-3}$& $7\times10^{-4}$ &$2\times10^{3}$& 4 & \change{$-2.5\times10^{-2}$ }& 0.2 \\
       ABC &1024&$1.1\times10^{-2}$& $7\times10^{-4}$ &$6\times10^{2}$& 16& \change{$-2.5\times10^{-2}$ } & 0.2 \\
       OT  &512 &$7\times10^{-4}$& $7\times10^{-4}$ & $9\times10^{3}$& 1 & 0.1 & $2\times10^{-9}$ \\
       OT  &1024&$7\times10^{-4}$& $7\times10^{-4}$ & $9\times10^{3}$& 1 & 0.1 & $2\times10^{-9}$ \\
       OT  &1024&$2.8\times10^{-4}$& $7\times10^{-4}$ &$2\times10^{3}$& 4 & 0.1 & $2\times10^{-9}$ \\
       OT  &1024&$1.1\times10^{-2}$& $7\times10^{-4}$ & $6\times10^{2}$& 16& 0.1 & $2\times10^{-9}$ \\
      \hline
   \end{tabular}
   \caption{Parameters of simulations we analyse: initial flow, resolution, \change{kinematic viscosity $\nu$,} resistivity $\eta$, \change{Reynolds number,} Prandtl number $\nu/\eta$, cross and magnetic helicities. All the simulations start with an r.m.s. sonic Mach number $\mathcal{M}_s=4$ and an r.m.s. Alfvénic Mach number of 1, with a zero mean magnetic field.}
   \label{tab:SimuParam}
\end{table*}   
   
    The quantities computed in the code are dimensionless. They are normalised by physical scales set such that initially the average square velocity is $<u^2>=1$, the cubic domain size is $L=2\pi$ and the average density $<\rho>=1$ where the brackets denote averages over the whole simulated domain.
    The non-dimensional value of the isothermal speed $c$ thus controls the r.m.s. initial sonic Mach number as $\mathcal{M}_s=1/c$. The initial density is uniform and the initial magnetic field is scaled to obtain $<\frac{1}{4\pi}B^2>=<\rho><u^2>=1$ so that the effective r.m.s. initial Alfvénic Mach number is equal to 1, as well as the r.m.s. initial Alfvén speed ($c_\mathrm{A}$). Note that the mean magnetic field is zero over the computational domain.
    
    For example, imagine one wants to apply these results to a physical region of physical dimension $\ell$, of r.m.s. velocity $u_{\rm r.m.s.}$ and average density $\rho_{\rm av}$. Then dimensionless quantities in the code can be converted to physical quantities according to: $x_{\rm phys}=\ell/(2\pi).x$ for distances, $u_{\rm phys}=u_{\rm r.m.s.}.u$ for velocities, and $B_{\rm phys}=u_{\rm r.m.s.}\sqrt{4\pi\rho_{\rm av}}.B$ for magnetic fields.
    
    As in \cite{Momferratos2014}, we consider a periodic box with initial conditions based either on the Arnol'd-Beltrami-Childress flows \citep[ABC, see][for example]{ABC2013} or on the Orszag-Tang vortex \citep[OT,][]{OT79}.
    For the ABC flow, the velocity field is set by a superposition of sines and cosines:
    \begin{multline}
         \boldsymbol{u}_\mathrm{ABC}=(A\sin (kz)+C\cos(ky),B\sin (kx)+A\cos(kz),\\
         C\sin (ky)+B\cos(kx)),
    \end{multline}
    where A, B, C are coefficients  chosen for the three smallest wave numbers $k$ (largest scales) from a uniform number generator in the interval $[-1,1]$. For smaller scales, a random field $u_\mathrm{E}$ is added, with energy spectrum 
    \begin{equation}
        E(k)= C_\mathrm{E} k^{-3}\exp{\left( -2(k/k_\mathrm{c}) ^2\right)},
    \end{equation}
    where $k_\mathrm{c}=3$, and $C_\mathrm{E}$ is chosen so that $<u_E^2>=1$. This random field is set in Fourier space with the amplitude of the complex coefficients prescribed by the above spectrum and the phase of each coefficient is drawn from a uniform distribution in the interval $[0,2\pi]$. The perturbed initial ABC velocity field $\boldsymbol{u}=\alpha(\boldsymbol{u}_\mathrm{ABC}+\boldsymbol{u}_\mathrm{E})$ is rescaled so that $<u^2>=1$ by properly setting $\alpha$. The initial magnetic field for the ABC runs is set with a random field drawn in a similar way as $\boldsymbol{u}_\mathrm{E}$. \change{The power spectrum of the initial random perturbation is a minimal seed to initiate a cascade, and let it develop naturally.  Indeed, we want our results to testify for our large scale initial conditions rather than for the added seed. We hence choose its logarithmic slope significantly steeper than the expected Kolmogorov ($k^{-5/3}$) or supersonic \citep[$k^{-2}$,see][]{Federrath2013,Federrath2021} spectra, with an additional exponential cut-off for safety.  }
    
    The OT vortex velocity is defined by 
    \begin{equation}
         \boldsymbol{u}_\mathrm{OT}=(-2\sin (y),2\sin (x),0),
    \end{equation}
     to which we also add random perturbations as in the ABC case.
        And the initial magnetic field for the OT vortex is set as
    \begin{multline}
         \boldsymbol{B}_\mathrm{OT}=(-2\sin (2y)+\sin(z),2\sin(x)+\sin(z),\\\sin(x)+\sin(z), 
    \end{multline}
   {\it without} additional perturbation. The velocity and magnetic fields are then rescaled so that $<u^2>=<\frac{1}{4\pi}B^2>=1$.
    
  Our ABC flows have a significant magnetic helicity ($\mathcal{H}_m=<\mathbf{A.B}>$ where $\vec{A}$ is the potential vector, $\vec{B}=\vec{\nabla} \times \vec{A}$ with Coulomb gauge $\mathrm{div}\mathbf{A}=0$) and an almost zero cross helicity ($\mathcal{H}_c=<\mathbf{u.B}>$). That means the magnetic field is topologically complex and there is no strong correlation between magnetic and velocity field. For OT initial conditions the situation is reversed, it has an almost null magnetic helicity and a non zero cross helicity (see table \ref{tab:SimuParam} for the values of helicities). 
 
 In addition to the initial conditions, we also investigate the resolution: our fiducial runs have a number of pixels $N=1024$ per side of the cubic computational domain and we degrade the resolution by a factor two to control the stability of our results. We also probe the effect of varying the Prandtl number $\Pm=\nu/\eta$. Table \ref{tab:SimuParam} summarises the parameter space we cover.
 
   \subsection{Dissipation recovery and control}
   \begin{figure}
       \centering
       \includegraphics[width=0.42\paperwidth]{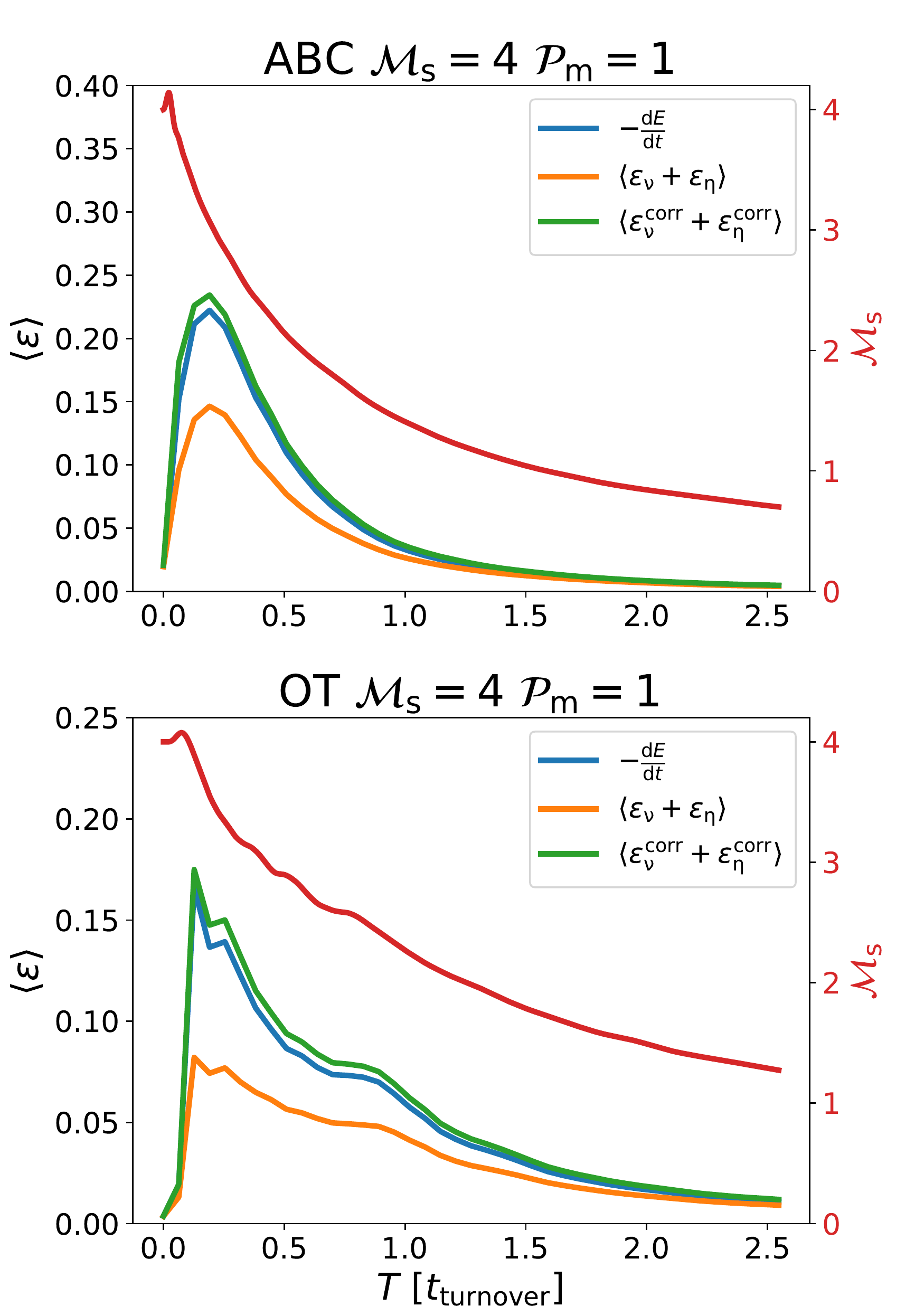}
          \caption{ Time evolution of volume integrated dissipation rates for the ABC and OT, $\mathcal{P}_\mathrm{m}=1$ runs. The blue line is the time derivative of the integrated  isothermal generalised mechanical energy $E=<\frac12 \rho u^2 + \frac{B^2}{8\pi} + p \log \rho>$. The orange curve is the sum of the physical viscous and ohmic dissipations computed from the velocity and magnetic fields (equations \ref{eq:viscous-dissipation} and \ref{eq:ohmic-dissipation} respectively). The green line is the volume integrated corrected dissipation field $\varepsilon_\mathrm{tot}^\mathrm{corr}$ determined by our recovery method (see text and Appendix \ref{sec:numerical-dissipation}).  Note that the time scale is in units of initial turnover time scale: we display $t/t_\mathrm{turnover}$ where $t_\mathrm{turnover}=L/\sqrt{<u^2>}=2\pi$. \change{We also show the evolution of the r.m.s. sonic Mach number $\mathcal{M}{\rm s}$ (red curve), its scale corresponds to the right axis.}}
             \label{fig:TimeDissComp}
    \end{figure}
   
   The numerical scheme we use (Godunov) implicitly introduces dissipation to evolve the ideal MHD equations, but as stated above, we incorporate additional explicit physical dissipation terms in our evolution equations. Indeed, it is important to retain some amount of physical viscosity as Godunov schemes do not provide an implicit viscosity in shear layers. Here, we discuss our methods to estimate the fraction of the dissipation which is due to the numerical scheme.
   
    We set values for the viscous and resistive coefficients $\nu$ and $\eta$ identical to those used by \cite{Momferratos2014} in pseudo-spectral simulations with 512$^3$ spectral elements: $\nu=\eta=7 \times 10^{-4}$ in the same non-dimensional units. This is motivated by the common belief that spectral codes are approximately twice more efficient as grid based codes. Our  study for shocks in appendix \ref{sec:numerical-dissipation} presents a more detailed picture. Figure \ref{fig:shock_convergence} shows the dissipation bump in a fiducial shock front at various resolutions.  For our chosen values for the dissipative coefficients and a resolution of $N=1024$, we see it is effectively spread up by nearly a factor of three, while one would have to increase the resolution by a factor 8 to fully resolve it. A resolution twice less would spread the front by a factor six and thus our current choice is a good compromise between accuracy and CPU efficiency.  
   
   In isothermal MHD, the integrated total isothermal generalised mechanical energy $E = <\frac12 \rho u^2 +\frac{1}{8\pi} B^2 + p \log \rho>$ decreases due to all irreversible processes taking place (see equation \ref{eq:method2}). Because our Godunov time integration scheme is conservative to round-off error, we can use its time derivative to estimate the global budget of dissipated energy:
   \begin{equation}
       -\frac {{\rm d} E}{{\rm d}t}=<\varepsilon_\mathrm{tot}>
       \label{equ:total-dissipation}
   \end{equation}
   where $<\varepsilon_\mathrm{tot}>$ is the total rate of irreversible heating integrated over the whole computational domain. Appendix \ref{sec:numerical-dissipation} presents and tests a new method to estimate {\it locally} the total irreversible heating $\varepsilon_\mathrm{tot}$. The chosen method has the additional advantage that it preserves to round-off error the validity of equation \eqref{equ:total-dissipation} when integrated over the whole domain. 
   
   We can now decompose the local total heating rate as 
\begin{equation}
    \varepsilon_\mathrm{tot}=\varepsilon_{\nu}+\varepsilon_{\eta}+\varepsilon_\mathrm{num}
\end{equation}
  where   
   \begin{equation}
   \label{eq:viscous-dissipation}
       \varepsilon_{\nu}=\rho \nu S_{ij}[\vec{u}] \partial_i u_j 
   \end{equation} and 
   \begin{equation}
   \label{eq:ohmic-dissipation}
       \varepsilon_{\eta}= 4\pi \eta J^2
   \end{equation} are the local viscous and resistive dissipative heating rates, and $\varepsilon_{\rm num}$ is the dissipation due to the numerical scheme.

     We can then estimate the local numerical dissipation rate simply by computing $\varepsilon_{\rm num}=\varepsilon_{\rm tot}-(\varepsilon_{\nu}+\varepsilon_{\eta})$ where we use well centered estimates for equations \eqref{eq:viscous-dissipation} and \eqref{eq:ohmic-dissipation}. \change{If our estimate for the local dissipation were perfect, this quantity would always be positive, because we are performing our simulations with a time step small enough for the scheme to be stable (it is set to 70\% of the shortest unstable time step). However, we are subject to truncation errors in both the $\varepsilon_{\rm tot}$ term (see appendix \ref{sec:numerical-dissipation}) and the $\varepsilon_{\nu}+\varepsilon_{\eta}$ terms (where a centered difference is used). The difference between the two terms can hence be negative due to these truncation errors.} We thus define $\varepsilon_\mathrm{tot}^\mathrm{corr}$ as a corrected local total dissipation rate which ensures the resulting estimate for $\varepsilon_\mathrm{num}$ is positive. It is equal to the total local dissipation $\varepsilon_{\rm tot}$ where the numerical dissipation is positive (i.e. where $\varepsilon_{\rm tot}>(\varepsilon_{\nu}+\varepsilon_{\eta})$), while it is equal to the total physical dissipation $\varepsilon_{\nu}+\varepsilon_{\eta}$ elsewhere. This ensures the corrected local numerical dissipation rate $\varepsilon_\mathrm{num}^\mathrm{corr}=\varepsilon_{\rm tot}^\mathrm{corr}-(\varepsilon_{\nu}+\varepsilon_{\eta})$ is always positive. In particular, the local corrected total dissipation rate $\varepsilon_\mathrm{tot}^\mathrm{corr}$ is always greater than $\varepsilon_\mathrm{tot}$. It is then shared between resistive and viscous natures in the same proportions as the physical terms we introduced to provide estimates for viscous and resistive dissipations including numerical dissipation:
    \begin{equation}
       \varepsilon_\mathrm{\nu}^\mathrm{corr} =\frac{\varepsilon_\mathrm{\nu}}{\varepsilon_\mathrm{\nu} +\varepsilon_\mathrm{\eta}}\varepsilon_\mathrm{tot}^\mathrm{corr}
   \end{equation}
   \begin{equation}
       \varepsilon_\mathrm{\eta}^\mathrm{corr}=\frac{\varepsilon_\mathrm{\eta}}{\varepsilon_\mathrm{\nu} +\varepsilon_\mathrm{\eta}}\varepsilon_\mathrm{tot}^\mathrm{corr}
   \end{equation}

     Figure \ref{fig:TimeDissComp} displays the temporal evolution of various total dissipation rates.  Thanks to the equality \eqref{equ:total-dissipation}, we can compute the exact total dissipation rate at each time step (blue curves), and we can compare it to the integrated local estimate $<\varepsilon_\mathrm{tot}^\mathrm{corr}>$ (green curves) which by construction is always greater. The difference between the two gives an estimate of the error we make on the estimation of the dissipation (on the order of a percent at most). It corresponds to the integrated estimated $\varepsilon_\mathrm{num}$ in all the pixels where it is negative. The orange curves show the integrated physical dissipative terms $<\varepsilon_{\nu}+\varepsilon_{\eta}>$. They amount to about two-thirds of the total, while the remainder is numerical dissipation by the scheme. 


   \subsection{The local frame of physical gradients}\label{GradientMethod}
   \begin{figure*}
    \centering
    \resizebox{\hsize}{!}{\includegraphics{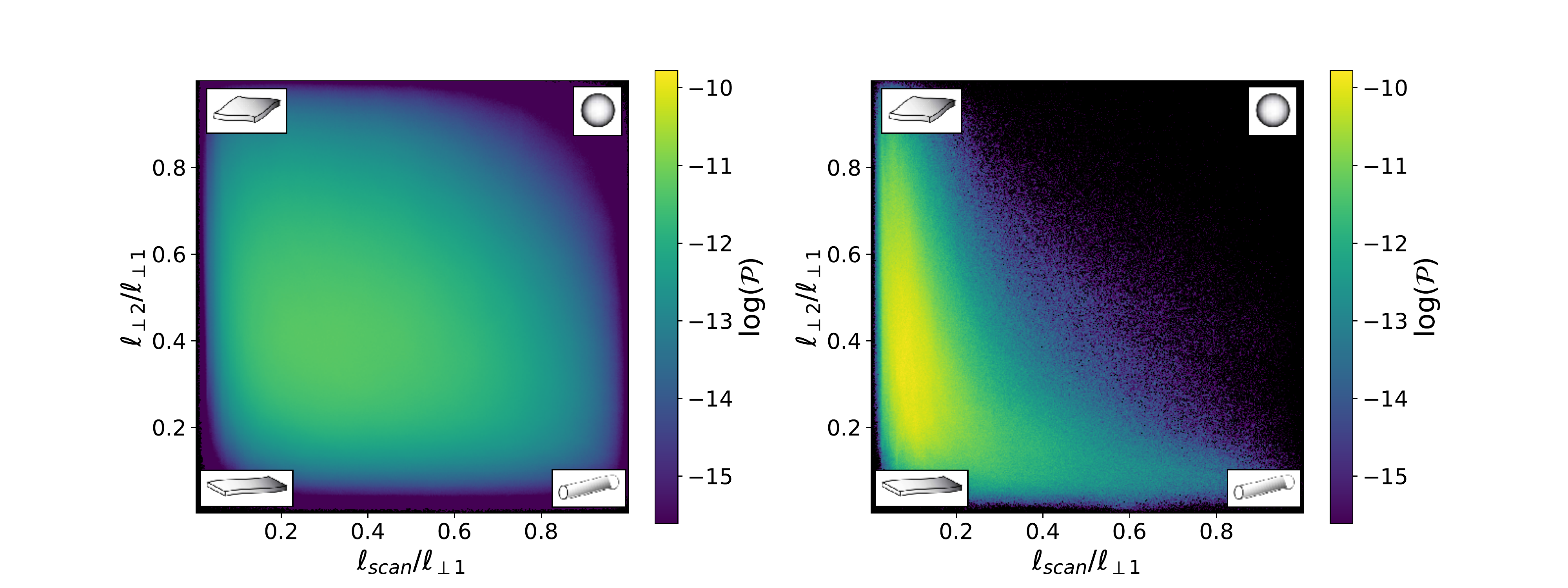}}
    \caption{The 2D joint probability density function of gradients aspect ratios (for the OT simulation at $\Pm=1$ at time $t=t_\mathrm{turnover}/3$). On the left, characteristic lengths are calculated for all the simulation cells. While on the right, the  domain is restricted to cells where $\epsilon_\mathrm{tot}^\mathrm{corr}\geq\left<\epsilon_\mathrm{tot}^\mathrm{corr}\right>+4\sigma_{\varepsilon_\mathrm{tot}^\mathrm{corr}}$. The color scale is logarithmic.}
              \label{FigPlanars}%
   \end{figure*}

   We know local intense dissipation events are caused by strong variations of some of the fluid state variables. Here, we want to identify regions where the fluid state varies strongly and to characterise its variations in each direction. 
   
   The fluid state is characterised by the seven (1+3+3) components of $\vec{W}=\left( \rho, \mathbf{u}, \mathbf{B} \right)$, which do not have the same physical dimensions. We want to put the variations of density, velocity and magnetic fields on equal footing. Hence, we need to rescale the gradient of each component of $\vec{W}$ to make them homogeneous to the same physical dimension. We now choose to define the rescaled gradient of $\vec{W}$ in a given direction $\vec{r}$ as 
    \begin{equation}
       \partial_{\vec{\hat{r}}} \vec{W} \equiv \left( 
       (\vec{\hat{r}}\cdot \vec{\nabla}) \log \rho, 
       \frac{1}{c} (\vec{\hat{r}}\cdot \vec{\nabla})\vec{u}, \frac{1}{c\sqrt{4\pi\rho}} (\vec{\hat{r}}\cdot \vec{\nabla}) \vec{B}  \right) 
    \end{equation}
    where $\vec{\hat{r}}=\vec{r}/r$ is the unit vector in the direction of $\vec{r}$. This rescaled gradient has the dimension of the inverse of a length scale, which represents the typical length scale over which the state variables vary in the direction $\vec{\hat{r}}$.

   The norm of this gradient will be large whenever there is a rapid change in one or several state variables. Its square can be expressed as 
   \begin{equation}
       ||\partial_{\vec{\hat{r}}} \vec{W}||^{2}=\alpha_{ij}\ \hat{r}_i\ \hat{r}_j,
       \label{GradientMatrix}
   \end{equation}
   where $\alpha_{ij}=\partial_i \vec{W}\cdot\partial_j \vec{W}$ is a 3$\times$3  matrix (and the dot product applies to the seven components vectors) with coefficients homogeneous to an inverse squared length. It is real symmetric, and therefore diagonal in an orthonormal basis. We can rewrite equation \eqref{GradientMatrix} in a more explicit form 
   \begin{equation}
       ||\partial_{\vec{\hat{r}}} \vec{W}||^{2}=\frac{1}{\ell_{\rm scan}^2} \left(\vec{\hat{r}}\cdot\vec{\hat{r}_{\mathrm{scan}}}\right)^2+
       \frac{1}{\ell_{\perp1}^2} \left(\vec{\hat{r}}\cdot\vec{\hat{r}_{\mathrm{\perp1}}}\right)^2+
       \frac{1}{\ell_{\perp2}^2} \left(\vec{\hat{r}}\cdot\vec{\hat{r}_{\mathrm{\perp2}}}\right)^2
       \label{EllipsForm}
   \end{equation}
   where $\ell_{\rm scan}^2$, $\ell_{\perp1}^2$, and $\ell_{\perp2}^2$ are the inverse of the eigenvalues associated to the eigenvectors $\vec{\hat{r}}_\mathrm{scan}$, $\vec{\hat{r}}_{\perp1}$, and $\vec{\hat{r}}_{\perp2}$ of the matrix $\alpha_{ij}$. Equation \eqref{EllipsForm} shows how the gradient of state variables depends on directions. A 3D polar plot of the norm of this gradient takes the form of an ellipsoid whose principal axes are in the three orthogonal eigenvalue directions of the above matrix:
   \begin{eqnarray}
       \ell_{\mathrm{scan}}=||\partial_{\vec{\hat{r}_{\mathrm{scan}}}}  \vec{W}||^{-1}, 
        &
       \ell_{\mathrm{\perp 1}}=|| \partial_{\vec{\hat{r}_{\mathrm{\perp 1}}}} \vec{W}||^{-1},
       &
       \ell_{\mathrm{\perp 2}}=||\partial_{\vec{\hat{r}_{\mathrm{\perp 2}}}} \vec{W}||^{-1}
       \label{CharLength}
   \end{eqnarray}
   with the three length scales ordered so that $\ell_{\rm scan}\leq\ell_{\perp1}\leq\ell_{\perp2}$. These three variation length scales and their associated orthogonal directions characterise the local geometry of the gradients of the fluid state variables.
   
   Figure \ref{FigPlanars} shows how the aspect ratios between these typical variation length scales are distributed in all cells of a simulation (left panel) and for only highly dissipating ones (four standard deviations over the mean, right panel). It shows that most fluid state variables vary primarily in one direction for extreme dissipation events, whereas aspect ratios span all possibilities if we consider the full simulation domain. We also notice a slight imbalance towards ribbons compared to sheets. When one variation direction is dominant ($\ell_{\rm scan}<<\ell_{\perp1}\leq\ell_{\perp2}$), quantities are essentially constant in the direction orthogonal to it and the local situation is hence nearly 1D plane-parallel. We thus define the planarity as the ratio $\ell_{\perp1}/\ell_{\rm scan}$ which is large whenever $\ell_{\rm scan}<<\ell_{\perp1}$, i.e. when the local geometry is close to plane-parallel. 
   
   This one-dimensional geometry of gradients for intense dissipation regions is consistent with the typical two-dimensional geometry of the structures found in MHD turbulence \citep{Uritsky2010,Zhdankin2013,Momferratos2014}. On intense dissipation structures, we should thus be able to capture most fluid variations by browsing those along the maximum gradient direction. In section \ref{IdentificationTitle} we use $\vec{\hat{r}_\mathrm{scan}}$ as a sampling direction to probe the variation of physical quantities around strong dissipation regions.
   
   \subsection{Gradient decomposition into MHD waves} \label{GradientDecomp}
   
   In the ideal case where the gradient would be strictly in one direction, the gas dynamics are governed by 1D plane-parallel MHD equations, and we show here how local gradients can be projected onto ideal MHD waves.
   
   We write $x$ the space coordinate along the direction of the gradient and $t$ the time coordinate. The requirement $\vec{\nabla}.\vec{B}=0$ implies that $\partial_x B_x=0$: the corresponding component of $\partial_x \vec{W}$ is thus zero. It turns out that the six non zero components of $\partial_x \vec{W}$ are spanned by the six ideal MHD waves, as we now turn to show.
  
    Wave solutions take the form $\vec{W}(x,t)=\vec{F}(x- \varv t)$ where $\varv$ is the traveling speed of the wave. We note that $\partial_t \vec{F}=-\varv\partial_x\vec{F}$ and plug this form into the {\it ideal} MHD part of the equations (without the dissipation terms). We arrive at a linear eigenvalue problem for which we can find six eigenvectors $\partial_x\vec{F}$ with eigenvalues $\varv$ corresponding to the six waves of ideal isothermal MHD\footnote{Note that formally, finding the gradient $\partial_x\vec{F}$ is equivalent to solving the amplitude for the linear wave problem when we identify $\partial_t\equiv i \omega$, $\partial_x \equiv i k$ and $\varv=\omega/k$ matches the phase velocity.}. We label them by their wave type $s,i,f$ for slow, intermediate and fast and their direction of propagation $R,L$ for right (or forward, $\varv>0$) and left (or backward, $\varv<0$). To within a multiplicative constant, the expressions for these eigenvectors are (see section 5.2.3 of \cite{goedbloed_MHD} or section 6.5 of \cite{gurnett_bhattacharjee_2005}, for example) : 
   \begin{itemize}
       \item for intermediate (Alfvén) waves
   \begin{equation}
       \partial_{\vec{\hat{x}}} \vec{F}^{R,L}_{i} \propto (0,\epsilon^{R,L}\vec{a}_t^{\perp},-\mathrm{sign}(a_x)\vec{a}_t^{\perp})
   \end{equation}
   where $\epsilon^{R,L}=-1$ for left-going (backward going) waves and $\epsilon^{R,L}=1$ for right-going (forward going) waves,  $\vec{a}=\vec{B}/\sqrt{4\pi\rho}$ is the Alfvén velocity vector, $\vec{a}_t$ is the transverse component of $\vec{a}$, $a_x$ is its $x$-component and $\vec{a}_t^{\perp}$ is $\vec{a}_t$ rotated by $\pi/2$ in the transverse plane. The first component of this gradient is zero, hence the density is uniform. And the transverse magnetic field has its gradient orthogonal to itself: it rotates along the scanning direction. The corresponding travelling speed is $c_i^{R,L}=\epsilon^{R,L}|a_x|$.
      \item for fast and slow magnetosonic waves
    \begin{equation}
       \partial_{\vec{\hat{x}}} \vec{F}^{R,L}_{s,f} \propto (-\frac{c}{c^{R,L}_{s,f}},\vec{\hat{x}}-\frac{a_x}{d}\vec{a}_t,\frac{c^{R,L}_{s,f}}{d}\vec{a}_t)
    \end{equation}
    where the propagation speed $c^{R,L}_{s,f}$ reads:
    \begin{equation}
        c^{R,L}_{s,f} = \epsilon^{R,L}\sqrt{(c^2+a^2)+\epsilon_{f,s}\sqrt{(c^2+a^2)^2-4a_x^2c^2}}
    \end{equation}
    with  $d=(c^{R,L}_{s,f})^2-a_x^2$ and $\epsilon_{f,s}=1$ for fast waves or -1 for slow waves. These waves are compressive (the density gradient is non-zero) and the gradient of transverse magnetic field is aligned with itself. In other words, the transverse magnetic field remains in the same direction, which also happens to be the same direction as the variation of the transverse velocity. Both the velocity and the magnetic field vectors thus remain in the plane defined by the scanning direction $\vec{\hat{x}}$ and the initial transverse field (a property sometimes referred to as the {\it coplanarity} of these waves).
   \end{itemize}
    These six gradients form an orthogonal basis which can be easily normalised to make it an orthonormal basis $\vec{\hat{e}}^{R,L}_{s,i,f}=\partial_{\vec{\hat{x}}} \vec{F}^{R,L}_{s,i,f}/||\partial_{\vec{\hat{x}}} \vec{F}^{R,L}_{s,i,f}||$ .
    
      Any gradient $\partial_{\vec{\hat{x}}} \vec{W}$ can now easily be decomposed into the six waves by computing the scalar product $\alpha^{R,L}_{s,i,f}=\hat{\vec{e}}^{R,L}_{s,i,f}.\partial_{\vec{\hat{x}}} \vec{W}$. Thanks to orthonormality, we have $\sum_{R,L,s,i,f}(\alpha^{R,L}_{s,i,f})^2=||\partial_{\vec{\hat{x}}} \vec{W}||^2$ and each coefficient $(\alpha^{R,L}_{s,i,f})^2/||\partial_{\vec{\hat{x}}}  \vec{W}||^2$ can be interpreted as a 0 to 1 coefficient which characterises how similar the gradient $\partial_{\vec{\hat{x}}} \vec{W}$ is to the corresponding ideal MHD wave. We call 'most representative wave' the wave with the largest coefficient in this decomposition. The most representative wave characterizes the local gradient as slow, intermediate or fast, each one in a left (backward) or right (forward) going version depending on the sign of its speed relative to the fluid $c^{R,L}_{s,i,f}$. Note also that this decomposition does not change if we add a constant vector to the velocity: it is independent from the choice of Galilean frame.
  
  Until now, we have only considered wave solutions of the {\it ideal} part of the MHD equations (without dissipation), while the gradients in our simulation result from the evolution of fully dissipative MHD. Let us now consider a non-linear wave solution of the 1D fully dissipative MHD $\vec{F}_\mathrm{full}(x-\varv_\mathrm{full}t)$, such as the isothermal shocks of appendix \ref{sec:steady-state-MHD-shocks}, for example. The profile of this wave continuously joins two uniform states related by the Rankine-Hugoniot relations (see subsection \ref{RHrelation}). These two states are separated by a region where dissipation occurs. Consider the gas state at the local maximum of dissipation: this is where the gradients of the state variables are the largest, and where the gradient of viscous and resistive stresses are likely to be small (because we are close to their maximum). At this position, the 1D dissipative physics behaves like the 1D ideal physics, and we can expect that the measured gradients fall along one of the ideal wave gradients we described above. As a result, the fully dissipative wave speed should be well approximated by its ideal estimate: $\varv_\mathrm{full} \simeq u_x + c^{R,L}_{s,i,f}$ where $u_x$ is the fluid velocity and $c^{R,L}_{s,i,f}$ applies to the most representative wave at the dissipation maximum. We will make use of this fact in the following to estimate the steady-state velocity of the structures we detect (see section \ref{sec:velocities}). Furthermore, we investigated the gradients of semi-analytic isothermal shock profiles (computed in appendix \ref{sec:steady-state-MHD-shocks}) and we noticed that gradients in slow shocks are dominated by slow magnetosonic waves all along their profiles. Similarly fast shocks gradients are dominated by fast magnetosonic waves. This result seems natural but we find it nevertheless surprising that dissipative physics does not affect more the nature of gradients and we did not yet find a satisfactory explanation for this behaviour.
  
     Finally, note that we can always decompose a gradient in a given direction, but it makes less sense if the 3D gradient is not strongly dominated by a single direction. By selecting intense dissipative cells, however, we are more likely to be in a situation where the gradient is well directed (see figure \ref{FigPlanars} and previous subsection).

\section{Dissipation structures}
   \subsection{Definition and visualization}\label{Def_visu}
   
   \begin{figure}
       \centering
       \includegraphics[width=0.42\paperwidth]{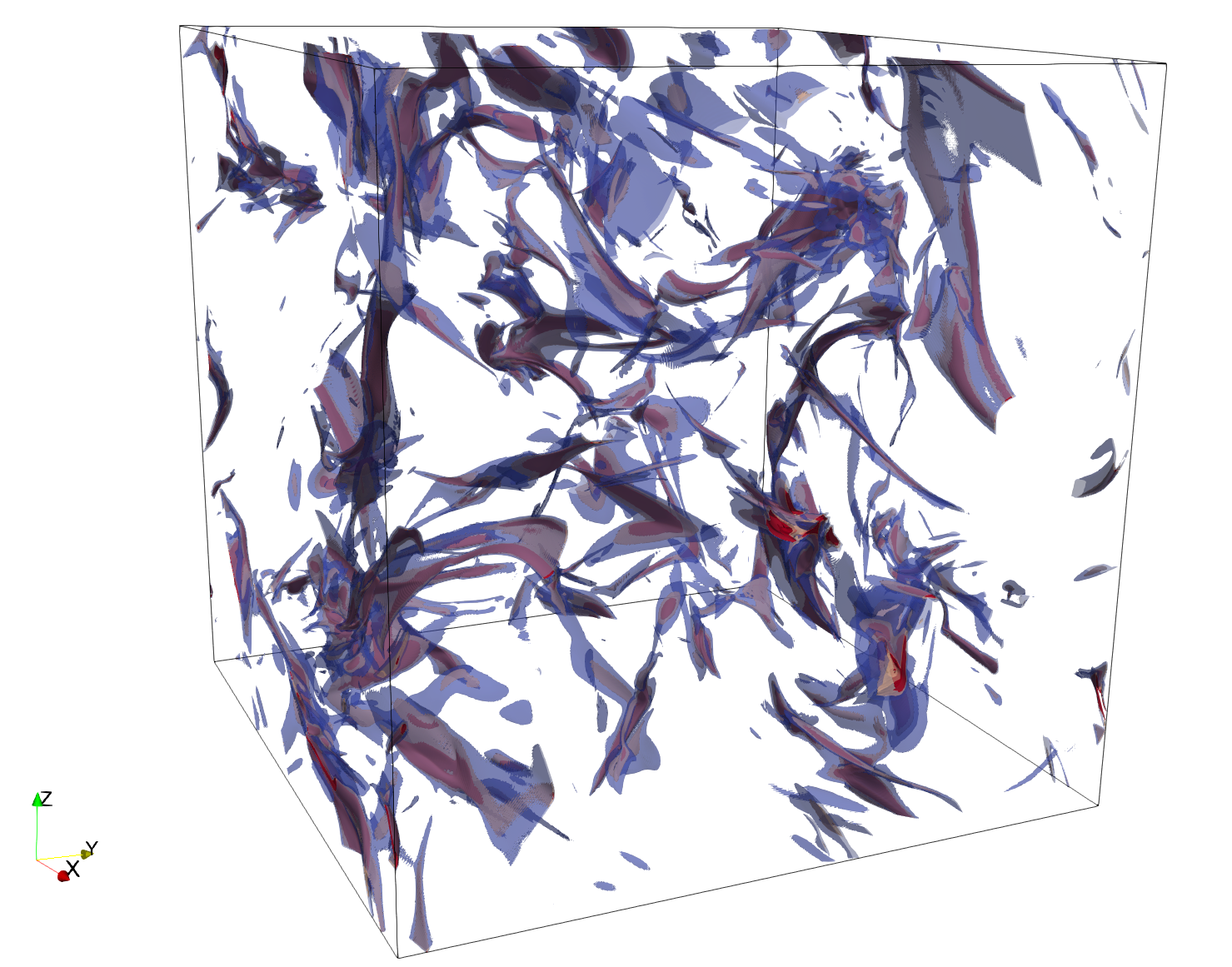}
          \caption{Intense dissipation structures extracted from an OT initial conditions simulation with $\Pm=1$. The time step of this output is $t\simeq 1/3 t_\mathrm{turnover}$. Structures are shown through dissipation isocontours. The first one in {\color{blue}blue} is set at $\varepsilon_\mathrm{tot}^\mathrm{corr} = \left<\varepsilon_\mathrm{tot}^\mathrm{corr}\right>+4\times \sigma_\mathrm{\varepsilon_\mathrm{tot}^\mathrm{corr}}$. The second in beige is at 8 times the standard deviation above the mean value and the last one, in {\color{red}red} is at 13.5 times. }
             \label{fig:fullcube}
    \end{figure}
    
    \begin{figure}
        \centering
        \resizebox{\hsize}{!}{\includegraphics{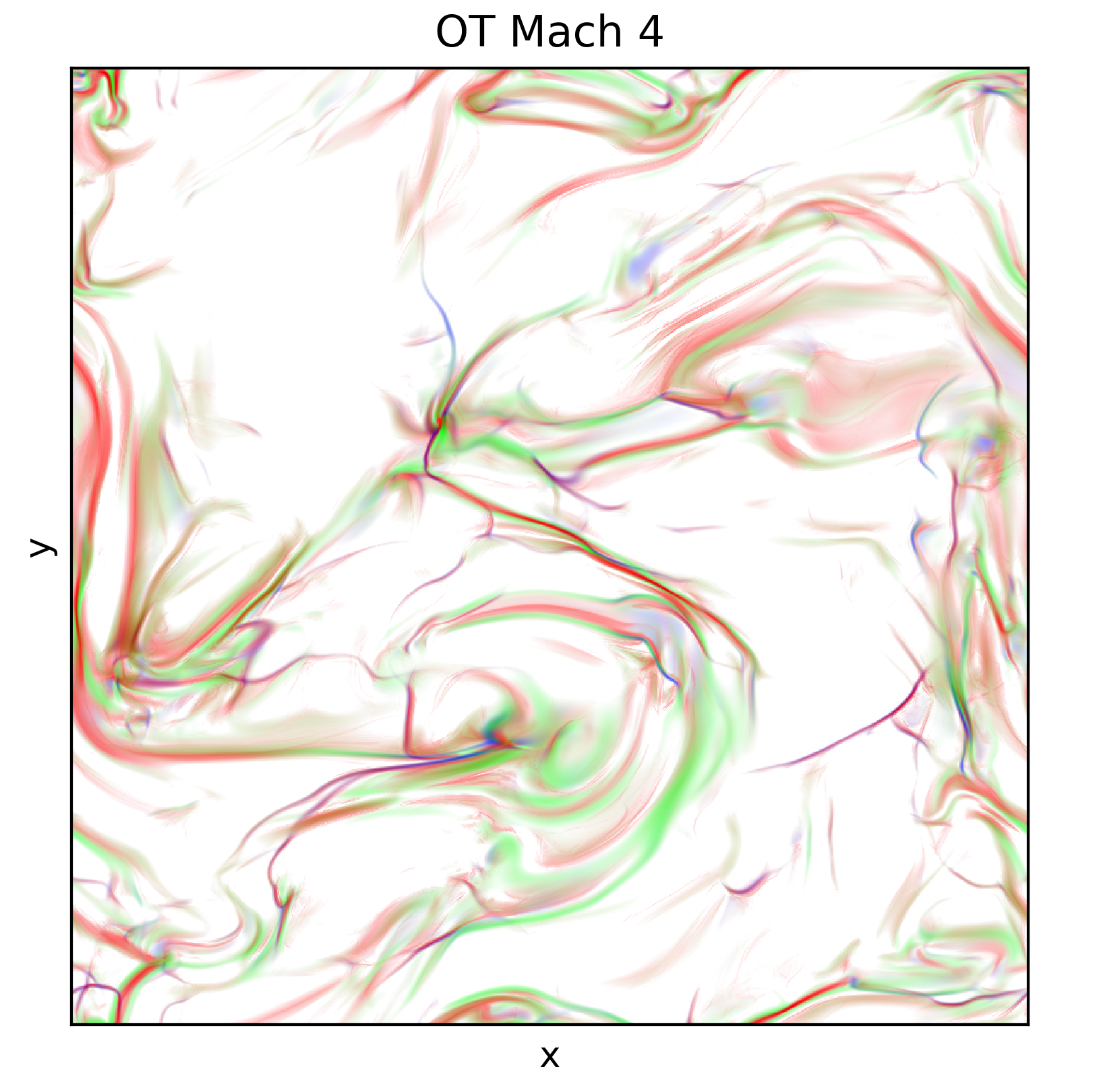}}
        \caption{Dissipation cut at time $t=1/3 t_\mathrm{turnover}$ for the OT initial conditions with $\Pm=1$. Lower and upper thresholds have been applied to the 3\% pixels with smallest and largest dissipation, the intensity scaling of the pixels is  logarithmic, while the colors code for Red: Ohmic dissipation $\varepsilon_{\eta}= 4 \pi \eta J^2$, Blue : compressive viscous heating $\varepsilon_\mathrm{comp}=4/3 \rho \nu \left(\mathbf{\nabla}\cdot \mathbf{u}\right)^2$, and Green : solenoidal viscous heating $\varepsilon_\mathrm{sol}=\rho \nu \left(\mathbf{\nabla}\times \mathbf{u}\right)^2$. Beware that $\varepsilon_\mathrm{tot}^\mathrm{corr}\ne \varepsilon_\mathrm{comp}+\varepsilon_\mathrm{sol}+\varepsilon_\eta$, both locally and globally (because we use a uniform $\nu$ and not a uniform $\rho \nu$, see \cite{CHEMSES}). Note that there is very little compressive heating (blue). }
        \label{fig:Dissipation-cut}
    \end{figure}

   In turbulent MHD flows, the bulk dissipation of kinetic and magnetic energy occurs in a small volume compared to the global scale of the flow. Dissipation has been analyzed and observed in several studies \citep[e.g.][]{Uritsky2010,Zhdankin2013,Momferratos2014} to be organized in coherent structures, ribbon-shaped or sheet-like.  
   
   Figure \ref{fig:fullcube} shows isocontours of the total dissipation rate $\varepsilon_\mathrm{tot}^\mathrm{corr}$. The dissipation rate in each cell is computed using the method described in appendices \ref{sec:steady-state-MHD-shocks} and \ref{sec:numerical-dissipation}. We follow previous work \citep{Uritsky2010} and define a connected dissipation structure as a connected set of cells where 
   \begin{equation}
        \varepsilon_\mathrm{tot}^\mathrm{corr} \geq \left<\varepsilon_\mathrm{tot}^\mathrm{corr}\right>+\lambda\times \sigma_\mathrm{\varepsilon_\mathrm{tot}^\mathrm{corr}},
   \end{equation}
   with $\lambda$ a parameter we use to tune the detection threshold, $\varepsilon_\mathrm{tot}^\mathrm{corr}$ the dissipation rate determined by our method and $\sigma_\mathrm{\varepsilon_\mathrm{tot}^\mathrm{corr}}$ the standard deviation of the dissipation rate distribution.  We choose $\lambda=4$ because we find that energy transfers are mainly due to events above $4\sigma$: we checked that the bulk of the third order structure function (responsible for energy transfers) is obtained from increments above 3-4 sigma. We also want the structure to be identifiable as clearly as possible and we expect such high dissipation structures to be associated with more intense gradients and a more clear-cut physical nature. 

   As already hinted at by local gradients (figure \ref{FigPlanars}), we see on figure \ref{fig:fullcube} that extracted dissipation structures are mainly sheets. Another way to see this is to look at a thin slice of the dissipation field in our OT simulation with $\Pm=1$ (figure \ref{fig:Dissipation-cut}) where the trace of the sheets appears as thin ridges. Compared to the same figure for the incompressible runs of \cite{Momferratos2014},  the viscous and Ohmic natures of dissipation are now much more entangled and sometimes even overlap. A close eye inspection of this figure (and of similar cuts at other time steps and initial conditions) reveals various sub-layering of Ohmic dissipation sheets (red striations or ohmic dissipation wrapped by shear) or isolated viscous and ohmic heating sheets (purple color, which hints at a mix of compressive viscous heating and ohmic heating). These are not the only situations which occur, but it reveals that the intense dissipation sheets are not always randomly positioned with respect to one another.
   
   A careful inspection of figure \ref{fig:fullcube} allows to witness a few small filament-like structures. Some of them may be traced on figure \ref{FigPlanars} by the low-probability tail in the bottom right hand corner of the right panel, where the aspect ratios of the gradients are such that $\ell_\mathrm{scan}\simeq\ell_\mathrm{\perp1}$ while $\ell_\mathrm{\perp1}>>\ell_\mathrm{\perp2}$. These tube-like structures will unfortunately be missed by our systematic investigation which focuses on locally planar structures, but we checked a posteriori that these structures only account for a very small fraction of the dissipation (less than one percent). 
   
    \begin{figure}
       \centering
       \includegraphics[width=0.42\paperwidth]{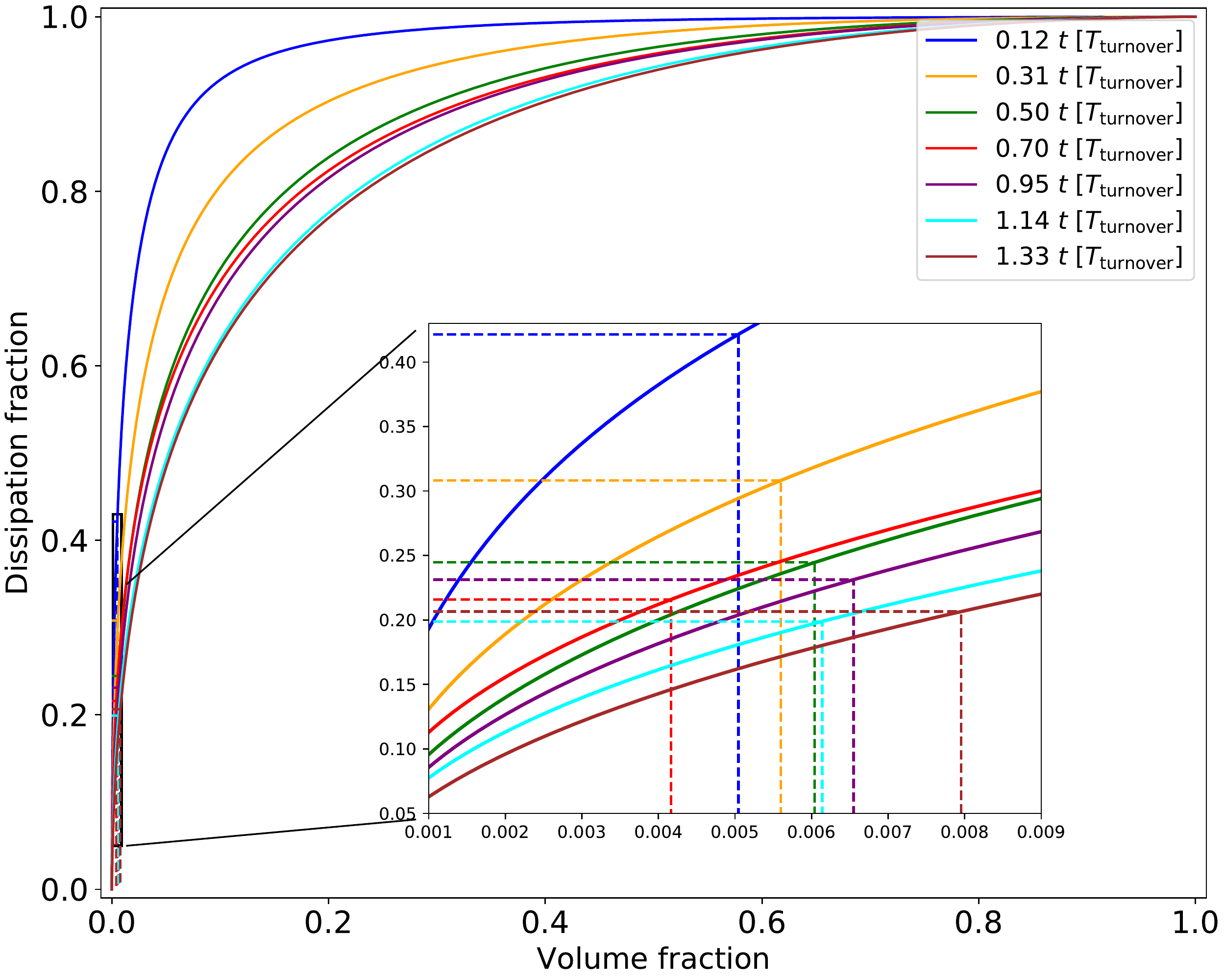}
          \caption{Dissipation filling factor for a simulation with OT initial conditions that started at $\mathcal{M}_\mathrm{s}=4$. Each solid colored curve gives the total dissipation corresponding to the fraction of the volume occupied by the most dissipative regions for different time steps. Vertical dashed lines mark the volume occupied by the selected threshold for the structure detection ($\varepsilon_\mathrm{tot}^\mathrm{corr} \geq \left<\varepsilon_\mathrm{tot}^\mathrm{corr}\right>+4\times \sigma_\mathrm{\varepsilon_\mathrm{tot}^\mathrm{corr}}$) and the horizontal ones the global dissipation fraction that it represent at each time step.}
             \label{fig:fracdiss_fracvol}
    \end{figure}

   Figure \ref{fig:fracdiss_fracvol} shows how the dissipation is distributed in volume: it gives the volume filling factor of the regions of large dissipation as a function of their dissipation fraction. This figure compiles several time steps up to $t=1.33 t_\mathrm{turnover}$, where $t_\mathrm{turnover}=L/\sqrt{<u^2>}=2\pi$ is the initial eddy turnover time. It shows that the intermittency of the dissipation decreases over time, as the r.m.s. sonic Mach number decreases,\change{ because we consider decaying turbulence simulations (figure \ref{fig:TimeDissComp} shows the rapid decline of the sonic Mach number)}. We see here that structures shown on figure \ref{fig:fullcube} (yellow lines for dissipation greater than four standard deviations above the mean) occupy $\simeq 0.8\%$ of the volume while they are at the origin of $\simeq 25 \%$ of the total dissipation rate.
    
    \change{As we use decaying simulations, the Mach number decreases rapidly as well as the dissipation. We have considered snapshots at two characteristic times.  The first snapshot is at 1/3 turnover time, when the first large dissipative structures form, and shortly after the dissipation peak. It is customary to think of the dissipation peak as a point similar to a steady state because the time derivative of the dissipation is zero. However, as we will show, this epoch still bears a strong imprint from the initial conditions. We therefore also consider a second snapshot at one turnover time. We do not consider much later times, as turbulence quickly decays and r.m.s. Mach numbers become much lower than at the beginning (see figure \ref{fig:TimeDissComp}).}
   
   \subsection{Identification of structures} \label{IdentificationTitle}

   \begin{figure*}
    \centering
    \resizebox{\hsize}{!}{\includegraphics{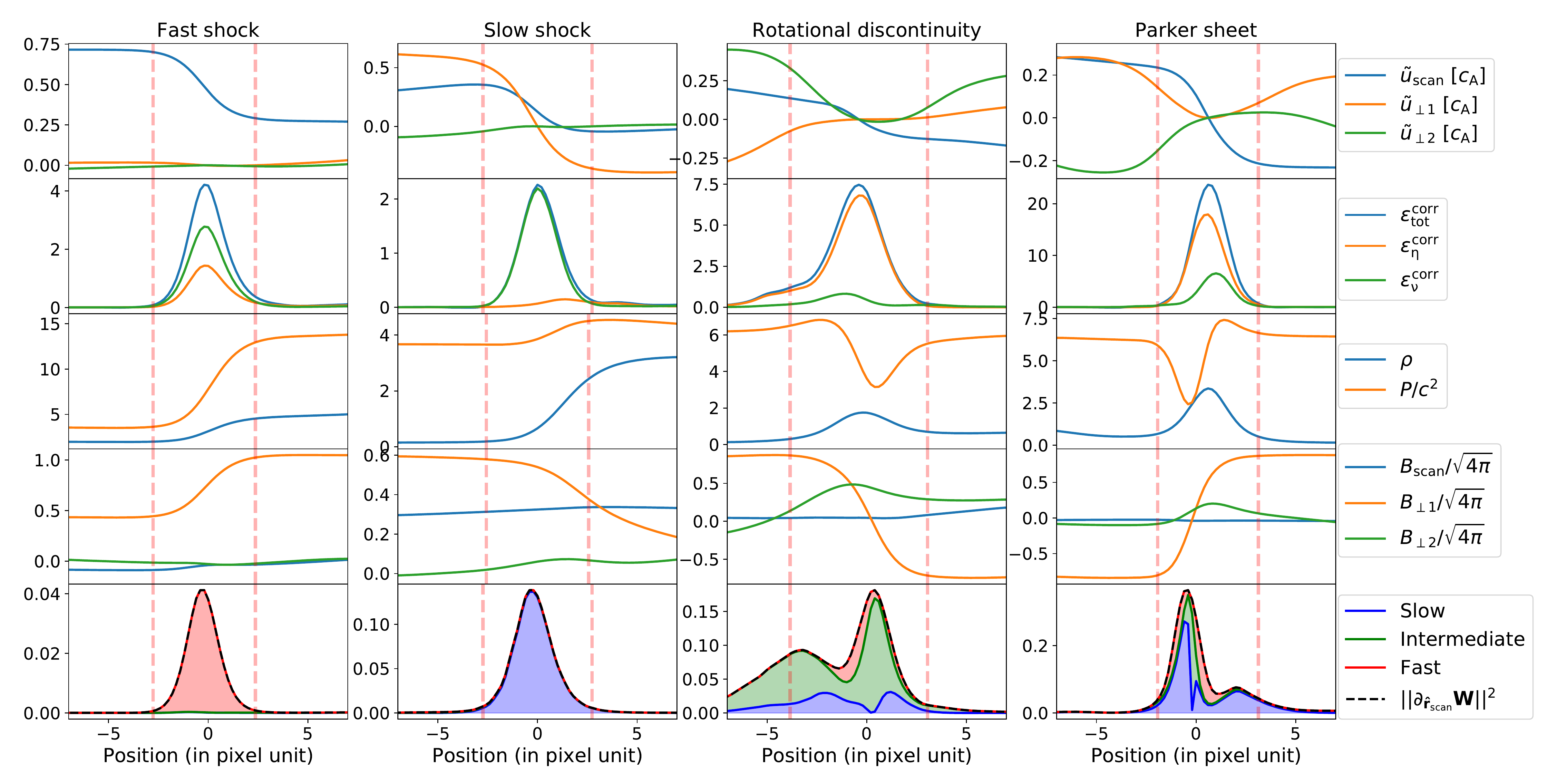}}
    \caption{Representative scan profiles used to identify the different kinds of dissipation structures in our simulations (here, for the ABC simulation at $\Pm=1$ at time $t=t_\mathrm{turnover}/3$). The four first rows of plots show respectively velocities (in the local velocity frame of the scan, and normalized by the initial r.m.s Alfvén speed), dissipation rates, density and total pressure, and magnetic field components profiles. The last row shows gradients decomposition into ideal waves. The colored surfaces in between the curves is proportional to the weight of each corresponding ideal wave (in the decomposition presented in section \ref{GradientDecomp}). Vertical dashed lines on each plot mark the positions of pre- and post-discontinuity that we define in section \ref{PrePost_pos}.}
    \label{1D_scans}%
   \end{figure*}
   
   We develop here the details of our procedure to identify scans along the connected structures.
   
\subsubsection{Scanned profiles}
   
   We consider each connected dissipation structure one at a time.  At a selection of cells (see our selection strategy in subsection \ref{sec:scanning-strategy}), we take $\hat{\vec{r}}_{\rm scan}$ as a scanning direction on which we sample the magnetic field, fluid velocity, density and total pressure $P=\rho c^2+\frac{1}{8\pi}||\mathbf{B_\perp}||^2$ where $\mathbf{B_\perp}$ is the magnetic field transverse to the scanning direction. Note that we do not include the contribution to the total pressure of the magnetic field component in the scanning direction, because it should remain uniform along this direction. We linearly interpolate their values every 0.2 cell side length (this is to avoid accuracy asymmetries resulting from the staggered position of the magnetic field components). As in SHOCK\_FIND \citep{Lehmann2016}, each value is then averaged over a 3-cell radius disk, orthogonal to the scanning direction. This smooths profiles and makes our identification less sensitive to the orientation of the scanning direction with respect to the cell edges. Four representative scans are displayed on figure \ref{1D_scans}.
   
\subsubsection{Pre- and post- positions}\label{PrePost_pos}

   To identify each side of the discontinuity causing the dissipation peak, we define reference positions pre- and post-discontinuity. To do so, we examine the total dissipation profile along the scan direction (figure \ref{1D_scans}, second row), and we estimate the local scale of variation of dissipation  $\ell_\varepsilon$ by fitting a parabola on $\log \varepsilon_\mathrm{tot}^\mathrm{corr}$ over two cell lengths. The resulting scale $\ell_\epsilon$ is usually between 2 and 4 cells length. We adopt +/- 3 $\ell_\varepsilon$ as a good compromise: not too close to the dissipative layer so that the dissipative terms are negligible and not too far away so that the dynamics are still dominated by the discontinuity. 
 
    To improve the reliability of our identification criteria, we allow ourselves to change the sign of the director vector $r_\mathrm{scan}$. We adopt the direction in which the total pressure and density increases from pre- to post-discontinuity. The sign of $r_\mathrm{\perp 1}$ is modified to keep a right-handed coordinates system. If density and total pressure variations are opposite, we then choose the direction of propagation of the dominant ideal wave in the gradient decomposition in ideal waves presented in section \ref{GradientDecomp}.  
    
    \subsubsection{Heuristic criteria}
    
     We first design three categories according to the classical MHD shock types classification that derive from Rankine-Hugoniot (RH) jump conditions: fast shocks, slow shocks and Alfvén discontinuities (see section \ref{RHrelation}). We define three heuristic criteria to sort the resulting profiles into these categories:
   \begin{itemize}
       \item H1. \textbf{Fast shock :} Total pressure rise and transverse magnetic field rise.
       \item H2. \textbf{Slow shock :} Density rise and transverse magnetic field decrease.   
       \item H3. \textbf{Alfvén discontinuity :} Density bump and transverse magnetic field trough. 
   \end{itemize}
   
   To determine the variation of the profiles we compare the values of the pre-discontinuity, peak dissipation and post-discontinuity positions, and each of these values is averaged over a 1-cell side window to avoid spurious variations. By 'rise' and 'decrease', we mean that the variation is monotonic across these three positions, while by 'bump' (resp. 'trough') we mean the central value is above (resp. below) the other two positions. 
   
   For shock identifications, the total densities and pressures must increase. However, for fast shocks, the jump in density is small compared to the jump in total pressure. In some cases, the uncertainty on the position of the post-shock could lead to a non-identification if the relaxation of the post-shock pressure to that of the ambient medium is fast enough. This is why we don't consider a density rise as a reliable criterion for fast shock identification. Slow shocks are the opposite case, the total pressure jump is small compared to the density jump. We then don't include the total pressure rise criterion to identify them.  
   
   Profiles that do not fall into any of the categories are flagged as unidentified.  
   
   \subsubsection{Gradient decomposition criteria}
   
    We now supplement these heuristic criteria by using the gradient decomposition method described in section \ref{GradientDecomp}. Gradient decomposition is another method to characterise locally the nature of the variations of gas state variables across discontinuities. The use of this technique on the analytical profiles of 1D isothermal fast and slow shocks (as computed in appendix \ref{sec:steady-state-MHD-shocks}) shows us that they decompose into almost pure fast and slow magnetosonic waves respectively. We have no prior information on the wave decomposition of Alfvén discontinuities, but we find that profiles corresponding to our heuristic criteria for 
    Alfvén discontinuities yield two exclusive cases: they either decompose mostly into intermediate waves, or they decompose mostly into slow magnetosonic waves. 
    
    For the specific case of a transverse magnetic field inversion, (i.e. the transverse magnetic fields are opposite of each other on the pre- and post- sides of the profile) we find there are two possible ways for it go from one side to the other side: it can rotate continuously until reaching the angle $\pi$, or it can use a co-planar path by shrinking until it vanishes and then expand in the other direction. These two situations cannot be distinguished from pre- and post-discontinuity values alone, as in the classical view of Rankine-Hugoniot relations. The difference resides in the internal structure of the discontinuity itself, with in one case a rotation (which has a gradient decomposition dominated by intermediate waves) and in the other case a co-planar variation of the transverse magnetic field (for which we find a gradient decomposition dominated by slow magnetosonic waves).
   
   For each scan, we thus estimate the relative weight of each type of ideal waves decomposition averaged over the scan as 
   
   \begin{equation}
       \mathcal{F}_{s,i,f}=\frac{\int_{x_\ppre}^{x_\post}dx \left[ (\hat{\vec{e}}^{R}_{s,i,f}.\partial_{\vec{\hat{x}}} \vec{W})^2+(\hat{\vec{e}}^{L}_{s,i,f}.\partial_{\vec{\hat{x}}}\vec{W})^2\right]} {\int_{x_\ppre}^{x_\post}dx||\partial_{\vec{\hat{x}}} \vec{W}||^2}
   \end{equation}
   where subscripts $s,i,f$ are for slow, intermediate or fast.  $x$ is the position along the scanning axis. $x_\ppre$ and $x_\post$ are the pre- and post- discontinuity positions respectively. Note that $\mathcal{F}_{s}+\mathcal{F}_{i}+\mathcal{F}_{f}=1$.
  
   Our identification criteria take into account the agreement between the heuristic and the ideal wave gradient decomposition methods. We therefore define as identified only those structures which show an agreement between the methods according to these criteria :
    \begin{enumerate}
       \item \textbf{Fast shock :} H1 and fast wave dominated gradients ($\mathcal{F}_f > \mathcal{F}_s $ and $\mathcal{F}_f > \mathcal{F}_i $)
       \item \textbf{Slow shock :} H2 and slow wave dominated gradients  ($\mathcal{F}_s > \mathcal{F}_f $ and $\mathcal{F}_s > \mathcal{F}_i $)
       \item \textbf{Rotational discontinuity :} H3 and intermediate wave dominated gradients ($\mathcal{F}_i > \mathcal{F}_f $ and $\mathcal{F}_i > \mathcal{F}_s $)
       \item \textbf{Parker sheet :} H3 and slow wave dominated gradients ($\mathcal{F}_s > \mathcal{F}_f $ and $\mathcal{F}_s > \mathcal{F}_i $).
   \end{enumerate}
   Representative example profiles of the four kinds of dissipative events we encounter are shown on figure \ref{1D_scans}. For some profiles the dominant wave weight does not correspond to the heuristic type: we flag them as misidentified. Finally, note that our divide between rotational discontinuities and Parker sheets may be arbitrary. We found no metric in which the statistics for these two classes clearly separate (i.e. with a gap between them), and there are on the contrary many indications that they just form the two sides of a continuum of Alfvén discontinuities.
   
   \subsubsection{Scanning strategy}
   \label{sec:scanning-strategy}
   We examine each connected dissipation structure one at a time.
   We sort the cells of a given structure by decreasing planarity ($\ell_{\perp 1}/\ell_{\mathrm{scan}}$)  to get the most reliable identification: the most planar cells are scanned first. To prevent overlap of integration domains and to save computation time, once a scan has been indentified, we remove cells around it from the remaining cells to be identified. We remove all the cells that belong to a rectangle parallelepiped, whose square faces are orthogonal to the scan axis and have a side length of 20 cells side length. We then examine the next most planar cell in the remainder of the structure, until we exhaust all cells for that structure. Once we have considered all available structures in the computational domain, we have a list of scans with their identification, for which we discuss the statistics in section \ref{sec:Results}.

   \subsection{Rankine-Hugoniot validations}\label{RHrelation}
   
   Rankine-Hugoniot (RH) relations express jump conditions across discontinuities in their stationary frame \citep{Rankine1870,gurnett_bhattacharjee_2005}. RH relations hold in a very specific situation where the fluid is stationary, with a plane parallel symmetry and homogeneous conditions on either side of a discontinuity. Nothing seems further away than our fully turbulent decaying turbulence simulations. Nevertheless, we want to check if our structure identification allows to recover some of the properties expected from the RH relations. If they hold, it would bring more weight to the selection criteria we have devised, and it would generalise to 3D MHD the results of \citet{CHEMSES} who found that in 2D decaying unmagnetised turbulence, 1D steady-state shocks could be used to model the strongest dissipation structures. In 1D steady-state isothermal MHD, conservation of mass, momentum and magnetic field read 
   
   \begin{equation}\label{Masscons}
       \left[ \rho \vec{u}\cdot\vec{n}\right]_{\ppre}^{\post}=0,
   \end{equation}
   \begin{equation}\label{eqn:MomFluxCons}
       \left[ \rho\vec{u} \left(\vec{u}\cdot\vec{n}\right)+\left(p+\frac{B^2}{8\pi}\right)\vec{n}-\frac{\left(\vec{B}\cdot\vec{n}\right)\vec{B}}{4\pi}\right]_{\ppre}^{\post}=0,
   \end{equation}
   \begin{equation}\label{eqn:BnCons}
       \left[ \vec{B}\cdot\vec{n}\right]_{\ppre}^{\post}=0,
   \end{equation}
   \begin{equation}\label{eqn:ContinuityElec}
       \left[ \vec{n}\times\left(\vec{u}\times\vec{B}\right)\right]_{\ppre}^{\post}=0,
   \end{equation}
   where $\left[\right]_{\ppre}^{\post}$ denotes the difference between the states at pre- and post- discontinuity. $\vec{n}$ is the normal to the discontinuity. In our study, we take $\vec{n}=\hat{\vec{r}}_{\rm scan}$, which contains most of the gradient for highly dissipative cells (see figure \ref{FigPlanars}). In other words, the planar region hypothesis which subtends RH relations is well verified for the most intense dissipative regions. 
   
    Across the discontinuity, the velocity of the fluid transitions from above to under a characteristic speed set by the three MHD linear wave speeds ($c^{R}_{s,i,f}$, see section \ref{GradientDecomp}). This leads to the traditional MHD velocity regimes classification \citep{Delmont2011}, where numbers designate up- and downstream states, and $u_n=\vec{u.n}$: 
   \begin{itemize}
       \item 1- superfast $u_n \geq c^R_f$ 
       \item 2- subfast/super-alfvénic $c^R_i \leq u_n \leq c^R_f$
       \item 3- sub-Alfvénic/superslow $c^R_s \leq u_n \leq c^R_i$
       \item 4- subslow $u_n \leq c^R_s$
   \end{itemize}

   The discontinuity type is labelled by $i \rightarrow j$, where $i \geq j$. These discontinuity types show different behavior for the transverse magnetic fields :  
   \begin{itemize}
       \item $1 \rightarrow 2$ are fast shocks. Magnetic field is refracted away from the shock normal, which yields a transverse magnetic field increase. Fast shocks efficiently convert kinetic to transverse magnetic energy. 
       \item $3 \rightarrow 4$ are slow shocks. Magnetic field is refracted toward the shock normal, which yields a transverse magnetic field decrease. Slow shocks are efficient at compressing the gas.
       \item $1 \rightarrow 3$, $1 \rightarrow 4$, $2 \rightarrow 3$, $2 \rightarrow 4$ are intermediate shocks. The transverse magnetic field flips across the shock normal.
       \item $2=3 \rightarrow 2=3$ are called Alfvén discontinuities or rotational discontinuities. The norm of the transverse magnetic field is unchanged between pre- and post-discontinuity regions, and only its direction changes in the plane parallel to the discontinuity. Alfvén discontinuities are believed to be efficient at reconnecting the field lines \citep{Zweibel1997,Zhdankin2013}.  
   \end{itemize}
   Density and total pressure profiles also show different signatures. In the first three cases, these profiles are jumps whose amplitude depends on the parameters of the shock. In the case of Alfvén discontinuities, these quantities must be identical on both sides of the discontinuity. 
   
    \subsubsection{Transverse magnetic field}
       \begin{figure}
       \centering
       \includegraphics[width=0.42\paperwidth]{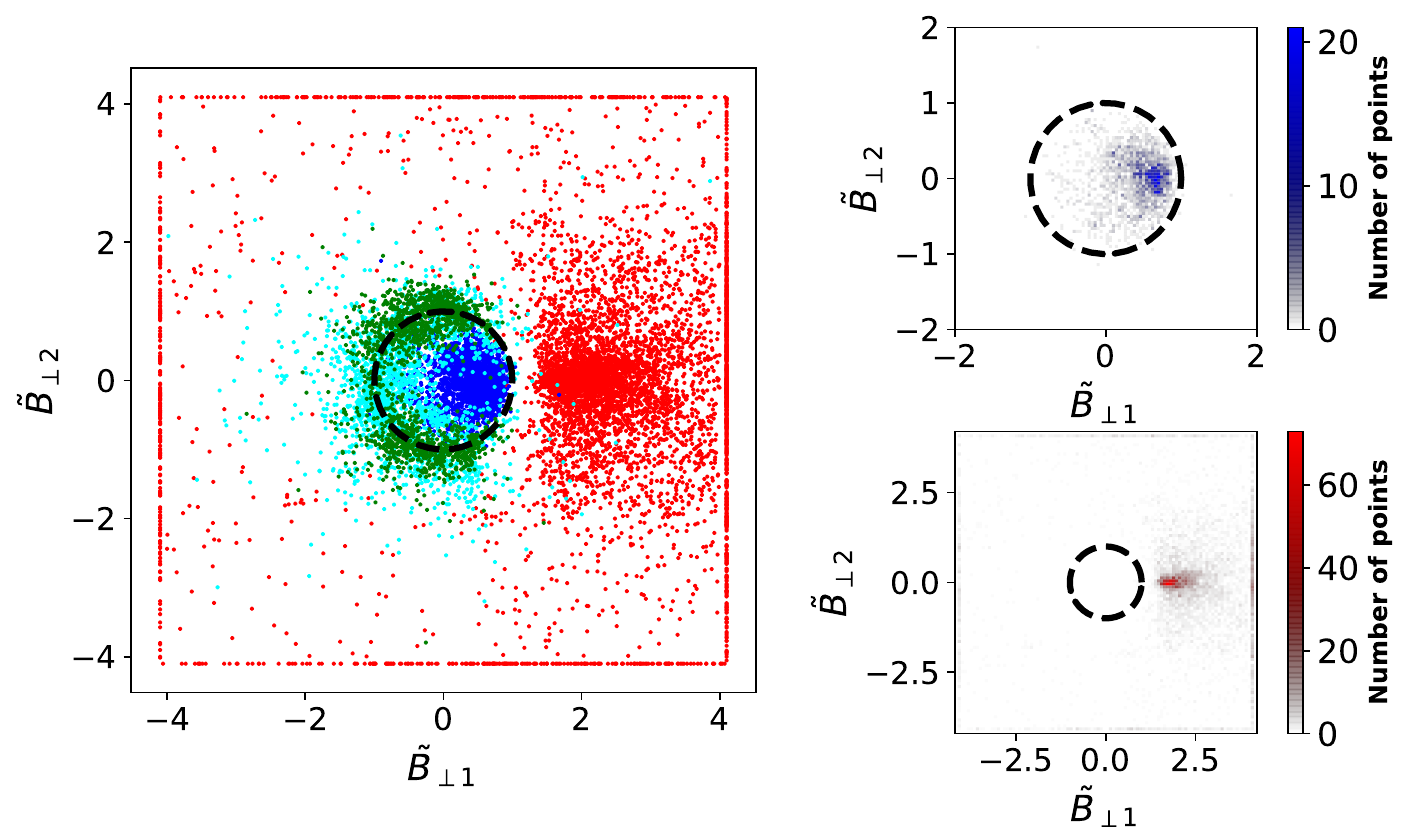}
          \caption{OT $\Pm=1$ run near dissipation peak (at time $t=1/3 t_\mathrm{turnover}$). Hodogram in which the pre-shock magnetic field is normalised and rotated such that $\tilde{B}_{\perp1}=1$ and $\tilde{B}_{\perp2}=0$. The post-shock magnetic field is plotted according to this rotation and normalization. Red dots denote fast shocks, blue ones slow shocks, green is for rotational discontinuities and cyan is for Parker sheets. On the right the top plot is the probability density function of the number of blue dots from the left plot. The bottom right one is for red dots. Dashed line is the unity radius circle that separates discontinuities where the transverse magnetic field increases (outside the circle) and those where it decreases (inside the circle).}
             \label{OThodo}
       \end{figure}  
    
       \begin{figure}
       \centering
       \includegraphics[width=0.42\paperwidth]{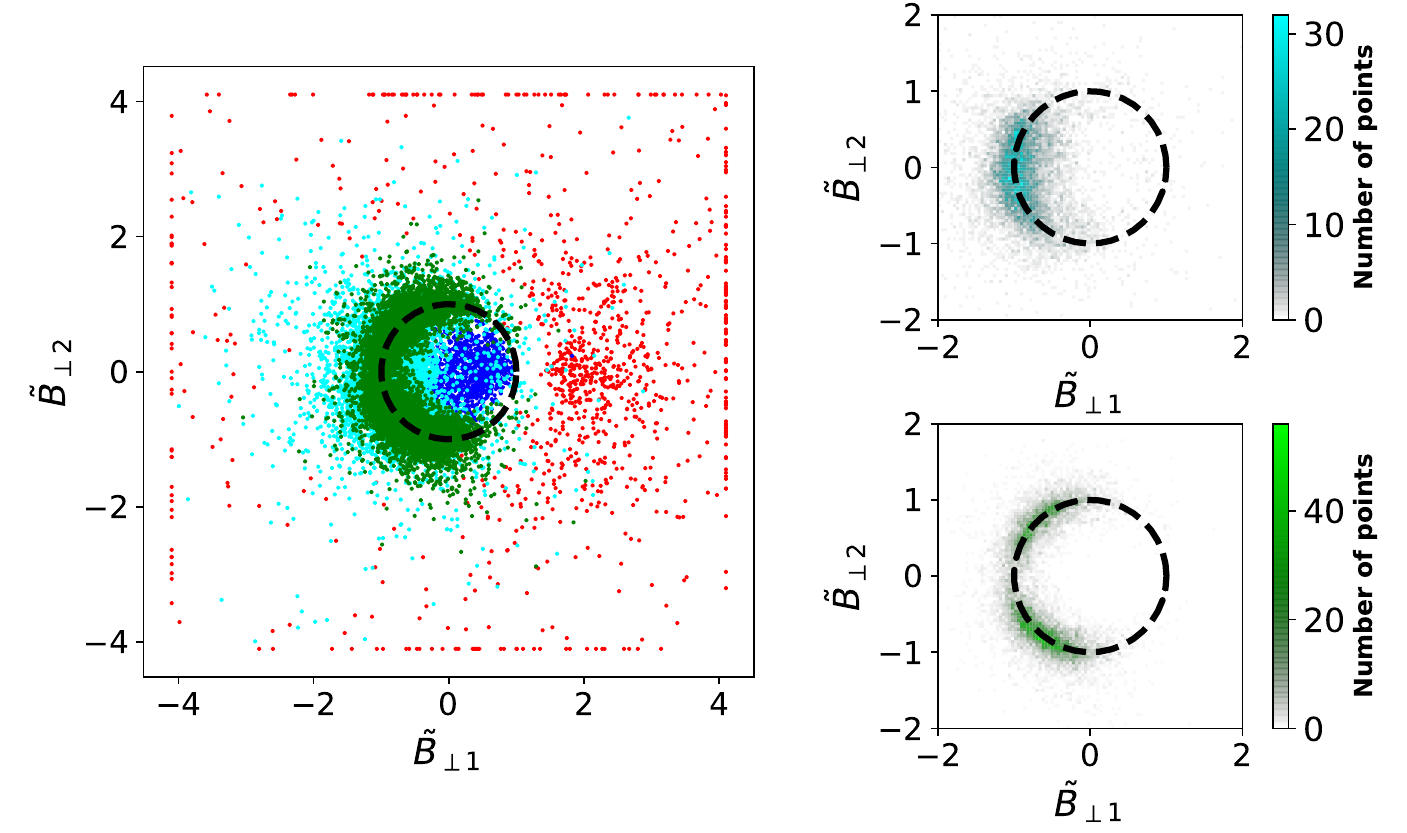}
          \caption{ABC $\Pm=1$ run near dissipation peak (at time $t=1/3 t_\mathrm{turnover}$). The left plot is identical to the one presented on figure \ref{OThodo}. Top right plot shows the number of dots PDF for Parker sheets. Bottom right is for rotational discontinuities. }
             \label{ABChodo}
       \end{figure}  
       
    Each type of RH discontinuity exhibits a different signature in the transverse magnetic field evolution from pre- to post- discontinuity. Our heuristic criteria to identify structures with 1D profiles use only the norm of the transverse magnetic field. We now examine the behaviour of the {\it direction} of the field to check its consistency with the RH relations, and we plot each structure in the form of an hodogram. We normalize the pre-discontinuity magnetic field and rotate our frame so that every scan has the same starting point. Applying the same rotation and normalization to post-discontinuity magnetic field allows us to see relative variations in norm and angle of the transverse magnetic field across the discontinuity : 

    \begin{equation}\label{eqn:Normalization}
        \begin{pmatrix}
         \tilde{B}_{\perp1}\\ 
         \tilde{B}_{\perp2}
        \end{pmatrix}
        =\frac{1}{|\mathbf{B}_{\perp,\mathrm{pre}}|^2}
        \begin{pmatrix}
         B_{\perp1,\mathrm{pre}} & B_{\perp2,\mathrm{pre}}\\ 
         -B_{\perp2,\mathrm{pre}} & B_{\perp1,\mathrm{pre}}
        \end{pmatrix}
        \begin{pmatrix}
         B_{\perp1,\post} \\ 
         B_{\perp2,\post} 
        \end{pmatrix}        
    \end{equation}
    where $\mathbf{B}_{\perp}=\left(B_{\perp1},B_{\perp2}\right)$ is the transverse magnetic field in the frame defined by the local gradient method (section \ref{GradientMethod}). The rotation matrix and the normalisation coefficient applied to the magnetic field depend only on the pre-discontinuity magnetic field components in this frame. $\mathbf{\tilde{B}}_{\perp}=\left(\tilde{B}_{\perp1},\tilde{B}_{\perp2}\right)$ is the post-discontinuity magnetic field that has been rotated and normalized. In the following, the subscript $n$ refers to the component orthogonal to the discontinuity plane, $B_n=\vec{B.n}$ for the magnetic field and $u_\mathrm{n}=\vec{u.n}$ for the velocity field. 
    
    Alfvén discontinuities are characterized by $u_n\ne0$ and $\left[\rho \right]_{\ppre}^{\post}=0$, so equation \eqref{Masscons} leads to $\left[u_n \right]_{\ppre}^{\post}=0$. Conservation of momentum flux equation \eqref{eqn:MomFluxCons} in the normal direction then yields $\left[B_{\perp}^2 \right]_{\ppre}^{\post}=0$. The transverse magnetic field norm is conserved which,  with our normalization, results in Alfvén discontinuities remaining on the circle $\tilde{B}_\perp = \sqrt{\tilde{B}_{\perp1}^{2}+\tilde{B}_{\perp2}^{2}}=1$.
    
    Shocks are characterized by a fluid flow across the discontinuity, $u_n\ne 0$, and a non-zero density jump, $\left[\rho\right]_\mathrm{pre}^\mathrm{post}\ne 0$. Mass flux conservation equation \eqref{Masscons} gives $\left[\rho u_\mathrm{n} \right]_\mathrm{pre}^\mathrm{post}=0$, and with $\left[B_\mathrm{n}\right]_\mathrm{pre}^\mathrm{post}=0$ (equation \ref{eqn:BnCons}) it allows us to rewrite the transverse momentum flux conservation as
    
    \begin{equation}\label{eqn:TransMom}
        \rho u_\mathrm{n} \left[\mathbf{u_\perp}\right]_\mathrm{pre}^\mathrm{post} - \frac{B_\mathrm{n}}{4\pi}\left[\mathbf{B_\perp}\right]_\mathrm{pre}^\mathrm{post}=0
    \end{equation}
    
    and the jump condition \eqref{eqn:ContinuityElec} becomes
    
    \begin{equation}
        \rho u_\mathrm{n} \left[\mathbf{\frac{B_\perp}{\rho}}\right]_\mathrm{pre}^\mathrm{post} - B_\mathrm{n}\left[\mathbf{u_\perp}\right]_\mathrm{pre}^\mathrm{post}=0.
    \end{equation}
    
    We first notice that $\left[\mathbf{\frac{B_\perp}{\rho}}\right]_\mathrm{pre}^\mathrm{post}$, $\left[\mathbf{B_\perp}\right]_\mathrm{pre}^\mathrm{post}$ and $\left[\mathbf{u_\perp}\right]_\mathrm{pre}^\mathrm{post}$ are all co-linear. Solving the second equation for $\left[\mathbf{u_\perp}\right]_\mathrm{pre}^\mathrm{post}$ and substituting it into the first equation then gives 
    
    \begin{equation}
        \left(\rho u_\mathrm{n}\right)^2 \left[\mathbf{\frac{B_\perp}{\rho}}\right]_\mathrm{pre}^\mathrm{post} - B_\mathrm{n}^2\left[\mathbf{B_\perp}\right]_\mathrm{pre}^\mathrm{post}=0
    \end{equation}
    that can be rewritten in the form 
    
    \begin{equation}
        \mathbf{B_{\perp,\mathrm{pre}}}\left(\frac{\left(\rho u_\mathrm{n}\right)^2}{\rho_\mathrm{pre}}-\frac{B_\mathrm{n}^2}{4\pi}\right)=\mathbf{B_{\perp,\mathrm{post}}}\left(\frac{\left(\rho u_\mathrm{n}\right)^2}{\rho_\mathrm{post}}-\frac{B_\mathrm{n}^2}{4\pi}\right).
    \end{equation}
    
    It is clear from these equations that pre- and post- shock magnetic fields must be co-linear. On an hodogram, with the normalization and rotation we apply to our post-discontinuity magnetic field (see equation \ref{eqn:Normalization}), all the shocks must remain at $\tilde{B}_{\perp2}=0$, while $\tilde{B}_{\perp1}>1$ for fast shocks, $0<\tilde{B}_{\perp1}<1$ for the slow ones and $\tilde{B}_{\perp1}<0$ for intermediate shocks.

    On figures \ref{OThodo} and \ref{ABChodo} hodograms are shown for respectively OT and ABC initial conditions. The two PDFs of figure \ref{OThodo} show that the vast majority of the points indeed cluster around the horizontal axis, where RH relations predict that fast and slow shocks should lie. Individual fast shocks that seem very far from coplanarity correspond to switch-on shocks, a limiting case of fast shocks, were the pre-shock transverse magnetic field is null:  the normalisation we introduced with respect to the pre-shock field sends the finite post-shock magnetic fields to infinity... Nevertheless, the finite spread along the $\mathbf{\tilde{B}_{\perp2}}$ axis for slow and fast shocks is an indication that there are deviations from the 1D RH relations. We conjecture that the origin of this discrepancy is due to violation of the 1D mass flux conservation for a large number of scans. This can originate from a leak of material in the plane of the shock (small deviations from the pure plane-parallel case) and/or through the difficulty to accurately probe mass flux conservation compared to other quantities, as noted in appendix \ref{sec:numerical-dissipation}.
    
    The second hodogram in the ABC case (figure \ref{ABChodo}) highlights Parker sheets (cyan dots) and rotational discontinuities (green dots). As for figure \ref{OThodo}, their 2D PDFs behave as expected from RH relations: the transverse magnetic field norm remains unchanged from pre- to post-shock, only the direction of the field changes. 
    Because Parker sheets are dominated by slow waves gradients, which are co-planar, they are hence constrained to perform a full $\pi$ rotation of the transverse field, which is indeed where the PDFs cluster. A surprising result highlighted by the PDFs is that rotational discontinuities have a lack of occurrences for such full $\pi$ rotations: inversions of the transverse magnetic field mostly occur through co-planar structures (which we call Parker sheets) rather than rotational discontinuities. The rotational discontinuities also show no rotation angle less than $\pi/2$. This is an effect of the threshold we apply in our method of detection of high dissipation structures. Structures with lower rotation of the transverse magnetic field dissipate less, and we do not detect them (we have checked that we see smaller angles when lowering that threshold to two standard deviations above the mean instead of four). 
    
    There are also important differences between the initial conditions ABC and OT concerning the distribution of the identifications of the different scans in the early times. These differences will be discussed in section \ref{ICimpact}.
    
    \subsubsection{Velocities estimates}
    \label{sec:velocities}
       \begin{figure}
        \centering
        \includegraphics[width=0.42\paperwidth]{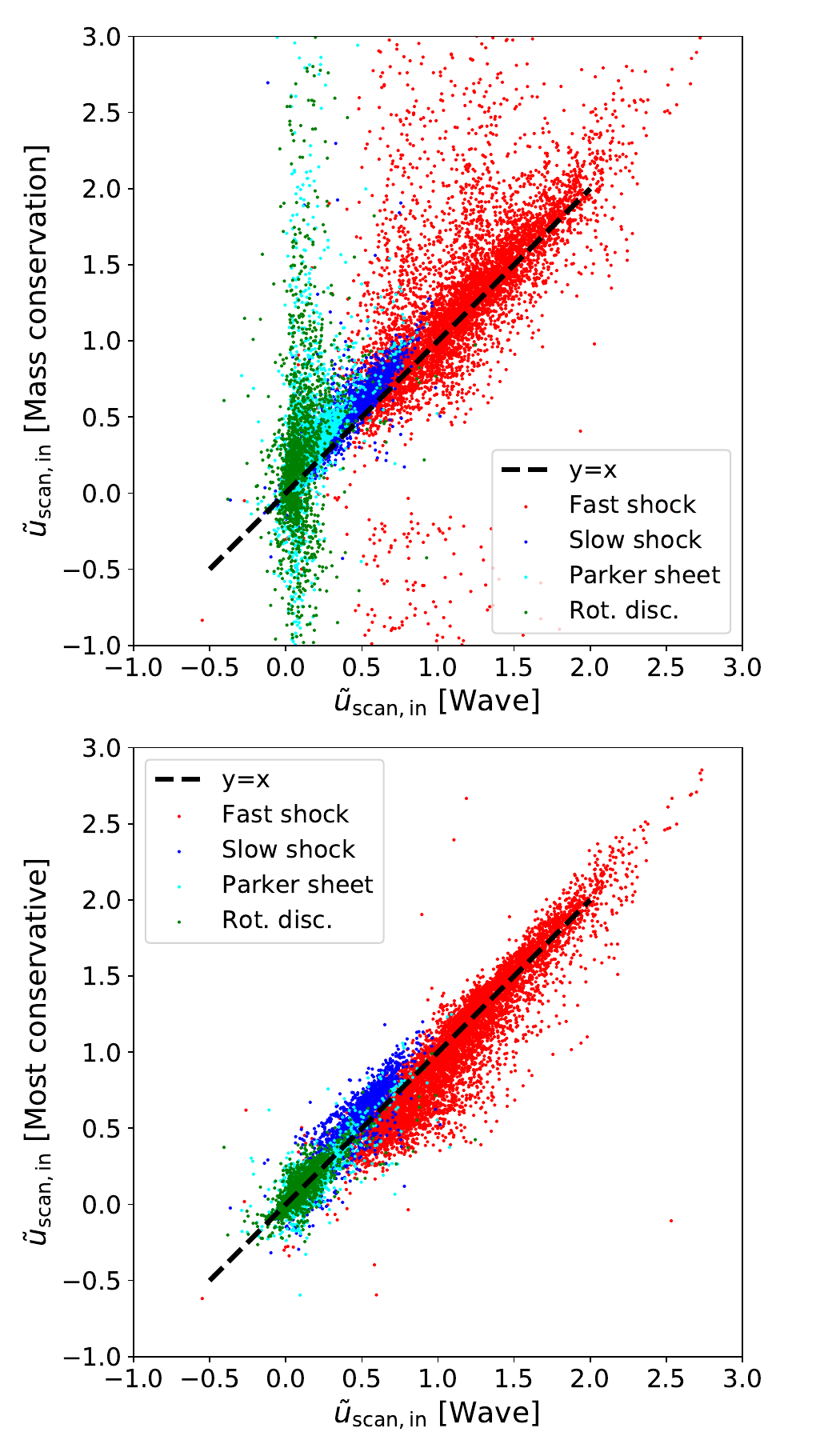}
          \caption{Comparison between different methods to access the velocity of the gas entering in the discontinuity in its co-moving frame (for the OT simulation at $\Pm=1$ at time $t=t_\mathrm{turnover}/3$). $\tilde{u}_\mathrm{scan,in}\  \left[ \mathrm{Wave}\right]$ is the gas velocity derived from the stationary wave frame (see subsection \ref{sec:velocities}). $\tilde{u}_\mathrm{scan,in}\  \left[\mathrm{Most\ conservative\ frame}\right]$ is determined in the frame that minimizes the violation of all fluxes conservation relations, mass flux excepted. $\tilde{u}_\mathrm{scan,in}\  \left[\mathrm{Mass\ conservation}\right]$ is the mass flux conserving frame. } \label{ABCVelComp}
       \end{figure}

    The velocity regimes pre- and post-discontinuity completely characterise discontinuity types. However, to estimate them, we must first determine the rest frame of the discontinuity with appropriate accuracy. We compare three independent methods to derive it :
    \begin{itemize}
        \item \textbf{Mass flux conservation:} To establish the stationary frame in the SHOCK\_FIND algorithm, \cite{Lehmann2016}  derive from equation \eqref{Masscons}   
        \begin{equation}
            u_\mathrm{ref}=\frac{u_\mathrm{scan,post}-\frac{\rho_{\ppre}}{\rho_{\post}}u_\mathrm{scan,pre}}{1-\frac{\rho_{\ppre}}{\rho_{\post}}}    
        \end{equation}
        were $u_\mathrm{ref}$ is the traveling velocity of the discontinuity in the frame of the computing domain. Note that when the density contrast is weak, the denominator goes to zero, making this estimate prone to large errors.
        \vspace{0.3cm}
        
        \item \textbf{Most conservative frame:} We use all the other conservation relations. We first introduce the traveling velocity $u_\mathrm{ref}$ with the frame change  $\Tilde{u}_n=u_\mathrm{scan}-u_\mathrm{ref}$. We then
      we consider the sum of he squared norms of the left hand sides of equations \eqref{eqn:MomFluxCons}, \eqref{eqn:BnCons} and \eqref{eqn:ContinuityElec}.  When $u_\mathrm{ref}$ is indeed the velocity of the discontinuity relative to the gas, this sum should be zero because all the conservation relations will be verified. We therefore estimate $u_\mathrm{ref}$ as the velocity which minimises the sum. Note that we drop mass conservation \eqref{Masscons} from the sum, because of mass leak through the working surface of the discontinuities which makes it less accurate.  This method is inspired from a more general technique described in \cite{Lesaffre2004b} to compute the local stationary frame in multi-fluid 1D simulations. 
        \vspace{0.3cm}
        
        \item \textbf{Stationary wave frame:} We use the propagation speed of the most representative wave given by the gradient decomposition at the dissipation peak (see section \ref{GradientDecomp}). Decomposition in slow and fast waves are always pure right or left going waves. We then simply choose the velocity at the dissipation peak corresponding to this wave. On the other hand, for intermediate waves, they are often right going on one side and left going for the other. In this case we take the average velocity weighted by the strength of the corresponding right and left going wave (the two averaged velocities usually turn out to be both small).      
    \end{itemize}
    
\begin{figure*}
        \centering
        \resizebox{\hsize}{!}{\includegraphics{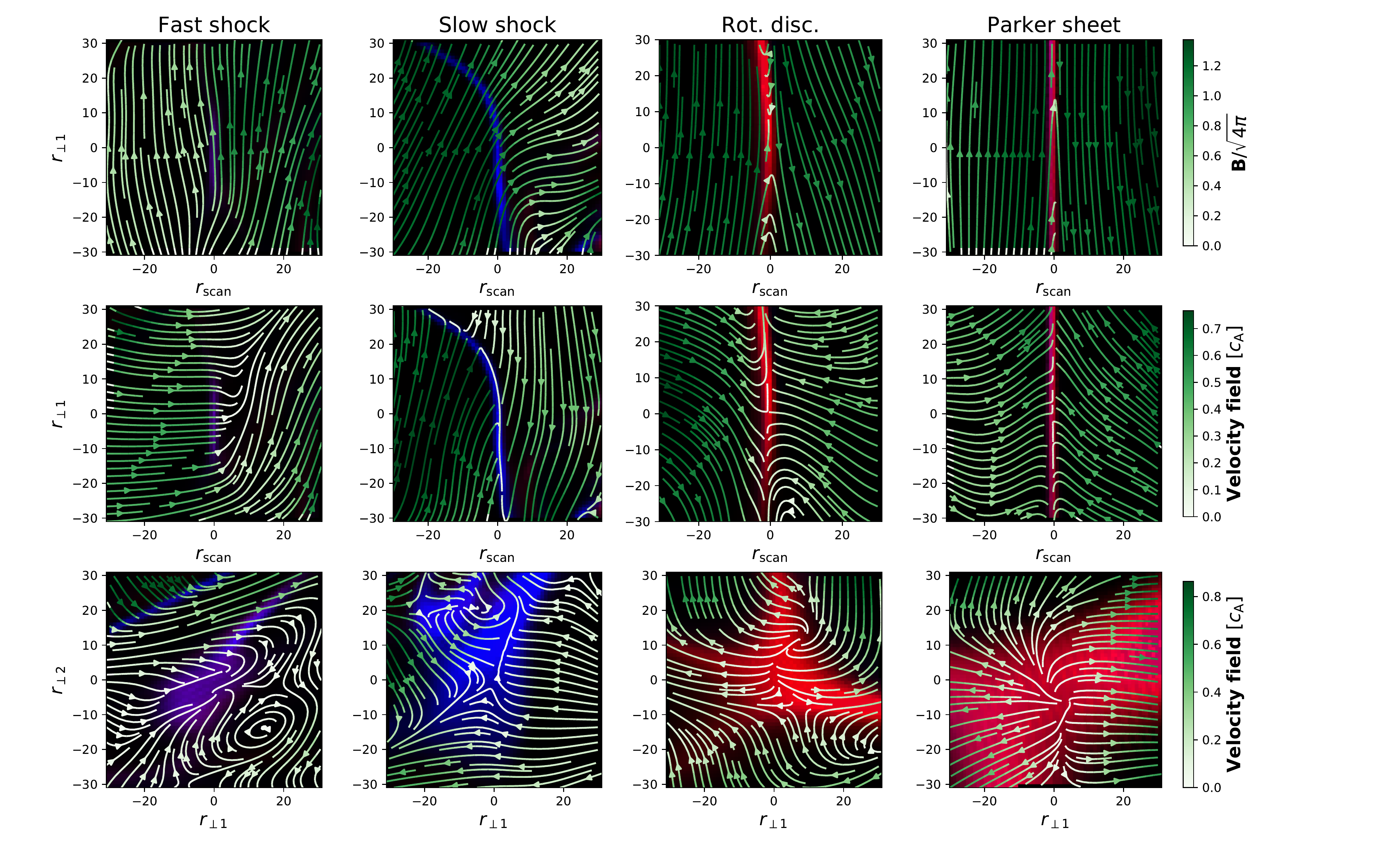}}
        \caption{ First two rows: cuts across the plane $r_\mathrm{scan}$-$r_{\perp1}$ for four examples of structures we identify (for the same four scans as in figure \ref{1D_scans}). Last row: cuts across the plane $r_{\perp1}$-$r_{\perp2}$ for the velocity stream lines. Top plots show magnetic field lines, while bottom ones show velocity stream lines. The frame of reference is set to be the stationary wave frame at the center of these images. The background is a two channel color map with red assigned to ohmic dissipation and blue to viscous.  }
        \label{fig:cutparkersheet}%
       \end{figure*}

    On figure \ref{ABCVelComp}, we compare the fluid velocity entering in the discontinuity by the pre-shock side ($\Tilde{u}_\mathrm{scan,in}=u_\mathrm{scan,pre}-u_\mathrm{ref}$), in the frame established with these three methods. On the top plot we notice that the mass flux conservation method is inconsistent with the stationary wave frame method for rotational discontinuities and Parker sheets and to a lesser extent for fast shocks. For Alfvén discontinuities, this is expected because of the weak density contrast which blows up the denominator in the mass flux conservation estimate. For shocks, the inaccuracy incurred by the mass flux conservation could be due to the difficulty to assess accurate mass conservation compared to other quantities, as noted in appendix \ref{sec:numerical-dissipation}. But it is more likely due to a genuine mass flow that occurs in the dissipating layer of the discontinuity, transverse to the propagation direction. The figure \ref{fig:cutparkersheet} illustrates this phenomenon clearly: stream lines are converging or diverging in the $(\vec{r}_{\perp 1},\vec{r}_{\perp 2})$ plane on the last rows. This was a known phenomenon for Parker sheets, where converging flows orthogonal to the reconnection zones are balanced by diverging flows in the plane of the current sheet. However, that this phenomenon is also present for shocks and rotational discontinuities is a discovery. In the case of shocks, we believe this provides the mechanism which allows the relaxation of the post-shock pressure toward that of the ambient medium. Furthermore, the fact that the SHOCKFIND estimate for shocks is biased towards higher values hints at mass loss in the direction transverse to the working surface (or diverging streamlines, opposite to the example case shown in figure \ref{fig:cutparkersheet} where it should however be noted that the velocities are really small so that this mass loss is almost insignificant). 
    
    The bottom plot of figure \ref{ABCVelComp} shows a relatively good agreement between the  other two independent methods. However, the stationary wave frame tends to give slightly higher velocities for fast shocks and slightly lower for slow shocks. For Alfvén discontinuities, the  agreement is optimal, no bias is observed. We choose to use the stationary wave frame in the following, because it gives pre- and post- velocity regimes more consistent with the RH nature of the discontinuities, which we now turn to check.

    \subsubsection{Velocity regimes}

        \begin{figure*}
           \centering
           \resizebox{\hsize}{!}{\includegraphics{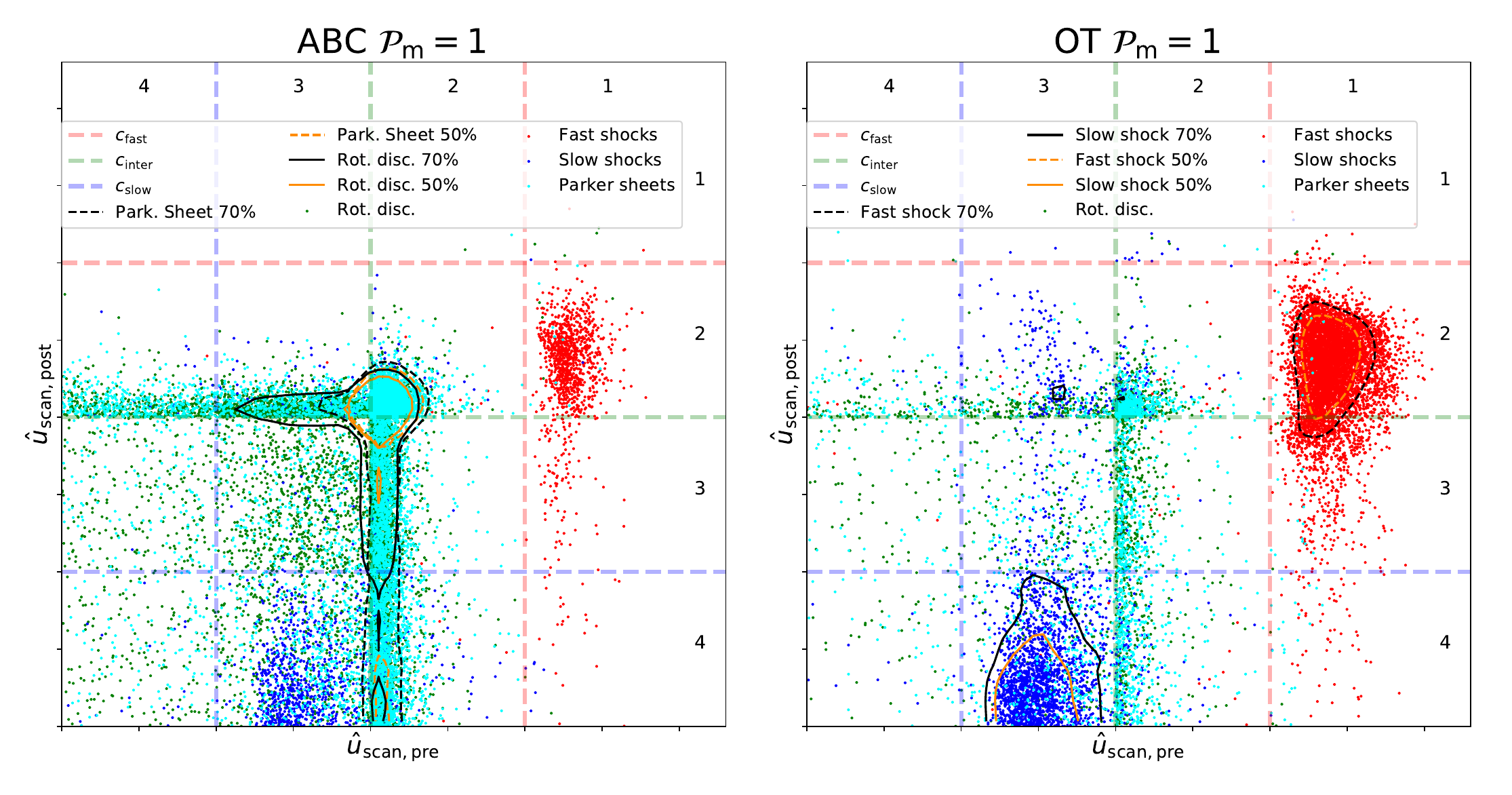}}
           \caption{The figure on the left is for ABC and the one on the right is for OT ($\Pm=1$), both near the dissipation peak. For each scan we compute pre- and post-shock slow, intermediate and fast velocities. We compare fluid velocities to these characteristic speeds in the stationary wave frame. We normalise velocities according to the pre- and post- regime following equations \eqref{eq:norm1} to \eqref{eq:norm4}. X-axis is the pre-shock regime and Y-axis the post-shock. Thus, each kind of discontinuity, in the classical MHD discontinuity classification, belongs to one box. Isocontours in solid and dashed lines are computed for the two most represented kinds of profile. The zones inside the contours delineate the densest area comprising respectively 70\% (black) and 50\% (orange) of the dots. }
           \label{fig:shockdiag}
        \end{figure*}
       
   \begin{figure*}
       \centering
       \resizebox{\hsize}{!}{\includegraphics{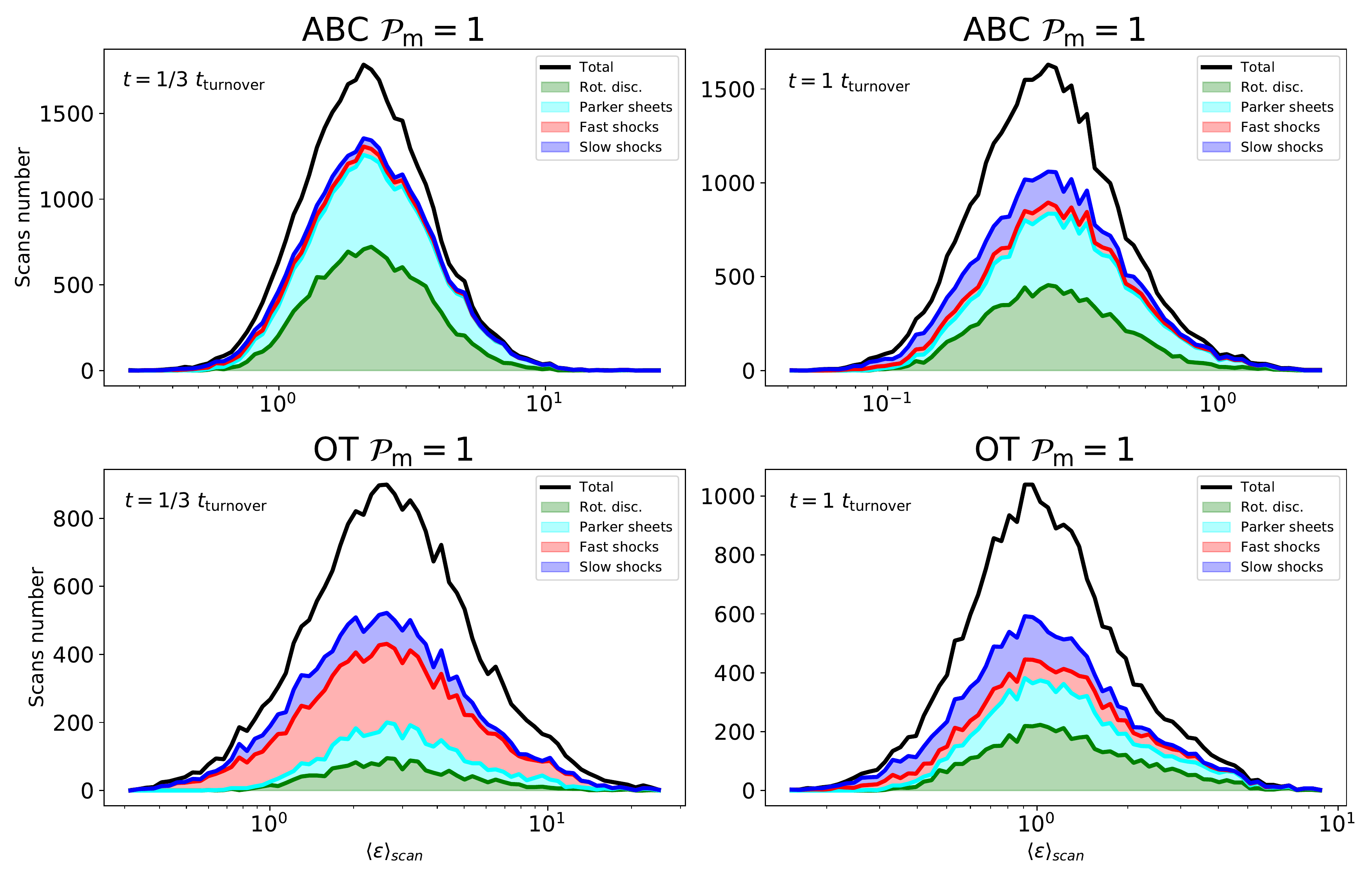}}
       \caption{Distribution of the different types of structure in terms of the mean scanned dissipation across the discontinuity. The black curve is the total distribution of scans, whereas coloured curves are identified structures contribution. The white area corresponds to unknown dissipation scans. Top plots are for ABC initial conditions, while bottom ones are for the OT flow. Distributions on the left are at an early time ($\simeq 1/3 t_\mathrm{turnover}$). The right panel shows the same distribution at  $t=t_\mathrm{turnover}$.}
       \label{fig:diss_comp}
   \end{figure*}
      \begin{figure*}
    \centering
    \resizebox{\hsize}{!}{\includegraphics{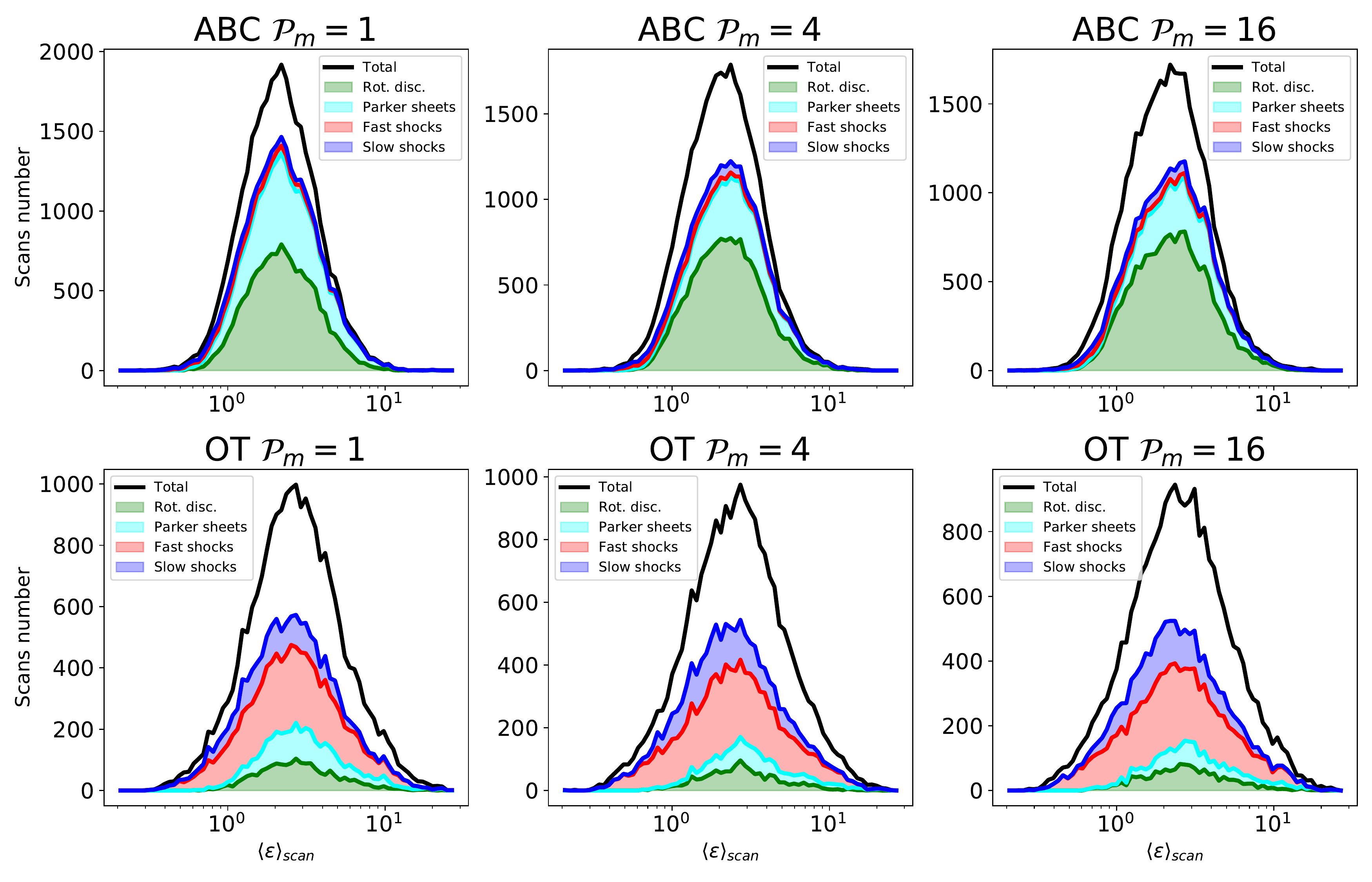}}
    \caption{dissipation structures distributions for our two initial conditions with varying magnetic Prandtl number from $\mathcal{P}_m =1$ on the left to $\mathcal{P}_m =16$ on the right. The time step shown here is at early time, near the dissipation peak.}
              \label{fig:Prandtl_distri}%
   \end{figure*}
   \begin{figure*}
    \centering
    \resizebox{\hsize}{!}{\includegraphics{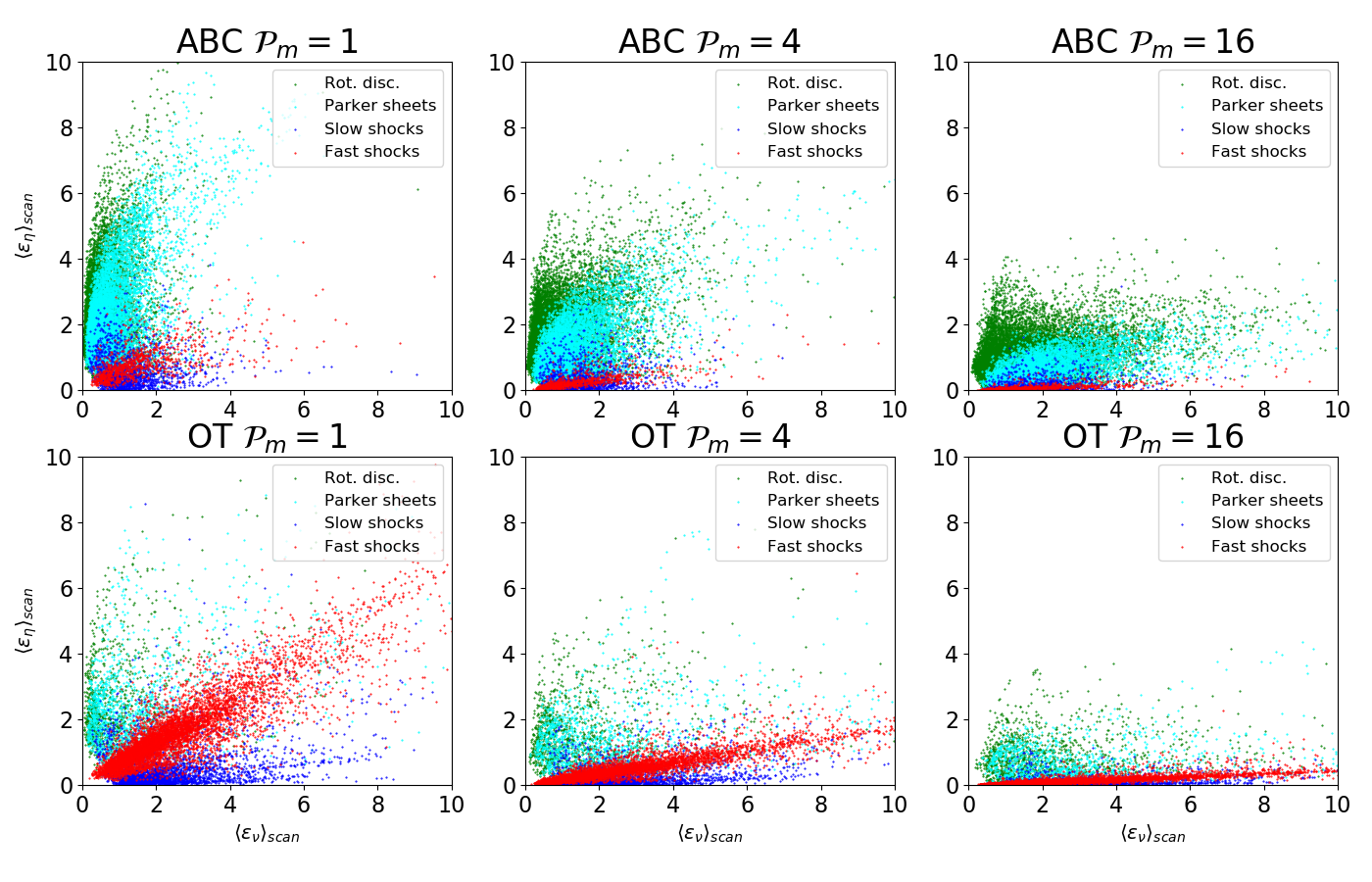}}
    \caption{Distributions of the Ohmic and viscous dissipations averaged within each scan for the ABC simulation above and OT below, according to the different identifications and for the values of the magnetic Prandtl ranging from $\mathcal{P}_\mathrm{m} =1$ on the left to $\mathcal{P}_\mathrm{m} =16$ on the right.}
              \label{fig:Prandtl_diss}%
   \end{figure*} 

    With the proper frame set, we can now study the velocity regime transitions. In order to represent up- and downstream states for all identified scans we use a scatter plot with a normalisation conditioned by the regime: superfast(1) or subfast/super-alfvénic(2) or sub-alfvénic/superslow(3) or subslow(4) : 
    
    \begin{equation} \label{eq:norm1}
        1\rightarrow    \hat{u}_\mathrm{scan} = 3+ |\tilde{u}_\mathrm{scan}|-c^\mathrm{R}_\mathrm{f} ,
    \end{equation}
    \begin{equation}
        2\rightarrow    \hat{u}_\mathrm{scan} = 2+\frac{|\tilde{u}_\mathrm{scan}|-c^\mathrm{R}_\mathrm{i}}{c^\mathrm{R}_\mathrm{f}-c^\mathrm{R}_\mathrm{i}},
    \end{equation}
    \begin{equation}
        3\rightarrow    \hat{u}_\mathrm{scan} = 1+\frac{|\tilde{u}_\mathrm{scan}|-c^\mathrm{R}_\mathrm{s}}{c^\mathrm{R}_\mathrm{i}-c^\mathrm{R}_\mathrm{s}},
    \end{equation}
    \begin{equation} \label{eq:norm4}
        4\rightarrow    \hat{u}_\mathrm{scan} = \frac{|\tilde{u}_\mathrm{scan}|}{c^\mathrm{R}_\mathrm{s}}
    \end{equation}
    
    where $\tilde{u}_\mathrm{scan}$ is the velocity plotted on the diagram and $c^\mathrm{R}_\mathrm{s,i,f}$ is the local positively signed slow, Alfvén/intermediate or fast speed. Note that the usual integers characterising the velocity regimes are in reverse order compared to our renormalised number $\hat{u}_\mathrm{scan}$. Figure \ref{fig:shockdiag} shows the resulting diagrams. Coloured dashed lines delimit regions specific to the gas velocity regime of a particular discontinuity type (see section \ref{RHrelation}). Columns give  pre-discontinuity regimes, from left to right subslow, sub-alfvénic/superslow, subfast/super-alfvénic and superfast. Whereas rows give access to the post-discontinuity velocity regime, with the same sorting from bottom to top. 
    
    The diagram on the right of figure \ref{fig:shockdiag} shows the results for the simulation with OT initial conditions. Because the points density makes distributions difficult to appreciate at the densest regions, we also compute 2D PDFs for shocks which we use to highlight contours at the value of the median pixel (orange lines) and the third decile one. Scans that are identified as fast shocks are located in the expected region for the most part, identified as $1\rightarrow2$ discontinuities. And the distribution of scans identified as slow shocks are indeed $3\rightarrow4$ discontinuities. However, we note that the slow shocks often have negative velocities out of the shock (not shown). We believe this reflects the fact that the post-shock state is affected by the mass loss in the plane of the shock. The determination of the stationary reference frame by the "most conservative" method gives more positive post-shock velocities (consistent with lower pre-shock velocities as seen on figure \ref{ABCVelComp}, bottom panel).  

    The ABC initial conditions case is shown on the left of figure \ref{fig:shockdiag}, where PDFs contours are now used for  Parker sheets and rotational discontinuities. Those two are very peaked at $2=3\rightarrow2=3$, where we expect Alfvén discontinuities in the traditional MHD shocks classification \citep{Delmont2011}.
    
     For both OT and ABC initial conditions, the  distributions of Alfvén discontinuities are stretched along the horizontal and vertical directions.  We find in these trailing populations an over-representation of scans in which $B_\mathrm{scan}\simeq0$ on one side of the discontinuity and not the other. These structures are hence not perfectly plane-parallel because the magnetic field normal to the discontinuity should be conserved. A nearly zero magnetic field on one side implies that $c_\mathrm{i}^\mathrm{R}\simeq c_\mathrm{s}^\mathrm{R} \simeq 0$, and as a result distances between green dashed lines and zeros are artificially expanded by the graphs normalisation relations \eqref{eq:norm1} to \eqref{eq:norm4}. A small error in the determination of the frame velocity or/and the position of the pre- or post-discontinuity positions leads to exaggerated distances between the expected and the actual position of the dots.

    The match between our identification criteria and the pre- and post-discontinuity velocity regimes is strongly dependent on the determination of the frame in which the structure is stationary. We checked our three methods, and we found that steady frame velocities obtained from the wave decomposition yield the best consistency between the types and the expected state transition for each type of structure.

\section{Results}\label{sec:Results}

    In this section, we use our identification algorithm to extract  statistical results from our simulation set. We study the impact of some of our input parameters on dissipation structures. Our set is composed of simulations with different magnetic Prandtl number ($\mathcal{P}_m$), and two different velocity field and magnetic field configurations (see table \ref{tab:SimuParam}).

   \subsection{Impact of initial conditions}\label{ICimpact}
   
    We consider here the effect of the initial conditions of the simulations on the nature of the structures formed.
    Figure \ref{fig:diss_comp} shows distributions of the identification scans as a function of the mean dissipation rate in the volume probed by each scan, and table \ref{tab:fracID} summarises these results averaged over all scans. The time step chosen for the two graphs on the left half of this figure (as well as for the left half of table \ref{tab:fracID}) is 1/3 of the initial turnover time, shortly after the dissipation peak, when the first and most intense dissipation structures form. The time step chosen for the right hand side of figure \ref{fig:diss_comp} and table \ref{tab:fracID} is after one turn-over. 
   \begin{table}[h]
   \begin{tabular}{|c|c|c|c|c|}
      \hline
      & \multicolumn{2}{c|}{$\frac13 t_\mathrm{turnover}$} & \multicolumn{2}{c|}{$t_\mathrm{turnover}$} \\
       Number fraction & ABC & OT & ABC & OT\\
      \hline
       UnID             &  19\% & 29\% & 24\% & 29\% \\
       MisID            &   7\% & 12\% &  8\% & 11\% \\
       Fast shocks      &   3\% & 32\% &  4\% & 9\% \\
       Slow shocks      &   6\% & 15\% & 12\% & 14\% \\
       Rotational disc. &  36\% &  6\% & 27\% & 21\% \\
       Parker sheet     &  30\% &  7\% & 25\% & 16\% \\
       \hline
       Dissip. fraction & \multicolumn{4}{c|}{}\\
       \hline
       UnID+MisID       &  22\% & 42\% & 30\% & 38\% \\
       Fast shocks      &   3\% & 34\% &  4\% &  9\% \\
       Slow shocks      &   4\% & 11\% & 10\% & 11\% \\
       Rotational disc. &  36\% &  5\% & 28\% & 24\% \\
       Parker sheet     &  34\% &  8\% & 29\% & 18\% \\
      \hline
   \end{tabular}
   \caption{Identification fractions within our scans in number (top) and weighted by dissipation (bottom) for several snapshots and initial conditions in $\Pm=1$ simulations. 
   \label{tab:fracID}}
   \end{table}

   As described in section \ref{sec:simu} we use two types of initial flows, ABC and OT. The main difference between these two flows resides in their magnetic and cross helicities (see subsection \ref{sec:initial-conditions}). This initially yields very different types of structures.
   Early time, OT is dominated by shocks (mainly fast shocks) while ABC is dominated by Alfvén discontinuities (rotational discontinuities and Parker sheets).
   After one turnover time, the impact of the initial conditions on the formation of the dissipation structures seems to be erased.  The main dissipation mechanism is then through rotational discontinuities and Parker sheets for both ABC and OT.
   
   Interestingly, for both types of initial conditions, at early and late times, the distribution of physical natures of scans does not seem to depend on their level of dissipation. Intense dissipative scans and weak scans have about the same proportions of each nature.
  
   Table \ref{tab:fracID} also shows the amount of unidentified and misidentified scans. Between 58\% and 78\% of intense dissipation is identified by our technique, with a greater success rate for ABC than OT. Early time structures are also better identified than later times.
  
   \subsection{Impact of the Prandtl number}\label{sec:PRimpact}

   One critical parameter of dissipation is the magnetic Prandtl number, $\Pm=\nu/\eta$, the ratio of kinematic viscosity $\nu$ to magnetic diffusivity $\eta$    \citep[see e.g.][]{Brandenburg2019}.  We perform our dissipation structure analysis on simulations with a range of magnetic Prandtl numbers, from $\Pm=1$ to $\Pm=16$ (see table \ref{tab:SimuParam}). As dicussed in appendix \ref{sec:numerical-dissipation}, the intrinsic numerical dissipation from the scheme causes the effective Prandtl number to be slightly different from the input one. Thanks to semi-analytical solutions (computed in appendix \ref{sec:steady-state-MHD-shocks}) we probe the effective Prandtl number of our scheme in 1D MHD shocks. Figure \ref{fig:shock_Pm} shows that for moderate velocity shocks (or $u_0\leq 1$) our scheme is already converged while the highest shock velocities we find in simulations are at $u_0 \simeq 5$. We thus remain confident that the effective Prandtl number in our simulations is overall close to the input one, at least as regards shocks. 
   
   Figure \ref{fig:Prandtl_distri} shows OT and ABC identification distributions for input $\Pm=1,4,16$: the magnetic Prandtl does not seem to have any impact on the distribution of structures. The only noticeable difference is a slight increase in the number of rotational discontinuities at the expense of Parker sheets for ABC initial conditions. 
   
   It was shown by \cite{Brandenburg2019} that an increase in $\mathcal{P}_\mathrm{m}$ causes an increase in $<\epsilon_v>/<\epsilon_\eta>$, a result which we confirm and make more precise here. If there is no statistical difference in high dissipation structures distributions, it must be differences in the internal structure of the dissipation layers that leads to differences in dissipation rates. On figure \ref{fig:Prandtl_diss} we show scatter plots of viscous versus Ohmic dissipation rates, where each dot marker expresses the mean of the corresponding dissipation rate within each scan. At $\Pm=1$, it is clear that rotational discontinuities and Parker sheets are dominated by magnetic energy dissipation. Fast shocks are an intermediate case, with a more balanced share between Ohmic and viscous dissipation rates.  Slow shocks are dominated by viscous dissipation. Each type of structure is more or less characterised by a given slope in these graphs (i.e. the ratio $<\varepsilon_v>_\mathrm{scan} / <\epsilon_\eta>_\mathrm{scan}$ is within a more or less well defined sector for each type of structure, regardless of initial conditions OT or ABC). 
   In particular for shocks (both slow and fast),  when $\Pm$ varies, this ratio simply scales as $\Pm$ : the behaviour of dissipation within fast shock scans follows the rule  $<\varepsilon_v>_\mathrm{scan} / <\epsilon_\eta>_\mathrm{scan} \propto \Pm$. This scaling is less clear for the other types of structures: Alfvén discontinuities experience a wider range of ratios at fixed $\Pm$ which makes it less easy to assess if such a scaling is present (in fact, the envelope of green and cyan points suggests a different scaling, closer to $\Pm^{1/2}$). 
   Provided the global dissipation is reflected by intense events, this could explain why the global average ratio  $<\varepsilon_\nu>/<\varepsilon_\eta>$ is also found to scale approximately as $\Pm$ in our simulations.
      We therefore conclude that in our simulations the increase in the viscous over Ohmic fraction when $\Pm$ rises comes not from a difference in the nature of the dissipation structures that form, but from a modification in the way each of these types of structure dissipates internally.
   
   \subsection{Statistics of entrance parameters}
  
\begin{figure*}
    \centering
    \resizebox{\hsize}{!}{\includegraphics{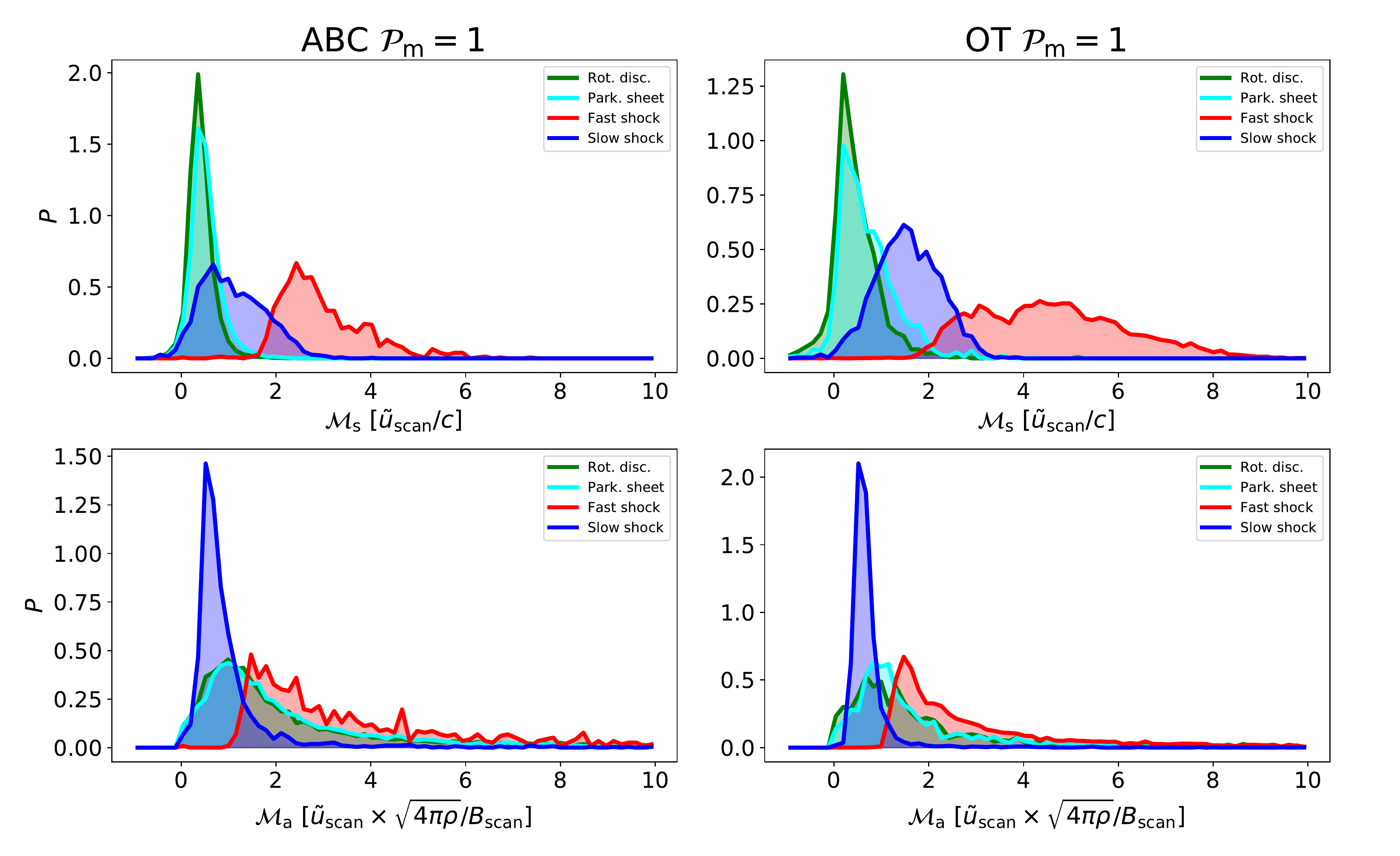}}
    \caption{PDFs of entrance sonic (top row) and Alfvénic (bottom row) Mach numbers in the $\Pm=1$ ABC (left) and OT (right) simulations at time $t=t_\mathrm{turnover}/3$.
              \label{fig:MsMaComp}%
              }
   \end{figure*}

   In this section we consider values of the state variables at the pre- and post- positions on either side of the discontinuities, and we look for statistical differences between the various natures of discontinuities. The distributions of entrance sonic Mach numbers (see figure \ref{fig:MsMaComp}) show without surprise that the entrance velocities are very small for Alfvén discontinuities, moderate for slow shocks and on the order of the r.m.s. Mach number for fast shocks, but with a wide spread distribution. Previous work by \cite{Lehmann2016} has shown the distributions of fast and slow shocks display exponential tails at large velocities. We have less statistics but our data is also consistent with this picture. The bottom row of figure \ref{fig:MsMaComp} displays the distributions for the entrance orthogonal Alfvénic Mach number $\mathcal{M}_\mathrm{a}=\tilde{u}_\mathrm{scan}/(B_\mathrm{scan}/\sqrt{4\pi\rho})=\tilde{u}_\mathrm{scan}/c_i^R$. Naturally, it is above 1 for fast shocks, and below 1 for slow shocks. Its distribution for Alfvén discontinuities is more surprising, though, with wide spread values ranging to values even above 1, while one would expect it to be close to zero. This comes from the fact that the entrance velocities in these discontinuities are on the order or below the sound speed, but the magnetic field happens to be almost transverse, thus yielding very small values for the intermediate speed $c_i^R$. Finally, we also looked at the statistical distributions of the density on the pre-discontinuity side, and found these distributions were independent of the nature of the discontinuity considered.
   
    \begin{figure*}
    \centering
    \resizebox{\hsize}{!}{\includegraphics{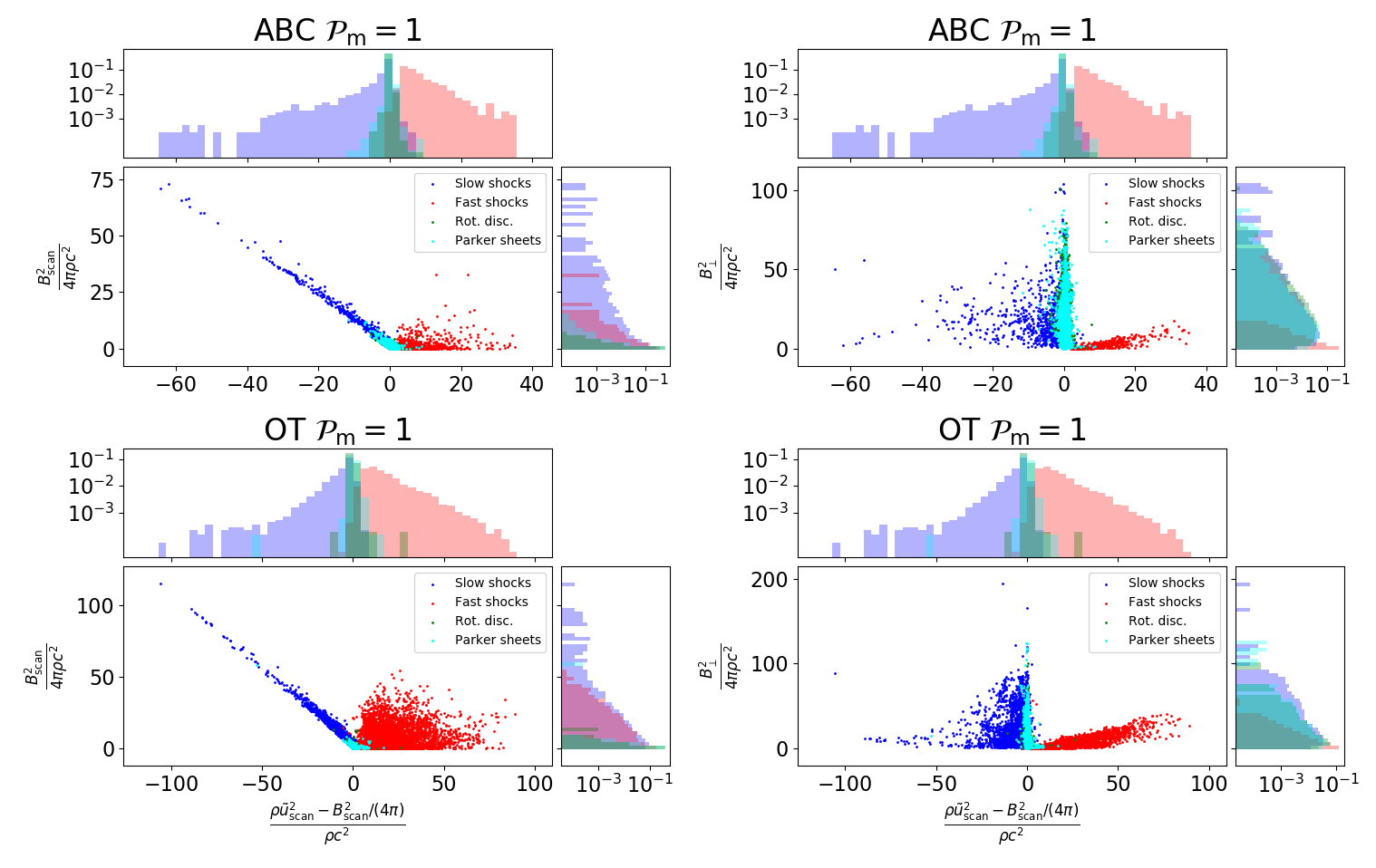}}
    \caption{Entrance (pre-discontinuity) parameters for each scans we identify at time $t=t_\mathrm{turnover}/3$. The position of the pre-discontinuity is defined in section \ref{RHrelation} (as three times the dissipation length $\ell_\varepsilon$ before the dissipation peak). Top plots are for ABC initial conditions and bottom plots are for OT ones. The $x$-axis is the difference between ram and magnetic normal pressures. On left plots the $y$-axis represents the magnetic pressure along the scan direction and on the right ones it represents the transverse magnetic pressure.  All quantities are normalized by the thermal pressure ($p=\rho c^2$). Integrated PDFs are given on each side of the panels. 
              \label{fig:EntranceScatterComp}%
              }
   \end{figure*}

    The environment of each discontinuity is defined by 7 state variables on either side (pre- and post-) of the discontinuity, a total of 14 independent state variables. We can reduce this number by using the 7 conservation relations (mass, momentum and magnetic field, -7 independent variables), a normalization by the pre-discontinuity density (-1 variable), a rotation of the transverse axes to cancel one component of the pre-discontinuity magnetic field (-1 variable), as well as a transverse boost of the frame to cancel the transverse velocity components pre-discontinuity (-2 variables). The environment can thus be fully characterized by a remainder of 3 independent variables, which can all be considered on the pre-discontinuity side.

    For these three independent 'entrance parameters', we choose the normal $p_1$ and transverse $p_2$ magnetic pressures as well as the difference $p_3$ between the ram pressure and the normal magnetic pressure (all are normalised by the thermal pressure $p=\rho c^2$). This choice makes it easy to predict where each discontinuity should lie in the 3D parameter space, according to the sign of $p_3$: negative for slow shocks, zero for Alfvén discontinuities and positive for fast shocks, while $p_1$ and $p_2$ could be arbitrary positive numbers. 
    
    Figure \ref{fig:EntranceScatterComp} displays two projections of this 3D space onto the planes $(p_3,p_1)$ and $(p_3,p_2)$.
   In this parameter space the only forbidden region is set by the positivity of the ram pressure $\frac{B_\mathrm{scan}^2}{4\pi} \geq -  ( \rho u_\mathrm{scan}^2 - \frac{B_\mathrm{scan}^2}{4\pi} ) $, which translates as $p_1 \geq -p_3$. However, we find it is not the only region that is devoid of points: the entrance parameters of our structures do not explore fully the available parameter space. In fact, tight correlations for fast shocks are visible on the right hand panel of figure \ref{fig:EntranceScatterComp} (where $\rho u_\mathrm{scan}^2 \simeq \frac{B_\mathrm{scan}^2}{4\pi} + 0.5 \frac{B_{\perp}^2}{4\pi})$ and similarly  for the slow shocks on the left hand panel (where $\rho u_\mathrm{scan}^2 << \frac{B^2}{4\pi}$). It also appears that rotational discontinuities and Parker sheets are preferably parallel to the magnetic field ($B_\mathrm{scan}<<B_{\perp}$). The latter two discontinuity types are not distinguishable in this parameters space, as we consider here only the pre- and post-discontinuity states without taking into account the internal structure.     
     
      These statistical constraints on the three entrance parameters of the dissipation structures reduce to two the number of independent parameters. It will be the subject of future work to understand the origin of these correlations in MHD turbulence.

\subsection{Transverse velocity differences}

\begin{figure}
       \centering
       \includegraphics[width=0.42\paperwidth]{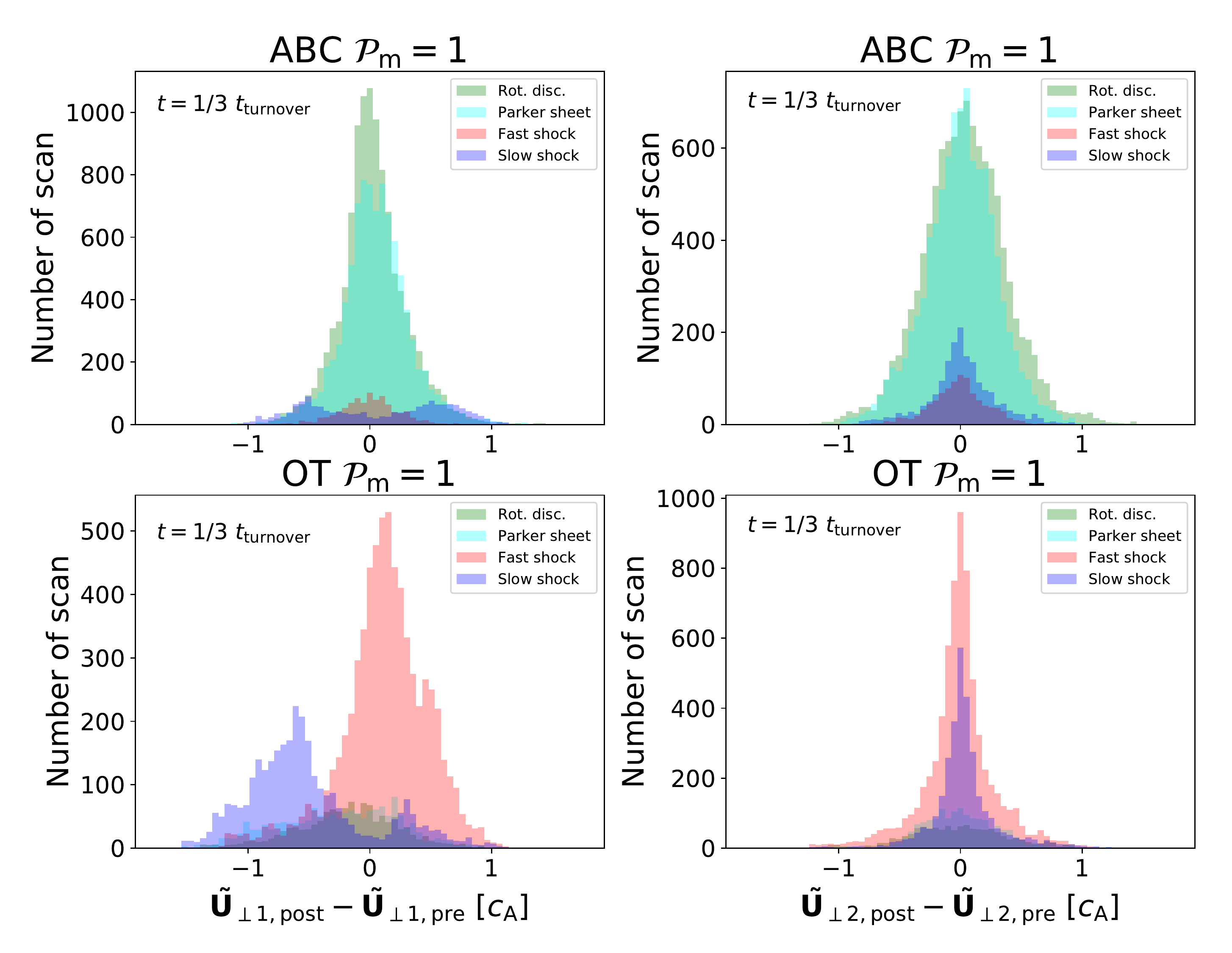}
          \caption{Transverse velocity differences along the two directions $\hat{\vec r}_{\perp 1}$ (left column) and  $\hat{\vec r}_{\perp 2}$ (right column) between pre- and post-discontinuity positions. Top panels are for ABC initial conditions and bottom ones  for OT initial conditions.  The time step of these outputs is $t\simeq 1/3 t_\mathrm{turnover}$.
             \label{fig:TransVel}
             }
\end{figure}

Molecular line observations probe the radial velocities across the plane of sky, while  we have access to the full three dimensional geometry of the intense dissipation regions in our simulations. When projected on the plane of sky, an intense dissipation sheet yields a salient filament-like feature on observable maps where the plane of the sheet is orthogonal to the plane of sky, i.e. where column-density is greatly enhanced through a caustic-like effect. On figure \ref{fig:TransVel} we hence display the velocity difference statistics projected on the two transverse directions $\hat{\vec r}_{\perp 1}$ and $\hat{\vec r}_{\perp 2}$, as a proxy to what an observer would measure for the velocity difference across the projected ridge of an intense dissipation sheet, in the two cases where the line of sight is along $\hat{\vec r}_{\perp 1}$ or $\hat{\vec r}_{\perp 2}$.  As expected due to rotation symmetry, rotational discontinuities have no noticeable difference depending on the direction, while fast and slow shocks are co-planar, hence the difference is greater in direction  $\hat{\vec r}_{\perp 1}$ than in direction  $\hat{\vec r}_{\perp 2}$. What is more surprising though is the fact that Parker sheets follow the same trend as rotational discontinuities: this is an indication that there is more continuity between the rotational discontinuity and Parker sheet classes than what our arbitrary division between the two would suggest. An interesting feature we also observe is the bimodality of the slow shocks compared to the fast shocks, which is linked to the fact that $B_\mathrm{scan}/\sqrt{4\pi\rho}$ needs to be greater than the typical velocity to yield a slow shock. 

We retained the sign of the velocity differences although they technically cannot be probed by observations, due to the unknown projection angle. In the OT case, the positive and negative velocity differences for slow and fast shocks along $\hat{\vec r}_{\perp 1}$ have markedly different statistics. This is due to the relative orientation between the fluid velocity and the magnetic field direction along the scanning vector. When they have the same orientation (i.e. $B_\mathrm{scan}=B_n>0$), transverse velocity differences have the same sign as transverse magnetic field differences (see equation \ref{eqn:TransMom}): in this case fast and slow shocks have respectively positive and negative velocity differences. For the OT initial conditions, the positive cross helicity results from a mean positive alignment between velocity and magnetic fields, hence a more likely positive velocity difference for fast shocks, or negative for slow shocks. For the ABC initial conditions, the mean cross helicity is zero, which yields symmetric statistics.

\section{Discussion}\label{sec:discussion}

\subsection{Resolution study}

   \begin{figure}
    \centering
    \protect\includegraphics[scale=0.25]{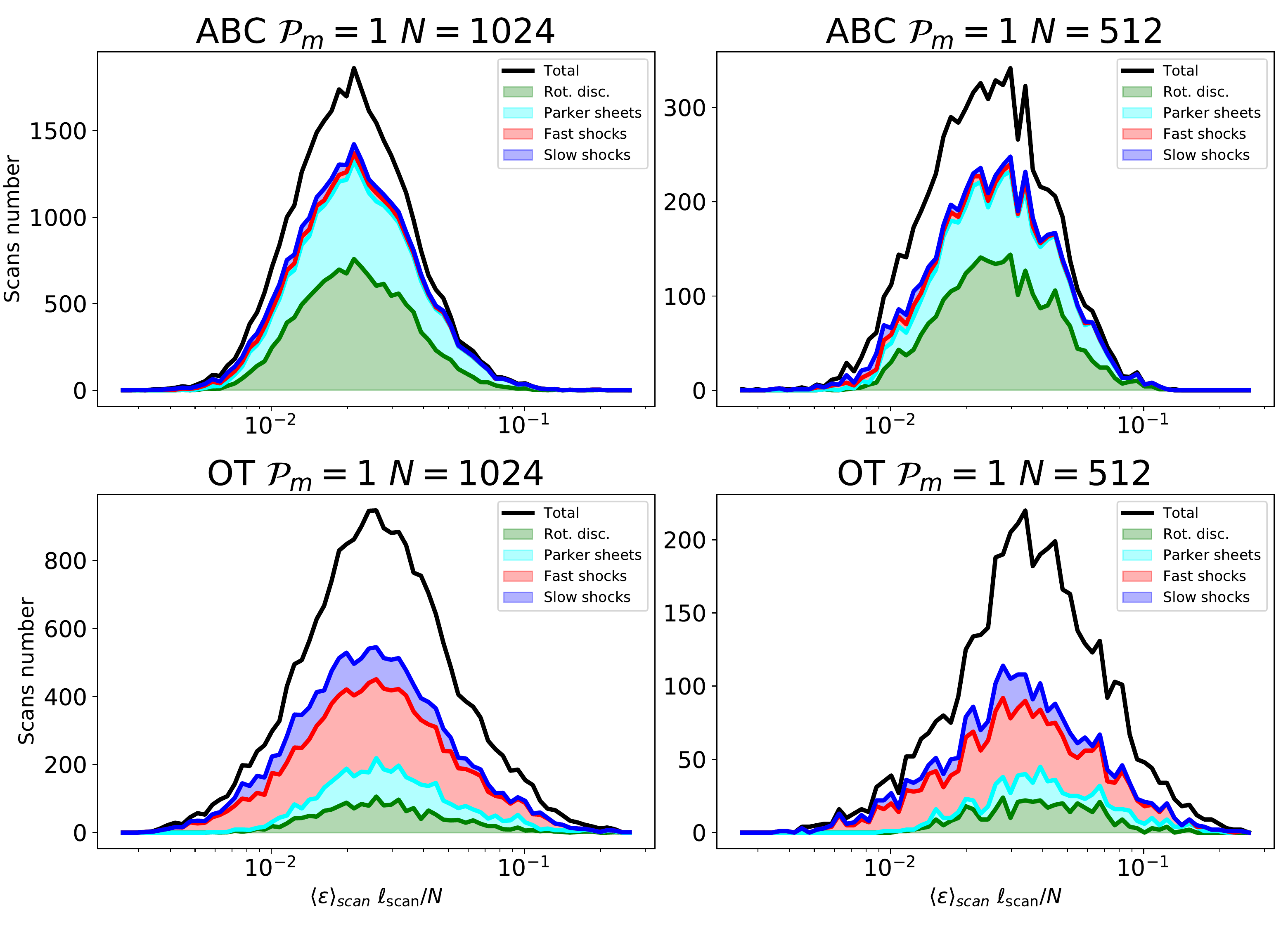}
    \caption{Nature by nature distributions of dissipation flux per scan in our $\mathcal{P}_\mathrm{m}=1$, 1024$^3$  simulations (left panels) and in our 512$^3$ simulations (right) at $t=1/3 t_\mathrm{turnover}$.
              \label{fig:Resolution-study}%
              }
   \end{figure}

We performed simulations at half the resolution (512$^3$ pixels) in order to check the stability our results. Note that the dimensions of our scanning cylinders are defined with respect to the pixel size, with 3 pixels lateral radius and a 6 $\ell_\varepsilon$ scanning length centered on the detected local maxima of dissipation. The appropriate quantity to consider is hence the average energy dissipation rate per unit surface, or the energy flux through the surface of the discontinuity. 

We consider on figure \ref{fig:Resolution-study} the statistics of these dissipation fluxes nature by nature for two corresponding runs at 512$^3$ and 1024$^3$. They are seen to match perfectly, except for statistical noise which jags the lower resolution results.  Indeed, we identify approximately four times less scans at low resolution, which is another indication that our dissipation structures are mainly sheets. 

We also find that for both $N=512$ and $N=1024$, $\ell_\varepsilon$ is on the order of 1.5 pixels. This is consistent with our findings on 1-D shocks in appendix \ref{sec:numerical-dissipation} that a twice lower resolution would yield an energy deposit twice more wide spread in the scanning direction, while keeping its integrated value constant (thanks to our method to recover numerical dissipation). It is also a hint that the same behaviour holds for Alfvénic discontinuities, which was not obvious.

Finally, this is an other indication that large scale dynamics set up the environmental characteristics of the discontinuities (the values of the state variables of the gas on either side of them), while the microphysics (physical and numerical dissipation) control the internal profiles of these discontinuities. The first evidence for this was uncovered with our Prandtl number study in section \ref{sec:PRimpact}.

  \subsection{Towards global dissipation fractions}
  \label{sec:globaldist}

   Previously (section \ref{sec:initial-conditions}, for example), we discussed the relative distributions within our {\it scans}. However, it is unfortunately difficult to relate it to the {\it global dissipation} in the computational domain, because the scans focus on the most intense events. 
   
   Nevertheless, we have seen in section \ref{Def_visu} that strong dissipative pixels considered in this study ($>4\sigma$) represent a large fraction ($\simeq 25\%$ for ABC near the dissipation peak and $\simeq 30\%$ for OT) of the global dissipation rate of the simulation time step. However, the dissipation rate exceeds this threshold only near the peak of each scan, so the dissipation in a given scan also accounts for some dissipation below that threshold. Hence, the dissipation within our scans must amount to more than these global fractions of dissipation. 
   
   Furthermore, if we had chosen a lower threshold to identify structures, we would have detected weaker scans closer to the lateral edges of the dissipation structures. Since the proportions in physical natures do not appear to depend much on the strength of the scan, we might expect to find the same proportions in these weaker scans. As a result the fractions we currently measure might apply to a more significant fraction of the global dissipation, although it is difficult to ascertain it (overlap between scans of neighbouring structures and lack of planarity for some of the weaker scans might moderate the above arguments). In particular, we are not able to assess whether there is a diffuse component of dissipation, outside the intense dissipative regions.  
\subsection{Distribution pixel by pixel}

  \begin{figure*}
    \centering
    \resizebox{\hsize}{!}{\includegraphics{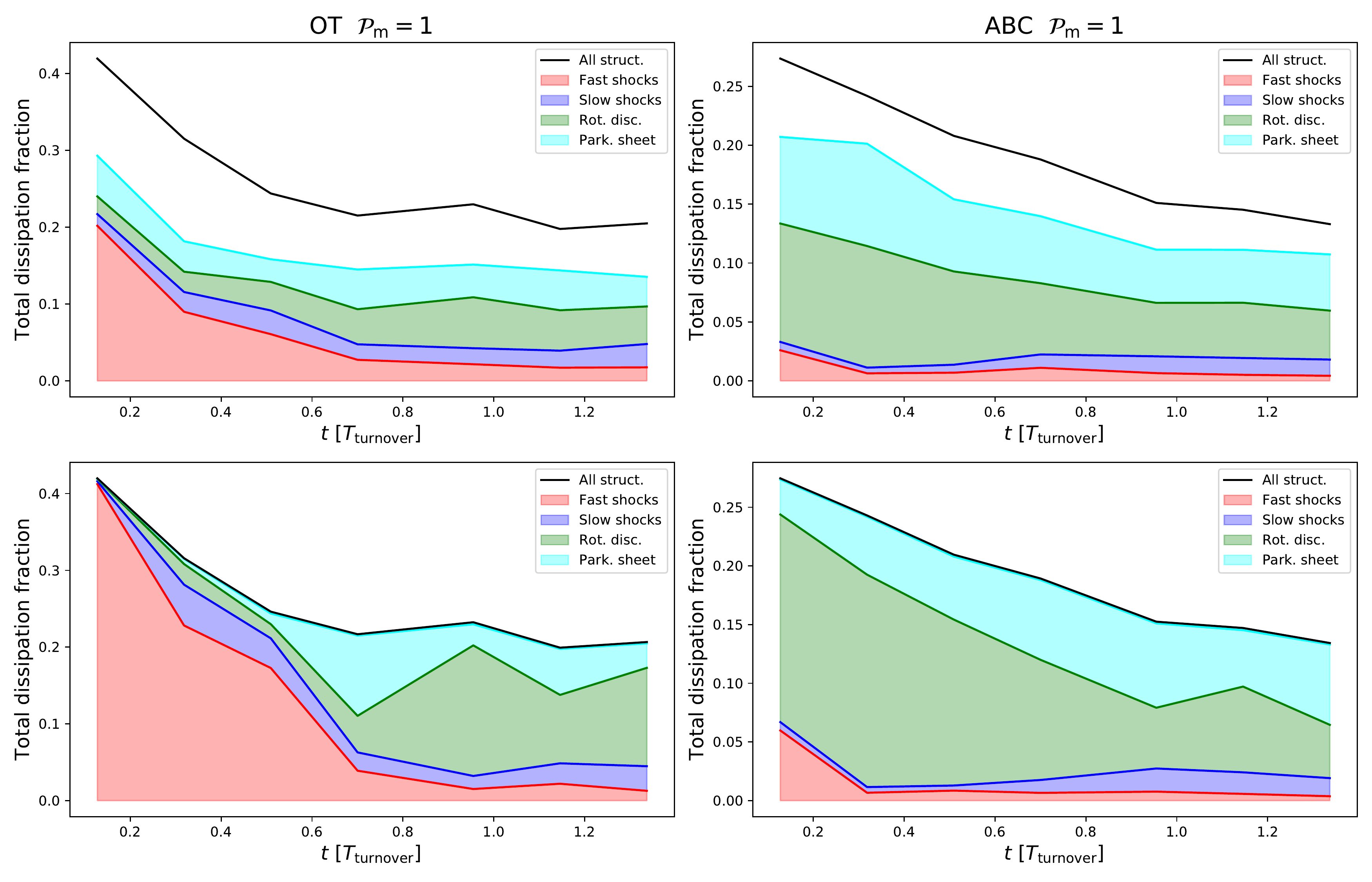}}
    \caption{\textbf{Top plots :} Fraction of the total dissipation high dissipation structure contribute to (black line) and the contribution of cells that belong to a scan that is identified (coloured areas) for OT and ABC initial conditions simulations. \textbf{Bottom plots :} We use the most represented identification in each structure and attribute the total dissipation rate of the structure to this kind. This method is supported by the fact that we find connected structures to be mainly made of one kind of discontinuity.}
    \label{fig:DissTimeEvoComp}%
   \end{figure*}
   
 Our method identifies a large majority of the scans we probed in each simulation and timesteps studied. The criteria for the identification of the scans are kept in this study voluntarily strict, with the aim to discover the structures basis causing the dissipation in the isothermal compressible MHD regime. Although restricted to the purest structures, we identify a significant fraction of the total dissipation of the simulations. We mentioned in the previous section \ref{sec:globaldist} the difficulty to relate total dissipation rates to the scan distribution. Here, we attempt to link the distributions of scans to the distribution of the pixels above a given threshold. To partly remedy to overlapping problems which might occur between scans, we flag cells above a given threshold of the simulation according to the first identification that contain them. The resulting dissipation rate identification is shown on the top plots of figure \ref{fig:DissTimeEvoComp}.
 
 The black line shows the fraction of dissipation captured by the threshold four standard deviations over the mean. The colors below show the fraction of each pixels above this threshold identified as each of the four main natures we find in our scans (the white space combines pixels which were never flagged because they always fall in unidentified or unknown scans). This time evolution graph clearly shows how the distributions differ at early time for our two different initial conditions, and how they stabilise after little bit less than one turnover time. This confirms the result that we found for scan distributions in section \ref{ICimpact}.
 
\subsection{Connectedness}

 \begin{figure}
       \centering
       \includegraphics[width=0.42\paperwidth]{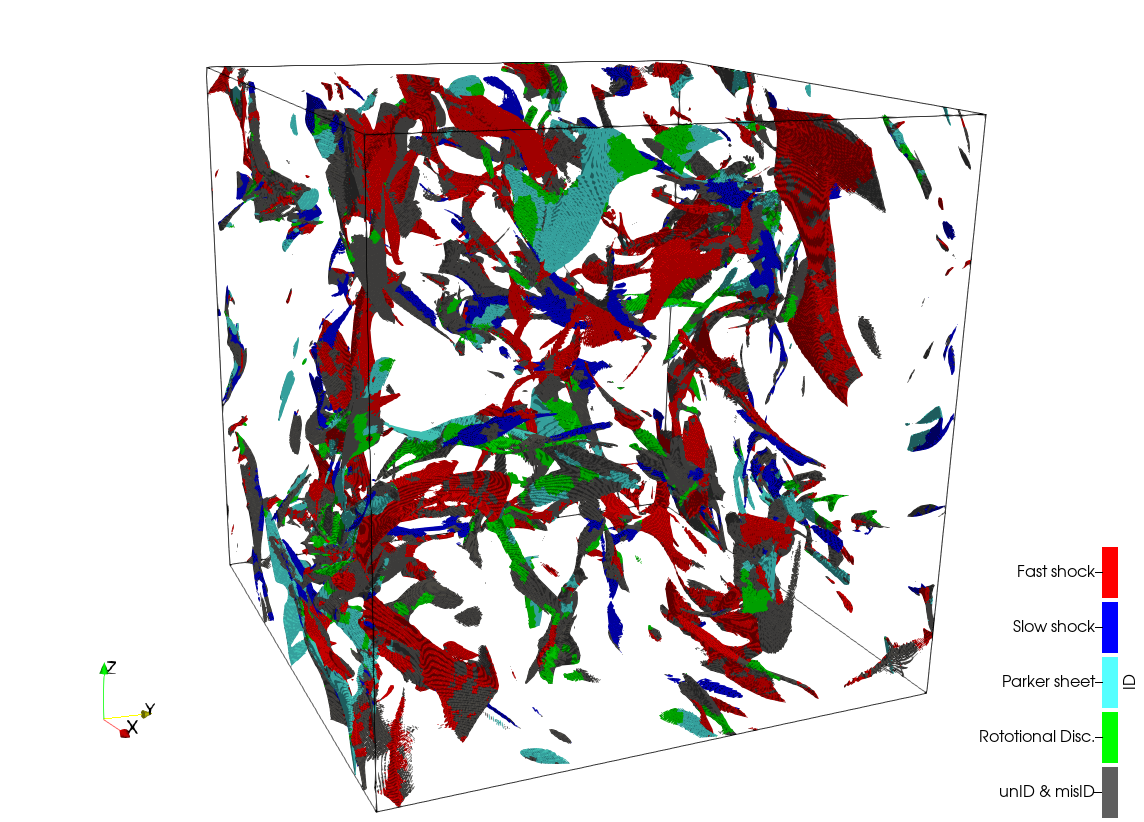}
          \caption{High dissipation structures extracted from an OT initial conditions simulation at $\Pm=1$. The time step of this output is $t\simeq 1/3 t_\mathrm{turnover}$. 
             \label{fig:NatureColoringOT}
             }
    \end{figure}

Furthermore, if we consider only the well identified scans, we observe that about 70\% of the related connected structures are identified by a single type of scan and 80\% have more than 75\% of their scans identified by the same nature (see figure \ref{fig:NatureColoringOT}).  We can therefore consider propagating the dominant type of a connected structure to the remainder of its cells. This allows  to avoid the edge effects of structures and increases the identified dissipation fraction. The result is shown on the bottom plots of figure \ref{fig:DissTimeEvoComp}. This graph tells us, first of all, that the fully unidentified connected structures, although representing a significant fraction of the studied structures in number, participate very little in the total dissipation of the cube. These are small fragmented events which also comprise the short filament-like structures seen on figure \ref{fig:fullcube}. Second, we notice that the unidentified scans often belong to structures dominated by rotational discontinuities, except for the OT simulations at an early time, when they are sometimes part of fast shocks (see figure \ref{fig:NatureColoringOT}). For ABC runs, until about 0.7 turnover times, the dissipation generated by the Parker sheets decreases slightly to the benefit of that produced by the rotational discontinuities (bottom right panel of figure \ref{fig:DissTimeEvoComp}). This implies that despite the uniqueness of the identifications within a related structure, a significant fraction of the Parker sheets are found within structures formed by rotational discontinuities, which relates to our previous remarks on the continuity between Parker-Sheets and rotational discontinuities: our divide between the two is rather arbitrary and these connected structures could probably be gathered into a single Alfénic discontinuity class.  
  
  We also tried to decrease the detection threshold of the dissipation structures to $\varepsilon_\mathrm{tot}^\mathrm{corr}=<\varepsilon_\mathrm{tot}^\mathrm{corr}>+2\sigma_\mathrm{\varepsilon_\mathrm{tot}^\mathrm{corr}}$ to increase the fraction of the total identified dissipation. In this case, the identification rate decreases only by a few percent compared to a threshold of $\varepsilon_\mathrm{tot}^\mathrm{corr}=<\varepsilon_\mathrm{tot}^\mathrm{corr}>+4\sigma_\mathrm{\varepsilon_\mathrm{tot}^\mathrm{corr}}$. The contribution of the different natures to the overall dissipation remains similar and the fraction of dissipation above threshold identified by the scans increases by about 10\% overall.

  \subsection{Unknown identifications}
  
  By using two sets of criteria which are rather independent, we have biased our identifications towards more false negative and less false positives. Thus, there remains many scans misidentified because they either do not fit any of our heuristic criteria (unidentified scans) or because the two sets of criteria do not match (misidentified scans).
  We list here some of the reasons why our identification criteria might miss a significant fraction of the scans.
  
  The main culprits are "edge" scans. These are scans at the periphery of structures where the main direction of the gradient is less well defined and therefore the scanning direction is less relevant. For instance, the scanning direction is irrelevant in the case of the small filament-like structures observed in figure \ref{fig:fullcube} and  \ref{fig:NatureColoringOT}, where scans probably fall at least in the misidenditified category. Also, when two structures are too close to each other, the heuristic part of the identification is confused, because bumps or jumps are less well defined. Note the wave decomposition suffers less from adjacent structures, because it is sensitive only at the cell scale. Some of the unidentifications could also be due to the presence of intermediate shocks (see section \ref{sec:intermediate-shocks} below) but we reckon they probably account for a small fraction  only. 
  
  Given the strong correlation between our two sets of criteria for identification (heuristic and ideal waves), one could suggest to use only ideal wave decomposition to greatly increase the identification rate. We would thus reach 100\% of identification, but our results would then be subject to caution, and would be biased toward false positives. Indeed, one should restrict this wave only decomposition to the most planar cells where the gradient approach makes sense. Also, the identification would then rely on the velocity regimes, which we have shown can be subject to caution depending on the method to estimate the travelling speed of the discontinuities. 
  
  \subsection{Intermediate shocks}\label{sec:intermediate-shocks}
  
  If intermediate shocks are present in our simulations, our heuristic criteria would voluntarily miss them, and they would fall in the unidentified category. Indeed, these shocks have either a density or a pressure jump, but \change{often} display a magnetic field \change{trough}. We chose not to add this criterion here because we would not have had an independent criterion on gradients to solidify this heuristic one. This might be a reason why we get less identification in the OT case at early times, which seems more prone to generate shocks.
  We have in fact attempted to target intermediate shocks, and have found some convincing cases. However, the uncertainty on our estimate for the steady state velocity of these discontinuities makes it difficult to validate the speed regimes of these shocks on a statistically significant population. We hence decided to postpone our investigation on these shocks. In any case, the fraction of unidentification that we publish here puts an upper limit on the fraction of intermediate shocks.

\change{\subsection{Driven vs. decaying turbulence}}
\change{Some astrophysical situations (a solar coronal ejection, a supernova, or a runaway star encountering a cloud, for example) can locally inject mechanical energy in a short amount of time. In these events, well defined initial conditions can suddenly be imposed, which will later develop into turbulence. In these contexts, decaying turbulence is probably a sensible approach.
However, one can question the applicability of decaying turbulence in many other realistic astrophysical situations. First, turbulence being ubiquitous, it is very unlikely that an initial set up would be entirely devoid of turbulent perturbations at the onset. Also, the fast decay of turbulence has led \cite{Stone1998} amongst other authors to advocate the necessity of persistent driving at the scales relevant for the interstellar clouds. Our experiments have shown that initial conditions can imprint significant changes in the early phases of the development of turbulence: similarly, we expect that a given forcing may impact at least to some degree the resulting turbulent cascade, and it will do so at all times. One should therefore be careful to properly characterise the properties of the forcing (such as helicity injection) used in the experiments of driven turbulence.}

\change{In particular, a lot of attention has been devoted to the importance of solenoidal vs. compressible forcing \citep{Federrath2010}. This has led to techniques to probe observationally the degree of compressibility of the gas: some regions of the ISM near the galactic center were found to be dominated by solenoidal  driving \citep{Federrath2016b} while others associated to stellar feedback were found to be more compressively driven \citep{Menon2021}. Intermediate situations were also witnessed in Orion B \citep[see][for example]{Orkisz2017}. These considerations are important, as the nature of star formation is believed to be strongly influenced by the driving of the turbulence \citep{Federrath2012}. In the present study, we have not attempted to check the influence of compressible vs. solenoidal initial velocity fields (our initial velocity fields are divergence free), but one can imagine that initially more weight would be given to shocks had we started with velocities including a compressive component. We still believe though that the tendency to evolve towards incompressible structures would persist (imagine starting the origin of times after a fraction of the turnover time: this might be an illustration of what could happen if we start the simulation with compressible initial conditions). Finally, an intermediate solution would be to start decaying turbulence with a snapshot of fully established driven turbulence, or to imprint on the initial conditions synthetic models of turbulence such as designed in \cite{Durrive2020}.
}

\section{Conclusions and prospects}
 The aim of the present study is to systematically characterise the physical nature of intense extrema of dissipation in MHD simulations of turbulence. We develop a technique to recover locally the total dissipation including the numerical losses. We tested the classic rule of thumb that grid based simulations need twice the resolution of similar spectral schemes: in this case, we find that numerical dissipation is indeed below a half of the total, but dissipative fronts are widened by a factor of about three. Since to get to the expected thickness would require an extra factor of ten in resolution, we feel the current usage provides a good compromise. 
 
 We devise a way to characterise the geometry and the physical nature of local intense variations of the state variables of the gas.  We find the non-linear waves associated with these large gradients and disclose their Rankine-Hugoniot category. We show that at the dissipation peak, the fully dissipative gradients must be close to an ideal MHD wave gradient. We observe that the nature of this gradient is surprisingly consistent throughout the profile of the dissipation structure. For example, fast shocks are composed of essentially fast wave gradients and we confirmed it with the 1D semi-analytical models of isothermal shocks of appendix \ref{sec:steady-state-MHD-shocks}. We use this property to our advantage and we design a method to classify the dissipation structures into fast shocks, slow shocks, Parker sheets and rotational discontinuities. We successfully identify a large majority of the intense dissipation, which allows us to draw statistical conclusions.   
 
   We show that initial conditions can strongly affect the nature of the dissipation structures at early times. However, early signatures of the initial conditions are quickly lost after about one turnover time. At this time, dissipation becomes dominated by weakly compressive structures (Alfvén discontinuities rather than shocks). This may be due to the sonic Mach number having decreased closer to one at this point, and we will investigate higher Mach numbers in the future as well as more compressive initial conditions.
 
   Despite the complexity of the magnetised 3D flows we investigated in this study, strongest dissipation structures are locally plane and steady and can be assimilated to Rankine-Hugoniot discontinuities. We noted unexpected correlations between the entrance parameters of these discontinuities (which can be reduced to a 2-parameters family): further work is needed to explain how these correlations arise in a turbulent medium. 
   
   We compare three methods to measure the traveling speed of these non-linear waves, and check the resulting velocity regimes are compatible with our identifications.
   The difficulty to accurately measure the traveling speed makes it impossible to assess the statistics of the elusive intermediate shocks, although we report we could find clear examples of them (not shown in this paper). 

   The access to an accurate traveling speed will facilitate the follow-up of structures in time, which will help discover if the statistical changes with respect to time are due to collisions (or breading) between structures, birth or death of given structures, possible changes in nature of a given structure in time or to the development of substructures and instabilities within a structure.

We do not find strong evidence for the slow shocks being more subject to corrugation instability as originally found by \cite{Park2019}. In general, connected structures appear equally fragmented regardless of their various natures (see figure \ref{fig:NatureColoringOT}), but a more quantitative study might conclude otherwise. It seems to us Alfvénic discontinuities are often found in parallel sub-layered systems, while fast shocks often occur in isolation, but here again a quantitative analysis might conclude otherwise.

One may challenge the applicability of such simplified isothermal MHD simulations to a medium as complex as the interstellar medium. However, the present study hints that to some extent the details of the microphysics matter only within the internal structure of discontinuities. For example, the statistics of the entrance parameters do not change when $\Pm$ is varied. This is reminiscent of the study by \cite{Brandenburg2014} who suggested that variations with $\Pm$ were controlled by the individual 1D structure of the shocks, and it is also echoed in the review of reconnection \cite{Zweibel2016} which focuses on the respective roles of global and local processes. If this holds, one could imagine post-processing the statistics from 3D simulations with more detailed 1D models 
including non-equilibrium chemistry, 
such as the Paris-Durham shock models \citep{Flower1985,Flower2015}, for example, as was demonstrated in \cite{CHEMSES} for 2D unmagnetised turbulence.

The ultimate objective is to estimate the turbulent dissipation rate in diffuse matter and its characteristics in the broad perspective of unravelling molecular cloud growth and star formation \citep[e.g.][]{Hennebelle2012}. The fall-off (or the absence of fall-off at small scales) of power spectra of a variety of tracers of diffuse interstellar matter \citep[e.g.][]{MivilleDeschenes2016} is a key information to be combined with  the kinetic information provided by high-spectral resolution observations of either atomic gas \citep[e.g.][]{Reach2021} or molecular lines  \citep{Hily-Blant2008,Falgarone2009}. This latter route is of course challenging because it requires the modelling of non-equilibrium chemistry driven by dissipation bursts.

Conversely, our multidimensional simulations suggest improvements to 1D traditional models. Indeed, although the structures we find are mostly plane-parallel, we find that the main deviation from 1D profiles is mass loss sideways into the dissipative sheet. In the future, we can imagine to refine 1D models by including such mass-loss, as did Parker in his fiducial Parker-sheet model, for example \citep{Parker1963}.
    
   Finally, we are convinced that the tools we put forward in this paper will give more ground to the view of developed turbulence as a statistical collection of coherent structures. For example, a careful series of works \citep[e.g.][]{Zhdankin2013,Zhdankin2014,Zhdankin2015,Zhdankin2016} on the dissipation structures in reduced MHD has led to new insights on the analytics of intermittency and turbulent dynamics by \cite{Schekochihin2017}. Density statistics deviations from log-normal were explained by an appropriate collection of shocks in \cite{RobertsonGoldreich2018}. Recent development in the theory of anisotropic compressible MHD turbulence use to their advantage the statistics of shocks \citep{Beattie2021}.
 In the meantime, Cluster satellites observations \citep{perrone2016,Perrone2017} have witnessed the signatures of both Alfv\'en and compressive coherent structures in the fast and slow components of the solar wind. Recent developments in solar wind observations may soon be able to constrain the statistics of the various individual types of dissipation structures \citep{Bruno2013}.
    
\begin{acknowledgements}
      \change{We thank our referee, Christoph Federrath, for his constructive comments which helped us improve the quality and scope of the manuscript.} The research leading to these results has received fundings from the European Research Council, under the European Community’s Seventh framework Programme, through the Advanced Grant MIST (FP7/2017-2022, No 742719). Computations were performed on the cluster Totoro funded by the MIST
      ERC and hosted by the computing center mesoPSL. We
      thank S. Fromang who provided us with his version of DUMSES.
      We thank Ben Snow for enlightnening discussions on intermediate shocks and Erwan Allys on dissipation scalings with $\Pm$ for each type of structure. The idea of gradient decomposition into linear waves is our generalisation to 3D MHD of a method communicated to us by Johann Carroll-Nellenback who previously used it on Astrobear 1D hydrodynamical simulations to detect shocks and contact discontinuities.
\end{acknowledgements}

\bibliographystyle{aa}
\bibliography{biblio}

\begin{thebibliography}{81}
\expandafter\ifx\csname natexlab\endcsname\relax\def\natexlab#1{#1}\fi

\bibitem[{{Appleton} {et~al.}(2013){Appleton}, {Guillard}, {Boulanger},
  {Cluver}, {Ogle}, {Falgarone}, {Pineau des For{\^e}ts}, {O'Sullivan}, {Duc},
  {Gallagher}, {Gao}, {Jarrett}, {Konstantopoulos}, {Lisenfeld}, {Lord}, {Lu},
  {Peterson}, {Struck}, {Sturm}, {Tuffs}, {Valchanov}, {van der Werf}, \&
  {Xu}}]{Appleton2013}
{Appleton}, P.~N., {Guillard}, P., {Boulanger}, F., {et~al.} 2013, \apj, 777,
  66

\bibitem[{{Beattie} {et~al.}(2021){Beattie}, {Mocz}, {Federrath}, \&
  {Klessen}}]{Beattie2021}
{Beattie}, J.~R., {Mocz}, P., {Federrath}, C., \& {Klessen}, R.~S. 2021,
  \mnras, 504, 4354

\bibitem[{{Bouya} \& {Dormy}(2013)}]{ABC2013}
{Bouya}, I. \& {Dormy}, E. 2013, Physics of Fluids, 25, 037103

\bibitem[{{Brandenburg}(2014)}]{Brandenburg2014}
{Brandenburg}, A. 2014, \apj, 791, 12

\bibitem[{Brandenburg \& Rempel(2019)}]{Brandenburg2019}
Brandenburg, A. \& Rempel, M. 2019, The Astrophysical Journal, 879, 57

\bibitem[{{Bruno} \& {Carbone}(2013)}]{Bruno2013}
{Bruno}, R. \& {Carbone}, V. 2013, Living Reviews in Solar Physics, 10, 2

\bibitem[{Delmont \& Keppens(2011)}]{Delmont2011}
Delmont, P. \& Keppens, R. 2011, Journal of Plasma Physics, 77, 207

\bibitem[{{Draine} \& {Katz}(1986)}]{DraineKatz1986}
{Draine}, B.~T. \& {Katz}, N. 1986, \apj, 310, 392

\bibitem[{{Durrive} {et~al.}(2020){Durrive}, {Lesaffre}, \&
  {Ferri{\`e}re}}]{Durrive2020}
{Durrive}, J.-B., {Lesaffre}, P., \& {Ferri{\`e}re}, K. 2020, \mnras, 496, 3015

\bibitem[{{Falgarone} {et~al.}(2009){Falgarone}, {Pety}, \&
  {Hily-Blant}}]{Falgarone2009}
{Falgarone}, E., {Pety}, J., \& {Hily-Blant}, P. 2009, \aap, 507, 355

\bibitem[{{Falgarone} {et~al.}(1995){Falgarone}, {Pineau des Forets}, \&
  {Roueff}}]{Falgarone1995}
{Falgarone}, E., {Pineau des Forets}, G., \& {Roueff}, E. 1995, \aap, 300, 870

\bibitem[{{Falgarone} {et~al.}(2005){Falgarone}, {Verstraete}, {Pineau Des
  For{\^e}ts}, \& {Hily-Blant}}]{Falgarone2005}
{Falgarone}, E., {Verstraete}, L., {Pineau Des For{\^e}ts}, G., \&
  {Hily-Blant}, P. 2005, \aap, 433, 997

\bibitem[{{Falgarone} {et~al.}(2017){Falgarone}, {Zwaan}, {Godard}, {Bergin},
  {Ivison}, {Andreani}, {Bournaud}, {Bussmann}, {Elbaz}, {Omont}, {Oteo}, \&
  {Walter}}]{Falgarone2017}
{Falgarone}, E., {Zwaan}, M.~A., {Godard}, B., {et~al.} 2017, \nat, 548, 430

\bibitem[{{Federman} {et~al.}(1996){Federman}, {Rawlings}, {Taylor}, \&
  {Williams}}]{Federman1996}
{Federman}, S.~R., {Rawlings}, J.~M.~C., {Taylor}, S.~D., \& {Williams}, D.~A.
  1996, \mnras, 279, L41

\bibitem[{{Federrath}(2013)}]{Federrath2013}
{Federrath}, C. 2013, \mnras, 436, 1245

\bibitem[{{Federrath}(2016)}]{Federrath2016}
{Federrath}, C. 2016, Journal of Plasma Physics, 82, 535820601

\bibitem[{{Federrath} \& {Klessen}(2012)}]{Federrath2012}
{Federrath}, C. \& {Klessen}, R.~S. 2012, \apj, 761, 156

\bibitem[{{Federrath} {et~al.}(2021){Federrath}, {Klessen}, {Iapichino}, \&
  {Beattie}}]{Federrath2021}
{Federrath}, C., {Klessen}, R.~S., {Iapichino}, L., \& {Beattie}, J.~R. 2021,
  Nature Astronomy, 5, 365

\bibitem[{{Federrath} {et~al.}(2016){Federrath}, {Rathborne}, {Longmore},
  {Kruijssen}, {Bally}, {Contreras}, {Crocker}, {Garay}, {Jackson}, {Testi}, \&
  {Walsh}}]{Federrath2016b}
{Federrath}, C., {Rathborne}, J.~M., {Longmore}, S.~N., {et~al.} 2016, \apj,
  832, 143

\bibitem[{{Federrath} {et~al.}(2010){Federrath}, {Roman-Duval}, {Klessen},
  {Schmidt}, \& {Mac Low}}]{Federrath2010}
{Federrath}, C., {Roman-Duval}, J., {Klessen}, R.~S., {Schmidt}, W., \& {Mac
  Low}, M.~M. 2010, \aap, 512, A81

\bibitem[{{Flower} \& {Pineau des Forets}(1998)}]{Flower1998}
{Flower}, D.~R. \& {Pineau des Forets}, G. 1998, \mnras, 297, 1182

\bibitem[{{Flower} \& {Pineau des For{\^e}ts}(2015)}]{Flower2015}
{Flower}, D.~R. \& {Pineau des For{\^e}ts}, G. 2015, \aap, 578, A63

\bibitem[{{Flower} {et~al.}(1985){Flower}, {Pineau des For\^ets}, \&
  {Hartquist}}]{Flower1985}
{Flower}, D.~R., {Pineau des For\^ets}, G., \& {Hartquist}, T.~W. 1985, \mnras,
  216, 775

\bibitem[{{Fromang} {et~al.}(2006){Fromang}, {Hennebelle}, \&
  {Teyssier}}]{DUMSES}
{Fromang}, S., {Hennebelle}, P., \& {Teyssier}, R. 2006, \aap, 457, 371

\bibitem[{{Gerin} \& {Liszt}(2021)}]{Gerin2021}
{Gerin}, M. \& {Liszt}, H. 2021, \aap, 648, A38

\bibitem[{{Godard} {et~al.}(2012){Godard}, {Falgarone}, {Gerin}, {Lis}, {De
  Luca}, {Black}, {Goicoechea}, {Cernicharo}, {Neufeld}, {Menten}, \&
  {Emprechtinger}}]{Godard2012}
{Godard}, B., {Falgarone}, E., {Gerin}, M., {et~al.} 2012, \aap, 540, A87

\bibitem[{{Godard} {et~al.}(2009){Godard}, {Falgarone}, \& {Pineau Des
  For{\^e}ts}}]{Godard2009}
{Godard}, B., {Falgarone}, E., \& {Pineau Des For{\^e}ts}, G. 2009, \aap, 495,
  847

\bibitem[{{Godard} {et~al.}(2014){Godard}, {Falgarone}, \& {Pineau des
  For{\^e}ts}}]{Godard2014}
{Godard}, B., {Falgarone}, E., \& {Pineau des For{\^e}ts}, G. 2014, \aap, 570,
  A27

\bibitem[{{Goedbloed} {et~al.}(2019){Goedbloed}, {Keppens}, \&
  {Poedts}}]{goedbloed_MHD}
{Goedbloed}, J.~P., {Keppens}, R., \& {Poedts}, S. 2019, Magnetohydrodynamics
  of Laboratory and Astrophysical Plasmas (Cambridge University Press)

\bibitem[{{Gry} {et~al.}(2002){Gry}, {Boulanger}, {Nehm{\'e}}, {Pineau des
  For{\^e}ts}, {Habart}, \& {Falgarone}}]{Gry2002}
{Gry}, C., {Boulanger}, F., {Nehm{\'e}}, C., {et~al.} 2002, \aap, 391, 675

\bibitem[{{Guillard} {et~al.}(2012){Guillard}, {Boulanger}, {Pineau des
  For{\^e}ts}, {Falgarone}, {Gusdorf}, {Cluver}, {Appleton}, {Lisenfeld},
  {Duc}, {Ogle}, \& {Xu}}]{Guillard2012}
{Guillard}, P., {Boulanger}, F., {Pineau des For{\^e}ts}, G., {et~al.} 2012,
  \apj, 749, 158

\bibitem[{Gurnett \& Bhattacharjee(2005)}]{gurnett_bhattacharjee_2005}
Gurnett, D.~A. \& Bhattacharjee, A. 2005, Introduction to Plasma Physics: With
  Space and Laboratory Applications (Cambridge University Press)

\bibitem[{{Hennebelle} \& {Falgarone}(2012)}]{Hennebelle2012}
{Hennebelle}, P. \& {Falgarone}, E. 2012, \aapr, 20, 55

\bibitem[{{Hily-Blant} {et~al.}(2008){Hily-Blant}, {Falgarone}, \&
  {Pety}}]{Hily-Blant2008}
{Hily-Blant}, P., {Falgarone}, E., \& {Pety}, J. 2008, \aap, 481, 367

\bibitem[{{Ingalls} {et~al.}(2011){Ingalls}, {Bania}, {Boulanger}, {Draine},
  {Falgarone}, \& {Hily-Blant}}]{Ingalls2011}
{Ingalls}, J.~G., {Bania}, T.~M., {Boulanger}, F., {et~al.} 2011, \apj, 743,
  174

\bibitem[{{Joulain} {et~al.}(1998){Joulain}, {Falgarone}, {Pineau des Forets},
  \& {Flower}}]{Joulain1998}
{Joulain}, K., {Falgarone}, E., {Pineau des Forets}, G., \& {Flower}, D. 1998,
  \aap, 340, 241

\bibitem[{{Kolmogorov}(1962)}]{Kolmogorov1962}
{Kolmogorov}, A.~N. 1962, Journal of Fluid Mechanics, 13, 82

\bibitem[{{Landau} \& {Lifshitz}(1959)}]{LandauLifschitz1959}
{Landau}, L.~D. \& {Lifshitz}, E.~M. 1959, {Fluid mechanics}

\bibitem[{Lehmann {et~al.}(2016)Lehmann, Federrath, \& Wardle}]{Lehmann2016}
Lehmann, A., Federrath, C., \& Wardle, M. 2016, Monthly Notices of the Royal
  Astronomical Society, 463, 1026

\bibitem[{Lesaffre \& Balbus(2007)}]{lesaffre_balbus_2007}
Lesaffre, P. \& Balbus, S.~A. 2007, Monthly Notices of the Royal Astronomical
  Society, 381, 319

\bibitem[{{Lesaffre} {et~al.}(2004){Lesaffre}, {Chi{\`e}ze}, {Cabrit}, \&
  {Pineau des For{\^e}ts}}]{Lesaffre2004b}
{Lesaffre}, P., {Chi{\`e}ze}, J.~P., {Cabrit}, S., \& {Pineau des For{\^e}ts},
  G. 2004, \aap, 427, 157

\bibitem[{{Lesaffre} {et~al.}(2007){Lesaffre}, {Gerin}, \&
  {Hennebelle}}]{Lesaffre2007}
{Lesaffre}, P., {Gerin}, M., \& {Hennebelle}, P. 2007, \aap, 469, 949

\bibitem[{{Lesaffre} {et~al.}(2013){Lesaffre}, {Pineau des For{\^e}ts},
  {Godard}, {Guillard}, {Boulanger}, \& {Falgarone}}]{Lesaffre2013}
{Lesaffre}, P., {Pineau des For{\^e}ts}, G., {Godard}, B., {et~al.} 2013, \aap,
  550, A106

\bibitem[{Lesaffre {et~al.}(2020)Lesaffre, Todorov, Levrier, Valdivia,
  Dzyurkevich, Godard, Tram, Gusdorf, Lehmann, \& Falgarone}]{CHEMSES}
Lesaffre, P., Todorov, P., Levrier, F., {et~al.} 2020, Monthly Notices of the
  Royal Astronomical Society, 495, 816

\bibitem[{{Levrier} {et~al.}(2012){Levrier}, {Le Petit}, {Hennebelle},
  {Lesaffre}, {Gerin}, \& {Falgarone}}]{Levrier2012}
{Levrier}, F., {Le Petit}, F., {Hennebelle}, P., {et~al.} 2012, \aap, 544, A22

\bibitem[{{Liszt} \& {Lucas}(1998)}]{LisztLucas1998}
{Liszt}, H.~S. \& {Lucas}, R. 1998, \aap, 339, 561

\bibitem[{{Lucas} \& {Liszt}(1996)}]{LucasLiszt1996}
{Lucas}, R. \& {Liszt}, H. 1996, \aap, 307, 237

\bibitem[{{Macquorn Rankine}(1870)}]{Rankine1870}
{Macquorn Rankine}, W.~J. 1870, Philosophical Transactions of the Royal Society
  of London Series I, 160, 277

\bibitem[{{Mallet} \& {Schekochihin}(2017)}]{Schekochihin2017}
{Mallet}, A. \& {Schekochihin}, A.~A. 2017, \mnras, 466, 3918

\bibitem[{{Meneveau} \& {Sreenivasan}(1991)}]{MeneveauSreenivasan1991}
{Meneveau}, C. \& {Sreenivasan}, K.~R. 1991, Journal of Fluid Mechanics, 224,
  429

\bibitem[{{Menon} {et~al.}(2021){Menon}, {Federrath}, {Klaassen}, {Kuiper}, \&
  {Reiter}}]{Menon2021}
{Menon}, S.~H., {Federrath}, C., {Klaassen}, P., {Kuiper}, R., \& {Reiter}, M.
  2021, \mnras, 500, 1721

\bibitem[{{Miville-Desch{\^e}nes} {et~al.}(2016){Miville-Desch{\^e}nes}, {Duc},
  {Marleau}, {Cuillandre}, {Didelon}, {Gwyn}, \&
  {Karabal}}]{MivilleDeschenes2016}
{Miville-Desch{\^e}nes}, M.~A., {Duc}, P.~A., {Marleau}, F., {et~al.} 2016,
  \aap, 593, A4

\bibitem[{{Moisy} \& {Jim{\'e}nez}(2004)}]{Moisy2004}
{Moisy}, F. \& {Jim{\'e}nez}, J. 2004, Journal of Fluid Mechanics, 513, 111

\bibitem[{Momferratos {et~al.}(2014)Momferratos, Lesaffre, Falgarone, \&
  Pineau~des Forêts}]{Momferratos2014}
Momferratos, G., Lesaffre, P., Falgarone, E., \& Pineau~des Forêts, G. 2014,
  Monthly Notices of the Royal Astronomical Society, 443, 86

\bibitem[{{Moseley} {et~al.}(2021){Moseley}, {Draine}, {Tomida}, \&
  {Stone}}]{Moseley2021}
{Moseley}, E.~R., {Draine}, B.~T., {Tomida}, K., \& {Stone}, J.~M. 2021,
  \mnras, 500, 3290

\bibitem[{{Myers} {et~al.}(2015){Myers}, {McKee}, \& {Li}}]{Myers2015}
{Myers}, A.~T., {McKee}, C.~F., \& {Li}, P.~S. 2015, \mnras, 453, 2747

\bibitem[{{Nehm{\'e}} {et~al.}(2008){Nehm{\'e}}, {Le Bourlot}, {Boulanger},
  {Pineau des For{\^e}ts}, \& {Gry}}]{Nehme2008}
{Nehm{\'e}}, C., {Le Bourlot}, J., {Boulanger}, F., {Pineau des For{\^e}ts},
  G., \& {Gry}, C. 2008, \aap, 483, 485

\bibitem[{{Orkisz} {et~al.}(2017){Orkisz}, {Pety}, {Gerin}, {Bron},
  {Guzm{\'a}n}, {Bardeau}, {Goicoechea}, {Gratier}, {Le Petit}, {Levrier},
  {Liszt}, {{\"O}berg}, {Peretto}, {Roueff}, {Sievers}, \&
  {Tremblin}}]{Orkisz2017}
{Orkisz}, J.~H., {Pety}, J., {Gerin}, M., {et~al.} 2017, \aap, 599, A99

\bibitem[{{Orszag} \& {Tang}(1979)}]{OT79}
{Orszag}, S.~A. \& {Tang}, C.~M. 1979, Journal of Fluid Mechanics, 90, 129

\bibitem[{{Park} \& {Ryu}(2019)}]{Park2019}
{Park}, J. \& {Ryu}, D. 2019, \apj, 875, 2

\bibitem[{{Parker}(1963)}]{Parker1963}
{Parker}, E.~N. 1963, \apjs, 8, 177

\bibitem[{{Perrone} {et~al.}(2016){Perrone}, {Alexandrova}, {Mangeney},
  {Maksimovic}, {Lacombe}, {Rakoto}, {Kasper}, \& {Jovanovic}}]{perrone2016}
{Perrone}, D., {Alexandrova}, O., {Mangeney}, A., {et~al.} 2016, \apj, 826, 196

\bibitem[{{Perrone} {et~al.}(2017){Perrone}, {Alexandrova}, {Roberts}, {Lion},
  {Lacombe}, {Walsh}, {Maksimovic}, \& {Zouganelis}}]{Perrone2017}
{Perrone}, D., {Alexandrova}, O., {Roberts}, O.~W., {et~al.} 2017, \apj, 849,
  49

\bibitem[{{Porter} {et~al.}(2015){Porter}, {Jones}, \& {Ryu}}]{Porter2015}
{Porter}, D.~H., {Jones}, T.~W., \& {Ryu}, D. 2015, \apj, 810, 93

\bibitem[{{Reach} \& {Heiles}(2021)}]{Reach2021}
{Reach}, W.~T. \& {Heiles}, C. 2021, \apj, 909, 71

\bibitem[{{Robertson} \& {Goldreich}(2018)}]{RobertsonGoldreich2018}
{Robertson}, B. \& {Goldreich}, P. 2018, \apj, 854, 88

\bibitem[{{Smith} {et~al.}(2000{\natexlab{a}}){Smith}, {Mac Low}, \&
  {Heitsch}}]{Smith2000b}
{Smith}, M.~D., {Mac Low}, M.~M., \& {Heitsch}, F. 2000{\natexlab{a}}, \aap,
  362, 333

\bibitem[{{Smith} {et~al.}(2000{\natexlab{b}}){Smith}, {Mac Low}, \&
  {Zuev}}]{Smith2000a}
{Smith}, M.~D., {Mac Low}, M.~M., \& {Zuev}, J.~M. 2000{\natexlab{b}}, \aap,
  356, 287

\bibitem[{{Stone} {et~al.}(1998){Stone}, {Ostriker}, \& {Gammie}}]{Stone1998}
{Stone}, J.~M., {Ostriker}, E.~C., \& {Gammie}, C.~F. 1998, \apjl, 508, L99

\bibitem[{{Teyssier}(2002)}]{RAMSES}
{Teyssier}, R. 2002, \aap, 385, 337

\bibitem[{{Toro}(1999)}]{Toro1999}
{Toro}, E. 1999, {Riemann Solvers and Numerical Methods for Fluid Dynamics},
  Vol.~10 ({Springer-Verlag Berlin Heidelberg}), 1038--1051

\bibitem[{Uritsky {et~al.}(2010)Uritsky, Pouquet, Rosenberg, Mininni, \&
  Donovan}]{Uritsky2010}
Uritsky, V.~M., Pouquet, A., Rosenberg, D., Mininni, P.~D., \& Donovan, E.~F.
  2010, Physical Review E - Statistical, Nonlinear, and Soft Matter Physics,
  82, 1

\bibitem[{{Valdivia} {et~al.}(2017){Valdivia}, {Godard}, {Hennebelle}, {Gerin},
  {Lesaffre}, \& {Le Bourlot}}]{Valdivia2017}
{Valdivia}, V., {Godard}, B., {Hennebelle}, P., {et~al.} 2017, \aap, 600, A114

\bibitem[{{White} \& {Rees}(1978)}]{WhiteRees1978}
{White}, S.~D.~M. \& {Rees}, M.~J. 1978, \mnras, 183, 341

\bibitem[{Yang {et~al.}(2015)Yang, Zhang, He, Tu, Wang, Marsch, Wang, Zhang, \&
  Feng}]{Yang_2015}
Yang, L., Zhang, L., He, J., {et~al.} 2015, The Astrophysical Journal, 809, 155

\bibitem[{{Zhdankin} {et~al.}(2016){Zhdankin}, {Boldyrev}, \&
  {Chen}}]{Zhdankin2016}
{Zhdankin}, V., {Boldyrev}, S., \& {Chen}, C. H.~K. 2016, \mnras, 457, L69

\bibitem[{{Zhdankin} {et~al.}(2014){Zhdankin}, {Boldyrev}, {Perez}, \&
  {Tobias}}]{Zhdankin2014}
{Zhdankin}, V., {Boldyrev}, S., {Perez}, J.~C., \& {Tobias}, S.~M. 2014, \apj,
  795, 127

\bibitem[{{Zhdankin} {et~al.}(2015){Zhdankin}, {Uzdensky}, \&
  {Boldyrev}}]{Zhdankin2015}
{Zhdankin}, V., {Uzdensky}, D.~A., \& {Boldyrev}, S. 2015, \prl, 114, 065002

\bibitem[{{Zhdankin} {et~al.}(2013){Zhdankin}, Uzdensky, Perez, \&
  Boldyrev}]{Zhdankin2013}
{Zhdankin}, V., Uzdensky, D.~A., Perez, J.~C., \& Boldyrev, S. 2013,
  Astrophysical Journal, 771

\bibitem[{{Zweibel} \& {Brandenburg}(1997)}]{Zweibel1997}
{Zweibel}, E.~G. \& {Brandenburg}, A. 1997, \apj, 478, 563

\bibitem[{{Zweibel} \& {Yamada}(2016)}]{Zweibel2016}
{Zweibel}, E.~G. \& {Yamada}, M. 2016, Proceedings of the Royal Society of
  London Series A, 472, 20160479

\end{thebibliography}

\begin{appendix}

\section{Steady-state 1D MHD shocks}
\label{sec:steady-state-MHD-shocks}

Here we consider the internal structure of a steady-state isothermal planar MHD shock. We can always operate a Galilean transformation to place ourselves in the frame moving along with the shock, so that the pre-shock velocity is along the normal of the working surface, which we define as the first space coordinate $x$. In addition, we can rotate this frame along the normal so that the second space coordinate $y$ is along the pre-shock transverse magnetic field, and so both third components $z$ of the magnetic field and the velocity are zero along the whole shock (thanks to the co-planarity property within shocks: this would not be the case in a rotational discontinuity). 

We write $u$ and $\varv$ for both the first and second coordinates of the velocity in this frame, and similarly we write $B_x$ and $B_y$ the coordinates of the magnetic field (orthogonal to the working surface and transverse). We finally write $\rho$ the mass density and $x$ the first space coordinate.

With these notations, the isothermal conservation of mass, momentum and magnetic field become:
\begin{align}
0=\,&\partial_x (\rho u)\\
0=\,&\partial_x[\rho u^2+\rho c^2+\frac{1}{8\pi} B_y^2+\frac43\mu\partial_x u]\\
0=\,&\partial_x[\rho u \varv+\frac{1}{4\pi}B_yB_x+\mu \partial_x \varv]\\
0=\,&\partial_x B_x\\
0=\,&\partial_x [uB_y-\varv B_x+\eta \partial_x B]
\end{align}
where we introduced the dynamical viscosity $\mu=\rho\nu$ and the resistivity $\eta$ coefficients as well as the isothermal sound speed $c$. We now affect subscripts 0 to the pre-shock quantities (except for the orthogonal magnetic field $B_x$ which is constant throughout the shock).
The mass conservation becomes $\rho u=\rho_0 u_0$. We define the quantity $a=B_y/\sqrt{4\pi\rho_0}$ which has the dimension of a velocity, and similarly the constant Alfvén speed $a_x=B_x/\sqrt{4\pi\rho_0}$ to arrive at the system of ordinary differential equations :
\begin{align}
\frac43 \frac{\mu}{\rho_0} \partial_x u=\,& u_0+\frac{c^2}{u_0}-(u+\frac{c^2}{u})+\frac12(\frac{a_0^2}{u_0}-\frac{a^2}{u_0})\\
\frac{\mu}{\rho_0} \partial_x \varv=\,& a_x\frac{a-a_0}{u_0}-\varv\\
\eta \partial_x a=\,&ua-u_0a_0-a_x \varv
\end{align}
to compute the internal structure of isothermal MHD shocks. 
  
The isothermal dynamical coefficient $\mu$ is a constant, but in the current application, we used a constant viscous coefficient $\nu$, so that $\mu=\nu \rho_0 u_0/u$. The typical viscous length scale of our simulated shocks is hence $\nu/u_0$. 
One can further simplify the above system by using non-dimensional quantitites $\tilde{x}=xu_0/\nu$, $\tilde{u}=u/u_0$, $\tilde{\varv}=\varv/u_0$, $\tilde{a}=a/a_0$ and $\mathcal{P}_{\rm m}=\nu/\eta$:

\begin{align}
\frac{4}{3\tilde{u}} \partial_{\tilde{x}} \tilde{u}=\,& 1-\tilde{u}+\mathcal{M}_s^{-2}(1-\frac{1}{\tilde{u}})+\frac12\mathcal{M}_a^{-2}(1-\tilde{a}^2) \\
\frac{1}{\tilde{u}} \partial_{\tilde{x}} \tilde{\varv}=\,& \tilde{u}\tilde{a}-1-\tilde{\varv}\\
 \partial_{\tilde{x}} \tilde{a}=\,&\mathcal{P}_{\rm m}[ \mathcal{M}_a^{-2}(\tilde{a}-1)\frac{a_x}{a_0}-\tilde{\varv}]
\end{align}
which shows that the intrinsinc structure of our shocks depends essentially on three non-dimensional parameters in the pre-shock: the sonic Mach number $\mathcal{M}_s=u_0/c$, the transverse Alfv\'enic Mach number $\mathcal{M}_a=u_0/a_0$ and the tangent of the angle of the magnetic field with respect to the shock working surface $a_x/a_0$. 

This system of ordinary differential equations (ODEs) can be integrated numerically between the preshock (at $\tilde{u}=\tilde{a}=1$ and $\tilde{\varv}=0$) and the post-shock. The stability analysis towards increasing $\tilde{x}$ of these two steady points yields three growing or decaying eigenvectors. We find fast shocks usually have three unstable eigenvectors at the pre-shock while they have three stable eigenvectors at the post-shock: one can simply integrate the system of ODEs from the post-shock to the pre-shock from a small perturbation of the post-shock opposite to the most stable eigenvector (which is the most unstable one in the direction of decreasing $\tilde{x}$). We find slow shocks usually have two unstable eigenvectors at the pre-shock while they have two stable eigenvectors at the post-shock: the solution leaves the pre-shock from its unstable plane, and reaches the post-shock in a stable plane. To find the solution, we use a boundary value solver with a request to be on both these planes at a small given distance from the two corresponding end points. 

We use the resulting solutions as reference models to benchmark the results of the dedicated experiments which are described in the section below. 

\section{Numerical dissipation in Godunov methods}
\label{sec:numerical-dissipation}
We report here on the method we used to recover the amount of numerical
dissipation present in our compressible simulations, and how we validated it using the above isothermal MHD shocks. 

In our compressible simulations, we adopt twice the resolution
of corresponding incompressible runs which we performed with pseudo-spectral methods in \cite{Momferratos2014}: 1024 vs 512, for the same dissipation
coefficients (viscosity and resistivity). Indeed, there is a common
belief that grid based methods need twice as many elements to obtain
a resolving power equivalent to Fourier elements. However, we will see that even in this case, the numerical scheme still affects considerably the dissipation in the code. 

\subsection*{Experimental set-up}
In order to check and control the dissipation in our configuration, we run 1D planar isothermal magnetized shocks with various resolutions, and compare them to the solutions devised in the previous section. To set up the computational experiments of this section, we first compute the Rankine-Hugoniot conditions for a magnetised shock in the frame of the shock and we set up initially the pre-shock and post-shock conditions in two halves of the computational box, with the jump in the middle. The outer box boundaries are inflow and outflow conditions on each side of the pre-shock and post-shock material, respectively. As the computation proceeds, the initial discontinuity smears out due to both numerical and physical dissipation but the discontinuity does not move in space thanks to the chosen set up. A steady-state is quickly reached, which we compare to semi-analytical solutions of the steady state as described in the previous subsection. 

\subsection*{Viscous spread in the shock}
Shocks have a viscous spread on the order of $\lambda_{v}=\nu/u_0$ (see section \ref{sec:steady-state-MHD-shocks}).
Our 3D simulations of decaying turbulence with $1024^{3}$ pixels have
box length $L_{{\rm box}}=2\pi$ and viscosity $\nu=0.7\times10^{-3}$ and
so the pixel size is nearly 9 times bigger than the viscous length for a $u_0=1$ shock: the viscous and resistive spread throughout these shocks is realised essentially by the grid. We showed in \cite{CHEMSES} that the  number of zones necessary to fully resolve the viscous spread of isothermal shocks is at least on the order of the Reynolds number $L.u_0/\nu\simeq9000$, way above what we can afford for a 3D computation. 



\subsection*{Dissipation natures}

 There are several sources of dissipation in our simulations: viscous and Ohmic dissipation due to the physical terms we have introduced in DUMSES, and numerical dissipation intrinsic to the scheme. Our main purpose is to locally recover the total amount of dissipation $\varepsilon_\mathrm{tot}$ produced by both the scheme and the physical dissipation terms. We design here several methods to  retrieve $\varepsilon_\mathrm{tot}$, by considering variants of the energy conservation equation.

\subsection*{Method 1}

Consider the evolution equation of kinetic and magnetic energy:

\begin{equation}
    \partial_{t}(\frac{1}{2}\rho u^{2}+\frac{1}{8\pi}B^{2})+\boldsymbol{\nabla.{\mathcal F}_{1}}+\boldsymbol{u.\nabla}(p)=-\varepsilon_\mathrm{tot} \label{eq:method1}
\end{equation}\

where $\varepsilon_\mathrm{tot}$ is the total irreversible heating and where the flux ${\mathcal F}_{1}$
reads:

\begin{equation}
    \boldsymbol{{\mathcal F}_{1}}=\mathbf{u}(\frac{1}{2}\rho u^{2})+\frac{1}{4\pi}\boldsymbol{(B\times u)\times B}-\nu\boldsymbol{S.u}+\eta\boldsymbol{J\times B}.
\end{equation}

We compute the left hand side of equation (\ref{eq:method1}) along
a replay of a time step of the simulation, using the flux estimates of each dissipative
half-step for the resistive and viscous contributions to $\boldsymbol{{\mathcal F}_{1}}$
and using a Lax-Friedrichs estimate for its non-dissipative part (evaluated
within the Godunov step). We estimate $\boldsymbol{u.\nabla}(p)$
at the middle of the time step thanks to the same TVD (total variation
diminishing) slopes used in the Godunov step. Finally, we recover
$\varepsilon_\mathrm{tot}$ simply by taking the opposite of the left hand side.

\subsection*{Method 2}

\begin{equation}
\label{eq:method2}
    \partial_{t}(\frac{1}{2}\rho u^{2}+\frac{1}{8\pi}B^{2}+p\log\rho)+\boldsymbol{\nabla.{\mathcal F}_{2}}=-\varepsilon_\mathrm{tot}
\end{equation}

where 

\begin{equation}
    \boldsymbol{{\mathcal F}_{2}}=\boldsymbol{{\mathcal F}_{1}}+\mathbf{u}p(\log\rho+1).
\end{equation}

We compute the flux as in method 1 (the additional contribution is
computed in the Godunov step using a Lax-Friedrichs estimate). This
method has the advantage that we recover exactly the total heating through the computational domain when we average the local resulting heating.

\subsection*{Method 3}

\begin{equation}
\partial_{t}(\frac{1}{2}\rho u^{2}+\frac{1}{8\pi}B^{2})+\boldsymbol{\nabla.{\mathcal F}_{3}}-p\boldsymbol{\nabla.u}=-\varepsilon_\mathrm{tot}
\end{equation}
where 
\begin{equation}
\boldsymbol{{\mathcal F}_{3}}=\boldsymbol{{\mathcal F}_{1}}+\mathbf{u}p.
\end{equation}
We evaluate $-p\boldsymbol{\nabla.u}$ as $\boldsymbol{u.\nabla}(p)$ in method 1, and retrieve $\varepsilon_\mathrm{tot}$ as in the previous two methods. 

\subsection*{Benchmark and comparison}

We checked that the implementations of the three methods on our shock experiments yield the same local total dissipation rate to within less than 1\% of the peak dissipation. This gives us confidence in our implementation of the three methods. We also checked on two actual snapshots of our simulations (ABC and OT runs after one turnover) that the statistics of the three methods are nearly identical for the distribution of positive values for the retrieved dissipation $\varepsilon_\mathrm{tot}$. However method 2 yields significantly less pixels with negative values, presumably because this methods does not require an estimate for terms like $p\boldsymbol{\nabla.u}$ and $-\boldsymbol{u.\nabla}(p)$ which are not divergences of fluxes. We further note that the means of method 2 and 3 are really close to one another (by less than 0.5\% of the standard deviation of $\varepsilon_\mathrm{tot}$), while 1 and 2 are a bit further apart (by less than 3\% of the standard deviation, though). We therefore adopt method 2 as the best compromise between methods 1, 2 and 3.

\subsection*{Numerical convergence}

\begin{figure}
\protect\includegraphics[scale=0.5]{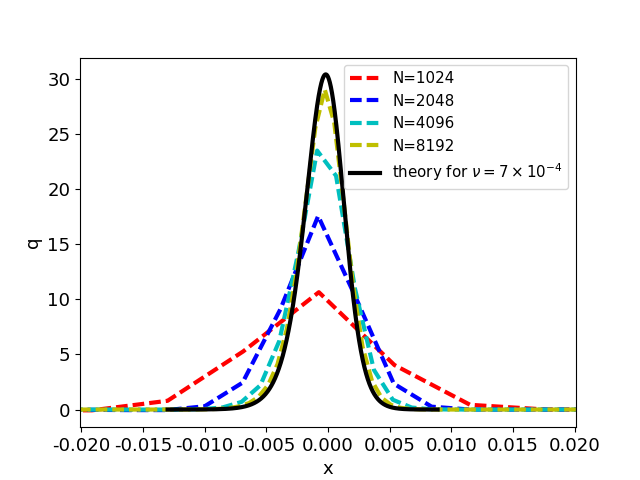}
\caption{\label{fig:shock_convergence} Dimension-less dissipation in a steady-state fast shock (with dimension-less parameters $u_0=1$, $B_x/\sqrt{4\pi}=0.2$, $B^y_0/\sqrt{4\pi}=0.3$, and $c=0.25$ with $\eta=\nu=0.7 \times 10^{-3}$) for various resolutions (dashed colored lines, $N$ is the number of pixels) compared to the analytical solution from the previous section (solid line).}
\end{figure}

Figure \ref{fig:shock_convergence} shows  the irreversible heating rates in
non-dimensional units at a close-up of a fast shock front. It illustrates the convergence of the total dissipation rate profile for increasing resolutions. We integrated the total dissipation across the shock and checked it matched the theoretical value obtained by computing the difference of the flux ${\mathcal F}_{2}$ between pre-shock and post-shock values. The integral of the total dissipation rate across the shock is thus always preserved. The effect of the resolution is only to smear out the dissipation profile without changing its total amount. 

Figure \ref{fig:shock_convergence} is similar to figure A2 of \cite{CHEMSES}, but here for magnetised isothermal shocks instead of hydrodynamic adiabatic shocks. 
It demonstrates that the resolution convergence for the heating rate is very slow and fully obtained only for $N=8192$ (see the dashed lines approaching the black solid line on Fig. \ref{fig:shock_convergence}).
The situation corresponding to our 3D simulations is the red
curve ($N=1024$): the viscous heating is largely underestimated and
spread out by about a factor 3. 

Note that for a factor two larger velocity, the analytical solution yields a twice thinner dissipation peak, so that the numerical spread would be even larger compared to what it should be. Had we used a constant dynamical viscous coefficient $\mu$, the viscous spread would respond to density in addition to velocity, and the situation would be even worse on the dense side of the shock, or for shocks penetrating denser material.
Finally, note that such a slow convergence rate (at most 30\% better accuracy for each doubling of the resolution) could lure an unaware numericist into thinking his/her simulations are converged...

\subsection*{Dissipative coefficients fit}

\begin{figure}
\protect\includegraphics[scale=0.5]{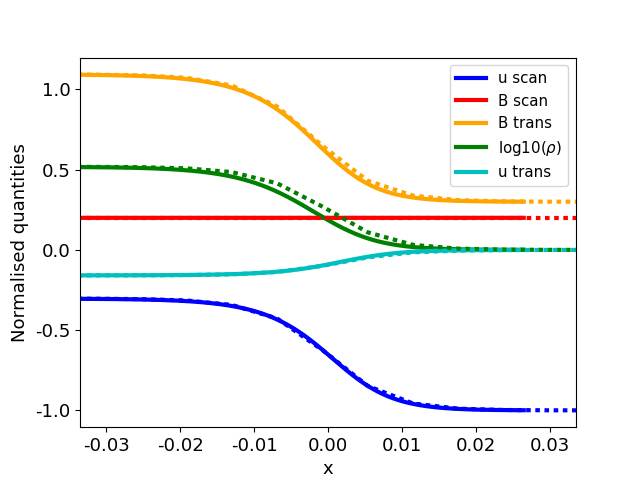}
\caption{\label{fig:shock_fit} 
Comparison of the profiles of various state variables of the gas for the same fast shock as Fig. \ref{fig:shock_convergence} between the results of our simulation at $N=1024$ (dotted lines) and the best-fit model (solid line). Best fit coefficients are  $\nu=2.2 \times 10^{-3}$ and $\eta=1.7 \times 10^{-3}$  (input coefficients are $\eta=\nu=0.7 \times 10^{-3}$).}
\end{figure}

We fit viscous isothermal MHD shock models of section \ref{sec:steady-state-MHD-shocks} to the velocity and magnetic field
profiles, and we recover best fit values for the viscosity and the resistivity coefficients which allow us to retrieve the effective viscous and resistive coefficients of our numerical scheme in the case of magnetised shocks. This is a complementary method to what \citet{lesaffre_balbus_2007} proposed for non-linear Alfvén waves. 

Figure \ref{fig:shock_fit} shows the comparison between the semi-analytical models of the previous section for the best fit $\eta$ and $\nu$ and the actual profile for the same shock as in Fig. \ref{fig:shock_convergence} and a resolution of $N=1024$ pixels. Note that the density is not as accurate as the other variables, and so we discarded it from the fit to retain only the velocity and magnetic fields components. This is because the mass flux conservation $\rho  u$ is estimated at interfaces, and the extrapolation of $\rho$ and $u$ which are one increasing while the other is decreasing makes it worse for the product. On the other hand, all other conserved quantities have a product of quantities either both increasing or decreasing, which renders the extrapolation more accurate for the product.

\begin{figure}
\protect\includegraphics[scale=0.5]{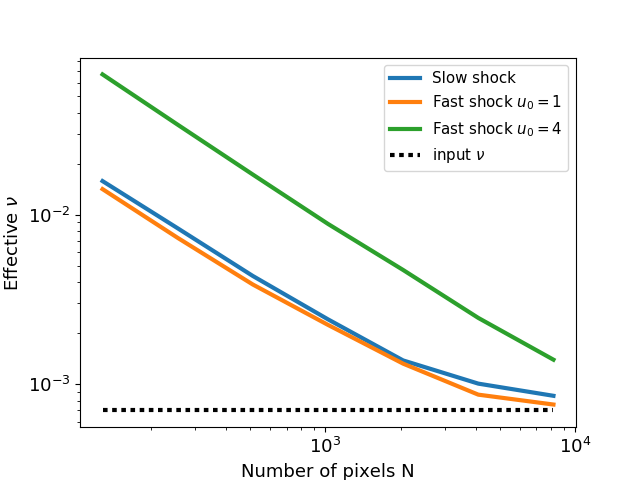}
\caption{\label{fig:shock_nu} Comparison between the fitted $\nu$ and the input $\nu$ (dotted black line) for various resolutions and three different shocks. A slow shock ($u_0=0.8$, $B_x/\sqrt{4\pi}=1$, $B^y_0/\sqrt{4\pi}=0.2$), a fast shock ($u_0=1$, $B_x/\sqrt{4\pi}=0.2$, $B^y_0/\sqrt{4\pi}=0.3$, same as Fig. \ref{fig:shock_fit}) and another fast shock 4 times faster ($u_0=4$, $B_x/\sqrt{4\pi}=0.2$, $B^y_0/\sqrt{4\pi}=0.3$).}
\end{figure}

We show on Fig. \ref{fig:shock_nu} an exploration of the effective viscosity thus recovered when varying the resolution. The effective viscosity tends to the actual input value when the resolution increases, which illustrates in an independent way the numerical convergence explored in the previous subsection. Because faster shocks have a smaller viscous spread, the effective viscosity is larger for faster shocks, with a required resolution proportional to the entrance velocity of the shock.
The type (slow or fast) of the shock does not seem to affect much the effective diffusivity of the scheme. At poor resolution, the effective viscosity is inversely proportional to the zone number. Our chosen resolution $N=1024$ corresponds to the end of this linear relation between resolution and scheme diffusion: higher resolution would yield a relatively lower increase of accuracy.

\begin{figure}
\protect\includegraphics[scale=0.5]{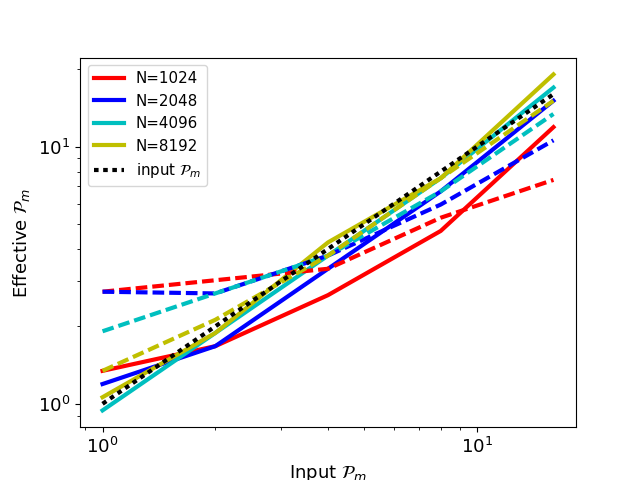}
\caption{\label{fig:shock_Pm} Comparison between the effective $\mathcal{P}_m$ and the input $\mathcal{P}_m$ (dotted black line) for various resolutions and the two fast shocks of figure \ref{fig:shock_nu} (solid lines for $u_0=1$ and dashed lines for $u_0=4$). $\mathcal{P}_m$ was varied by keeping the value of the resistive coefficient $\eta=0.7\times10^{-3}$ fixed while varying the value of the viscous coefficient accordingly $\nu=\eta \mathcal{P}_m$.}
\end{figure}

We also explored the capacity of the scheme to account for various Prandtl numbers $\mathcal{P}_m$ by increasing the viscous coefficient with respect to the resistive coefficient. Because the scheme increases the diffusivity, the overall span for the Prandtl number is not as wide as for the input physical value. The situation is even worse for the larger velocity shocks, but a resolution of 1024 pixels still allows to probe a comfortable range of $\Pm$.  Slow shocks at $\mathcal{P}_m>1$ are not sensitive to the Prandtl number, and so they could not be used to probe its effective value due to the numerical scheme. This is because when $\mathcal{P}_m>1$ in slow shocks, the magnetic fields profiles are dominated by the kinetic to magnetic energy transfers as the resistive terms becomes negligible.

\subsection*{Ohmic vs. viscous dissipation}

\begin{figure}
\protect\includegraphics[scale=0.5]{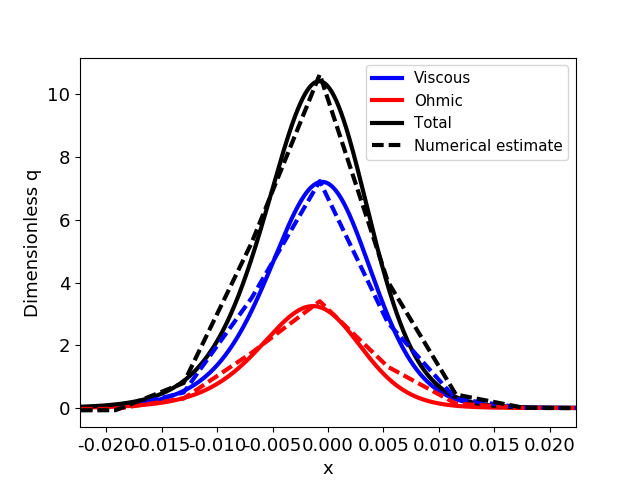}
\caption{\label{fig:shock_dissipation} Our numerical estimation (dashed) of the total (black) Ohmic (red) and viscous (blue) dissipation in the same fast shock as Fig. \ref{fig:shock_convergence} compared to the actual quantities in the best-fit model (solid lines with corresponding colors). Note that the correct share between Ohmic and viscous dissipation relies on the effective $\mathcal{P}_m$ being close to the actual input value of $\mathcal{P}_m$, so that the method we use is worse for larger velocity shocks.}
\end{figure}


Although thanks to our method we gained access to the total numerical
dissipation, we could not find an accurate way to separate the numerical
dissipation of magnetic fields from the numerical dissipation of kinetic
energy. In order to compute corrected values for the viscous heating
and the Ohmic heating, we simply shared between each of them the total
numerical heating in proportion to their relative physical values,
namely:
$\varepsilon_{v}^{{\rm corr.}}=\varepsilon_\mathrm{tot}\varepsilon_{v}/(\varepsilon_{v}+\varepsilon_{\eta})$
for viscous dissipation and conversely for Ohmic dissipation $\varepsilon_{\eta}^{{\rm corr.}}=\varepsilon_\mathrm{tot}\varepsilon_{\eta}/(\varepsilon_{v}+\varepsilon_{\eta})$. Whenever our estimate for the purely numerical dissipation is negative (i.e: $\varepsilon_\mathrm{tot}<\varepsilon_{v}+\varepsilon_{\eta}$),
we simply set $\varepsilon_{v}^{{\rm corr.}}=\varepsilon_{v}$ and
$\varepsilon_{\eta}^{{\rm corr.}}=\varepsilon_{\eta}$.
We then compute the viscous and Ohmic heatings in the best fit shock model and compare them to the above estimation on figure \ref{fig:shock_dissipation}. 

\subsection*{Summary }
We control the implementation of our dissipation rate recovery method by comparing several variants of it and we benchmark them against analytical solutions.
We find that we are able to recover the total dissipated energy within a localised shock to a very good precision. We use the benchmark models to estimate the diffusivity of the scheme and we find that the effective viscosity and resistivity are both enhanced due to lack of resolution, especially for the large velocity shocks. As a result, the effective Prandtl number is also affected. On
the other hand, the slow convergence to the shock solution justifies
our use of a moderate resolution associated to this dissipation recovering method:
we would not gain much by running our simulations at twice the current resolution, while we would have to increase the resolution more than ten-fold to dispense ourselves with this dissipation recovery method.

\subsection*{Prospects}
The numerically acute reader will have noted that our method is currently limited to Lax-Friedrichs implementations of the Riemann solver. Other Riemann solvers require that we design how to incorporate the additional components of the fluxes $\mathcal{F}_i$. We checked slow and fast shocks, and they seem equally well treated at equivalent velocities. But we did not check the effective diffusivities in rotational discontinuities (although non-linear Alfvén waves such as used in \citet{lesaffre_balbus_2007} may provide a good guess) or Parker sheets (these require at least a two-dimensional treatment).
\end{appendix}

%
%

\end{document}